\documentclass[aps,prd,twocolumn,showpacs,superscriptaddress,floatfix]{revtex4-2}
\usepackage{graphicx}
\usepackage{dcolumn}
\usepackage{bm}
\usepackage{hyperref}
\usepackage{orcidlink}
\usepackage{comment}
\usepackage{lineno}
\usepackage{placeins}
\usepackage{amsmath,amsfonts}
\usepackage{longtable}
\usepackage{multirow}
\hypersetup{
    colorlinks=true,
    linkcolor=blue,
    citecolor=blue,
    urlcolor=blue
}

\newcommand{\strutlike}{\rule{0pt}{1.1\normalbaselineskip}}

\begin{document}

\title{First double-differential cross-section measurements in proton multiplicity and kinematics for mesonless $\nu_{\mu}$ charged-current interactions on argon using the MicroBooNE detector}

\newcommand{\ANL}{Argonne National Laboratory (ANL), Lemont, IL, 60439, USA}
\newcommand{\Bern}{Universit{\"a}t Bern, Bern CH-3012, Switzerland}
\newcommand{\BNL}{Brookhaven National Laboratory (BNL), Upton, NY, 11973, USA}
\newcommand{\UCSB}{University of California, Santa Barbara, CA, 93106, USA}
\newcommand{\Cambridge}{University of Cambridge, Cambridge CB3 0HE, United Kingdom}
\newcommand{\CIEMAT}{Centro de Investigaciones Energ\'{e}ticas, Medioambientales y Tecnol\'{o}gicas (CIEMAT), Madrid E-28040, Spain}
\newcommand{\Chicago}{University of Chicago, Chicago, IL, 60637, USA}
\newcommand{\Cincinnati}{University of Cincinnati, Cincinnati, OH, 45221, USA}
\newcommand{\CSU}{Colorado State University, Fort Collins, CO, 80523, USA}
\newcommand{\Columbia}{Columbia University, New York, NY, 10027, USA}
\newcommand{\Edinburgh}{University of Edinburgh, Edinburgh EH9 3FD, United Kingdom}
\newcommand{\FNAL}{Fermi National Accelerator Laboratory (FNAL), Batavia, IL 60510, USA}
\newcommand{\Granada}{Universidad de Granada, Granada E-18071, Spain}
\newcommand{\IIT}{Illinois Institute of Technology (IIT), Chicago, IL 60616, USA}
\newcommand{\ICL}{Imperial College London, London SW7 2AZ, United Kingdom}
\newcommand{\Indiana}{Indiana University, Bloomington, IN 47405, USA}
\newcommand{\Kansas}{The University of Kansas, Lawrence, KS, 66045, USA}
\newcommand{\KSU}{Kansas State University (KSU), Manhattan, KS, 66506, USA}
\newcommand{\Lancaster}{Lancaster University, Lancaster LA1 4YW, United Kingdom}
\newcommand{\LANL}{Los Alamos National Laboratory (LANL), Los Alamos, NM, 87545, USA}
\newcommand{\Louisiana}{Louisiana State University, Baton Rouge, LA, 70803, USA}
\newcommand{\Manchester}{The University of Manchester, Manchester M13 9PL, United Kingdom}
\newcommand{\MIT}{Massachusetts Institute of Technology (MIT), Cambridge, MA, 02139, USA}
\newcommand{\Michigan}{University of Michigan, Ann Arbor, MI, 48109, USA}
\newcommand{\MSU}{Michigan State University, East Lansing, MI 48824, USA}
\newcommand{\Minnesota}{University of Minnesota, Minneapolis, MN, 55455, USA}
\newcommand{\Nankai}{Nankai University, Nankai District, Tianjin 300071, China}
\newcommand{\NMSU}{New Mexico State University (NMSU), Las Cruces, NM, 88003, USA}
\newcommand{\NotreDame}{University of Notre Dame, Notre Dame, IN, 46556, USA}
\newcommand{\Oxford}{University of Oxford, Oxford OX1 3RH, United Kingdom}
\newcommand{\Pitt}{University of Pittsburgh, Pittsburgh, PA, 15260, USA}
\newcommand{\QMUL}{Queen Mary University of London, London E1 4NS, United Kingdom}
\newcommand{\Rutgers}{Rutgers University, Piscataway, NJ, 08854, USA}
\newcommand{\SLAC}{SLAC National Accelerator Laboratory, Menlo Park, CA, 94025, USA}
\newcommand{\SDSMT}{South Dakota School of Mines and Technology (SDSMT), Rapid City, SD, 57701, USA}
\newcommand{\Maine}{University of Southern Maine, Portland, ME, 04104, USA}
\newcommand{\TelAviv}{Tel Aviv University, Tel Aviv, Israel, 69978}
\newcommand{\UTA}{University of Texas, Arlington, TX, 76019, USA}
\newcommand{\Tufts}{Tufts University, Medford, MA, 02155, USA}
\newcommand{\VTech}{Center for Neutrino Physics, Virginia Tech, Blacksburg, VA, 24061, USA}
\newcommand{\Warwick}{University of Warwick, Coventry CV4 7AL, United Kingdom}


\affiliation{\ANL}
\affiliation{\Bern}
\affiliation{\BNL}
\affiliation{\UCSB}
\affiliation{\Cambridge}
\affiliation{\CIEMAT}
\affiliation{\Chicago}
\affiliation{\Cincinnati}
\affiliation{\CSU}
\affiliation{\Columbia}
\affiliation{\Edinburgh}
\affiliation{\FNAL}
\affiliation{\Granada}
\affiliation{\IIT}
\affiliation{\ICL}
\affiliation{\Indiana}
\affiliation{\Kansas}
\affiliation{\KSU}
\affiliation{\Lancaster}
\affiliation{\LANL}
\affiliation{\Louisiana}
\affiliation{\Manchester}
\affiliation{\MIT}
\affiliation{\Michigan}
\affiliation{\MSU}
\affiliation{\Minnesota}
\affiliation{\Nankai}
\affiliation{\NMSU}
\affiliation{\NotreDame}
\affiliation{\Oxford}
\affiliation{\Pitt}
\affiliation{\QMUL}
\affiliation{\Rutgers}
\affiliation{\SLAC}
\affiliation{\SDSMT}
\affiliation{\Maine}
\affiliation{\TelAviv}
\affiliation{\UTA}
\affiliation{\Tufts}
\affiliation{\VTech}
\affiliation{\Warwick}

\author{P.~Abratenko\,\orcidlink{0000-0001-6945-5941}}\affiliation{\Tufts} 
\author{D.~Andrade~Aldana\,\orcidlink{0009-0008-3143-3374}} \affiliation{\IIT}\affiliation{\LANL}
\author{J.~Asaadi\,\orcidlink{0000-0001-6915-5279}}   \affiliation{\UTA}
\author{A.~Ashkenazi\,\orcidlink{0000-0002-1995-3851}}   \affiliation{\TelAviv}
\author{S.~Balasubramanian} \affiliation{\FNAL}
\author{B.~Baller\,\orcidlink{0000-0001-8731-9281}}  \affiliation{\FNAL}

\author{A.~Barnard\,\orcidlink{0000-0001-6117-1768}} \affiliation{\Oxford}

\author{G.~Barr\,\orcidlink{0000-0002-9763-1882}} \affiliation{\Oxford}
\author{D.~Barrow\,\orcidlink{0000-0001-5844-709X}} \affiliation{\Oxford}
\author{J.~Barrow\,\orcidlink{0000-0002-7319-3339}} \affiliation{\Minnesota} 
% ...

\collaboration{The MicroBooNE Collaboration}
\thanks{microboone\_info@fnal.gov}\noaffiliation

\begin{abstract}
    We present the first double-differential cross-section measurements of muon neutrino charged-current interactions on argon with no mesons and one or more protons ($N_{p} \geq 1$) in the final state as a function of proton multiplicity and kinematics. The analysis uses the full Booster Neutrino Beam dataset collected by the MicroBooNE detector at Fermi National Accelerator Laboratory, corresponding to a total of $1.30 \times 10^{21}$ protons on target. We measure cross sections as a function of the proton multiplicity, the momenta and muon-proton opening angles of the leading and subleading protons up to and including the fourth-leading proton, as well as the total available kinetic energy of the hadronic system. The treatment of the statistical and systematic uncertainties includes correlations between the different extracted distributions. The measured cross sections are compared to predictions from several event generators using different primary and final-state interaction models. The results reveal strong sensitivity to the modeling of final-state interactions, and none of the generators tested provides a fully adequate description of the data. Significant disagreement between data and predictions is observed in the proton multiplicity distributions, particularly at backward opening angles and for subleading proton kinematics.
\end{abstract}

\maketitle

\section{Introduction}
\label{sec:introduction}

Accelerator-based neutrino experiments are poised to address several of the most significant open questions in neutrino physics in the near future. The Deep Underground Neutrino Experiment (DUNE) \cite{DUNE:2020lwj} aims to make precision measurements of neutrino oscillation parameters, including the determination of the neutrino mass ordering and the search for leptonic charge-parity (CP) violation \cite{DeSalas:2018rby,Branco:2011zb}. The Short-Baseline Neutrino (SBN) program \cite{Machado:2019oxb} will investigate the short-baseline anomalies that may indicate the existence of sterile neutrino species \cite{Acero:2022wqg,Diaz:2019fwt}. Both of these programs use liquid argon time projection chambers (LArTPCs) \cite{Rubbia:1977zz,Majumdar:2021llu} as their primary detector technology, which provides detailed imaging of neutrino interaction final states. The success of both experiments relies on the accurate modeling of neutrino-argon interactions, as the associated uncertainties are among the leading sources of systematic error in their analyses. Correct reconstruction of the hadronic system in neutrino interactions, in particular the number and kinematics of final-state protons, is essential for accurate neutrino energy estimation and event classification. However, these are notoriously challenging to predict, especially for subleading protons. Among the relevant interaction channels, charged-current (CC) muon neutrino scattering producing no mesons and one or more protons (denoted as $\text{CC} 0\pi Np$) is the dominant topology at neutrino energies less than $1 ~ \mathrm{GeV}$, and remains an important channel at the higher energies relevant to the DUNE program. The absence of pions in the final state makes this channel particularly well-suited for direct measurements of the proton kinematics.

The final-state proton multiplicity in $\text{CC} 0\pi Np$ events is determined by several physics processes. Different primary interaction mechanisms, such as quasi-elastic scattering (QE), resonance production (RES), and deep inelastic scattering (DIS), lead to different numbers of protons exiting the nucleus after a neutrino interaction. Multi-nucleon emission mechanisms, such as meson-exchange currents (MEC), involve the neutrino interacting with correlated nucleons rather than a single one, ejecting them together and adding multiple nucleons to the final state. The initial multiplicities are further modified by final-state interactions (FSI), which govern how protons and other hadrons propagate through the nuclear medium. Differences in the modeling of these mechanisms across event generators can lead to significant discrepancies in predicted proton multiplicities. Previous measurements of this channel on argon have revealed tension between data and generator predictions, particularly when looking at correlations between kinematic distributions \cite{MicroBooNE:2024yzp}. However, the measurements did not distinguish between different proton multiplicities, leaving the role of FSI and multi-nucleon processes in shaping the hadronic final state largely unexplored.

We present the first double-differential cross-section measurements of $\text{CC} 0\pi Np$ interactions on argon as a function of proton multiplicity and leading and subleading proton kinematics using the full MicroBooNE dataset of $1.30 \times 10^{21}$ protons on target (POT) collected from the Booster Neutrino Beam (BNB) \cite{Stancu:2001cpa}. The analysis targets observables directly sensitive to proton production mechanisms, including the momenta and opening angles of up to and including the fourth-leading proton, as well as the total hadronic available energy. Compared to the previous MicroBooNE $\text{CC} 0\pi Np$ measurement \cite{MicroBooNE:2024yzp}, this work introduces proton multiplicity as the primary observable, uses less stringent requirements on the outgoing muon that result in higher selection efficiency, and benefits from nearly double the beam exposure. The unfolded cross sections are compared to predictions from several generators, including multiple FSI model variations within GENIE, to directly probe the sensitivity to the cascade treatment. These measurements provide a unique benchmark for testing and refining the modeling of final-state interactions and multi-nucleon processes in neutrino event generators, with direct relevance for the DUNE and SBN physics programs.

\section{Experimental setup}
\label{subsec:experimental_setup}

\subsection{Detector and beamline}
\label{subsec:detector}

The MicroBooNE detector \cite{MicroBooNE:2016pwy} is a surface LArTPC that operated at Fermilab from 2015 to 2021. It is situated on-axis relative to the BNB \cite{Stancu:2001cpa} $463 ~ \mathrm{m}$ downstream of the target. The detector has an active mass of $85$ metric tons of liquid argon within its field cage. It is instrumented with a three-plane wire anode with a $3 ~ \mathrm{mm}$ pitch. Two induction planes are placed at an angle $\pm 60^{\circ}$ from the vertical, while a third collection plane is vertically oriented. Behind the wire planes, an array of photomultiplier tubes with fast timing resolution serves as the light collection system.

Neutrinos from the BNB are produced by colliding an $8 \, \mathrm{GeV}$ proton beam with a beryllium target. The generated charged mesons are focused by means of a magnetic horn and subsequently decay in flight, producing neutrinos. After the decay region, a hadron absorber made of concrete and steel removes the undecayed particles. The flux at the MicroBooNE detector is dominated by muon neutrinos, with a mean energy of $800 ~ \mathrm{MeV}$. The beam also has a small muon antineutrino component, corresponding to $5.9\%$ of the beam content, as well as a combined $0.6\%$ of electron neutrinos and antineutrinos.

\subsection{Simulation}
\label{subsec:sim}

Monte Carlo (MC) simulations are used in MicroBooNE to develop the event selection, efficiency estimation, background characterization, and systematic uncertainty evaluation.

The process starts with the BNB flux prediction \cite{MiniBooNE:2008hfu}, which is based on Geant4 beamline simulations \cite{GEANT4:2002zbu,Allison:2006ve,Allison:2016lfl} developed for the MiniBooNE experiment and adapted to the MicroBooNE location. Neutrino interactions in the MicroBooNE detector are simulated using the GENIE neutrino event generator \cite{Andreopoulos:2009rq}, version 3.0.6. The baseline interaction model used is based on the \texttt{G18\_10a\_02\_11a} configuration of GENIE. This model configuration is tuned to external $\text{CC}0\pi$ data from the T2K experiment \cite{T2K:2016jor}. The simulated events are reweighted to adjust four CCQE and CCMEC cross-section model parameters, resulting in a prediction that agrees better with the MicroBooNE $\nu_{\mu}$ CC dataset. Further details on the GENIE MicroBooNE tune can be found in Ref.~\cite{MicroBooNE:2021ccs}.

The events generated with GENIE according to the predicted BNB flux are propagated through the MicroBooNE detector geometry using Geant4 \cite{GEANT4:2002zbu,Allison:2006ve,Allison:2016lfl}. LArSoft \cite{Snider:2017wjd} is the event-processing framework adopted by MicroBooNE for its readout simulation and reconstruction. A number of custom modules in LArSoft simulate the response of the electronics to the ionizing particles exiting the neutrino interaction. Effects such as light attenuation, electron diffusion, and electric field distortions caused by positive ions are included here \cite{MicroBooNE:2017qiu,MicroBooNE:2018swd,MicroBooNE:2018vro,MicroBooNE:2019koz,MicroBooNE:2020kca,MicroBooNE:2021icu}.

All the generated MC samples employ an overlay technique, in which simulated neutrino events from GENIE are superimposed onto actual beam-off data samples taken when the BNB was not in operation \cite{MicroBooNE:2018vxr}. This method ensures that event-by-event fluctuations in cosmic backgrounds and noise levels are realistically represented in the analysis. These beam-off samples are also used to estimate the rate of cosmic ray and other non-beam correlated backgrounds.


\subsection{Reconstruction}
\label{subsec:reco}

The reconstruction paradigm used for this analysis is the Pandora multi-algorithm pattern-recognition toolkit \cite{Marshall:2015rfa,MicroBooNE:2017xvs}. It uses topological and calorimetric information to reconstruct particles in the detector and identify neutrino interaction candidates. It aggregates the reconstructed hits from the waveforms in each of the wire planes into 2D clusters, which are then correlated across planes to form 3D particle candidates. These particles are grouped into slices, clusters of spatially and temporally correlated activity. Timing information is used to select neutrino candidates from coincidence with the BNB beam window.

The Pandora outputs include a reconstructed neutrino vertex and a particle hierarchy associated with it. Particles are assigned a track score that distinguishes shower-like from track-like objects. This is used to help reject background events containing electromagnetic showers, such as those from neutral pion decays. The reconstruction is complemented with additional algorithms for calorimetric calibration \cite{MicroBooNE:2019efx}, a log-likelihood ratio particle identification (PID) score used to separate muons from protons \cite{MicroBooNE:2021ddy}, and muon momentum estimation based on multiple Coulomb scattering (MCS) \cite{MicroBooNE:2017tkp}.

\section{Analysis methods}
\label{sec:methods}

\subsection{Signal definition}
\label{subsec:signal_definition}

This analysis focuses on charged-current muon neutrino interactions on argon producing one or more protons and no mesons in the final state. We require the momentum of the primary muon to be above $0.10 ~ \mathrm{GeV}/c$. The momentum of the leading final-state proton must lie within the interval $[0.25, \, 1.00] ~ \mathrm{GeV}/c$. Subleading protons are considered detectable if their momentum is above the $0.25 ~ \mathrm{GeV}/c$ threshold. Events in which the leading proton momentum exceeds $1.00  ~ \mathrm{GeV}/c$ are excluded from the signal definition, regardless of whether additional protons fall within the accepted range.

These phase-space limits are applied to select regions with good signal purity and resolution. The muon threshold of $0.10 ~ \mathrm{GeV}/c$ is motivated by a $10.0 ~ \mathrm{cm}$ track length requirement for the efficient selection of muon candidates. The $0.25 ~ \mathrm{GeV}/c$ lower threshold for the proton follows selection efficiency and energy reconstruction bias considerations. The $1.00 ~ \mathrm{GeV}/c$ upper limit on the proton momentum ensures their reliable identification, as high-momentum protons tend to be misclassified as muon-like.

The signal definition closely follows that of previous MicroBooNE $\text{CC} 0\pi Np$ analyses \cite{MicroBooNE:2020akw, MicroBooNE:2024yzp}. However, since we do not report cross sections as a function of muon momentum, we accept events with exiting muons despite their reduced momentum resolution compared to contained ones. Therefore, an upper limit on muon momentum is not used.

In the following, simulated events are divided into different categories based on their generator-level information:
\begin{itemize}
    \item[-] \textbf{Signal}: Events satisfying all signal requirements, with a neutrino vertex contained within the fiducial volume (FV) defined in Sec.~\ref{subsec:event_selection}. Signal events are further subdivided either by their true topology (i.e., the number of protons within the momentum range $[0.25, \, 1.00] ~ \mathrm{GeV}/c$ in the final state) into $1p$, $2p$, or $Mp$ with $M \geq 3$, or by the primary interaction mode: QE, MEC, RES, or other. The last subcategory is dominated by DIS events.
    \item[-] \textbf{Out FV}: Events with a true interaction vertex outside the FV. This category includes events from all different topologies which fail the FV cut.
    \item[-] $\bm{\nu_{\mu}}$ \textbf{CC}$\bm{0\pi 0p}$: Muon neutrino charged-current interaction events with no leading proton within the required momentum limits. Events with no protons at all, or leading protons below or above the specified range, are included.
    \item[-] $\bm{\nu_{\mu}}$ \textbf{CC}$\bm{N\pi}$: Muon neutrino charged-current interaction events with any number of charged or neutral pions in the final state.
    \item[-] \textbf{Other} $\bm{\nu_{\mu}}$ \textbf{CC}: Any other muon neutrino charged-current interaction events. This includes events containing mesons other than pions, or with a primary muon momentum below $0.10 ~ \mathrm{GeV}/c$.
    \item[-] $\bm{\nu_{e}}$ \textbf{CC}: Electron neutrino charged-current interaction events.
    \item[-] \textbf{NC}: Neutral-current neutrino or antineutrino interaction events.
	\item[-] \textbf{Beam-off}: Background events induced by cosmic rays.
	\item[-] \textbf{Other}: Any background events which do not fall into another category.
\end{itemize}

\subsection{Event selection}
\label{subsec:event_selection}

The selection of signal events uses the products of the automated Pandora reconstruction framework. It consists of two main steps: an inclusive $\nu_{\mu}$ CC preselection \cite{VanDePontseele:2020tqz} and the specific $\mathrm{CC} 0\pi Np$ selection part. The event selection strategy is shared with the previous $\text{CC} 0\pi Np$ analysis from Ref.~\cite{MicroBooNE:2024yzp}. However, because this analysis focuses on proton multiplicity and kinematics instead of muon-proton correlations, muon containment and quality cuts (the latter requiring agreement within $25\%$ between range-based and MCS-based muon momentum estimates \cite{MicroBooNE:2024yzp}) are not applied, consistent with the relaxed muon momentum requirements described in Sec.~\ref{subsec:signal_definition}. This choice results in a significantly higher selection efficiency. The same selection criteria are applied to both data and simulated events.

\begin{figure*}
  \begin{minipage}[b]{0.47\linewidth}
    \includegraphics[width=\linewidth]{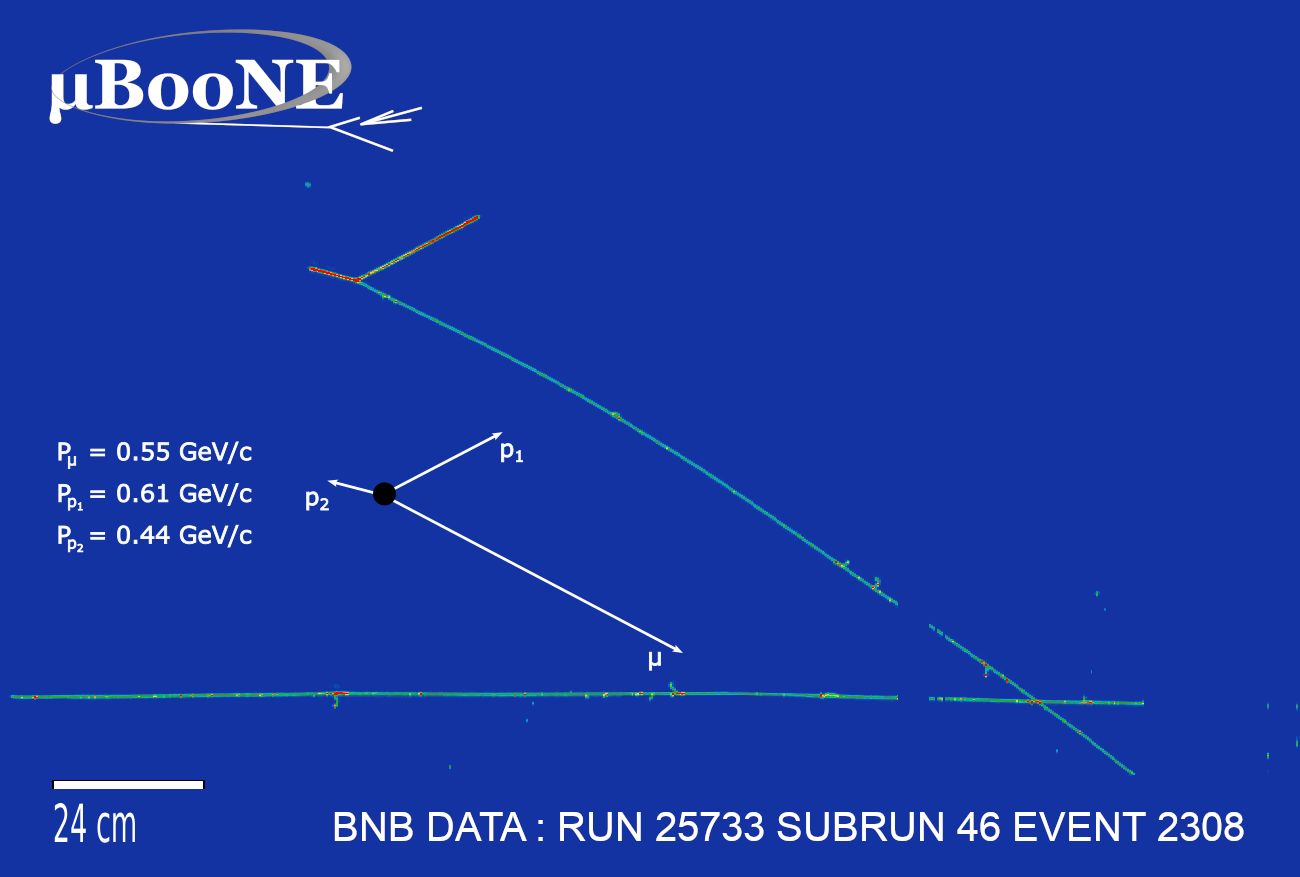}
    \centerline{(a)}
  \end{minipage}
  \hfill
  \begin{minipage}[b]{0.47\linewidth}
    \includegraphics[width=\linewidth]{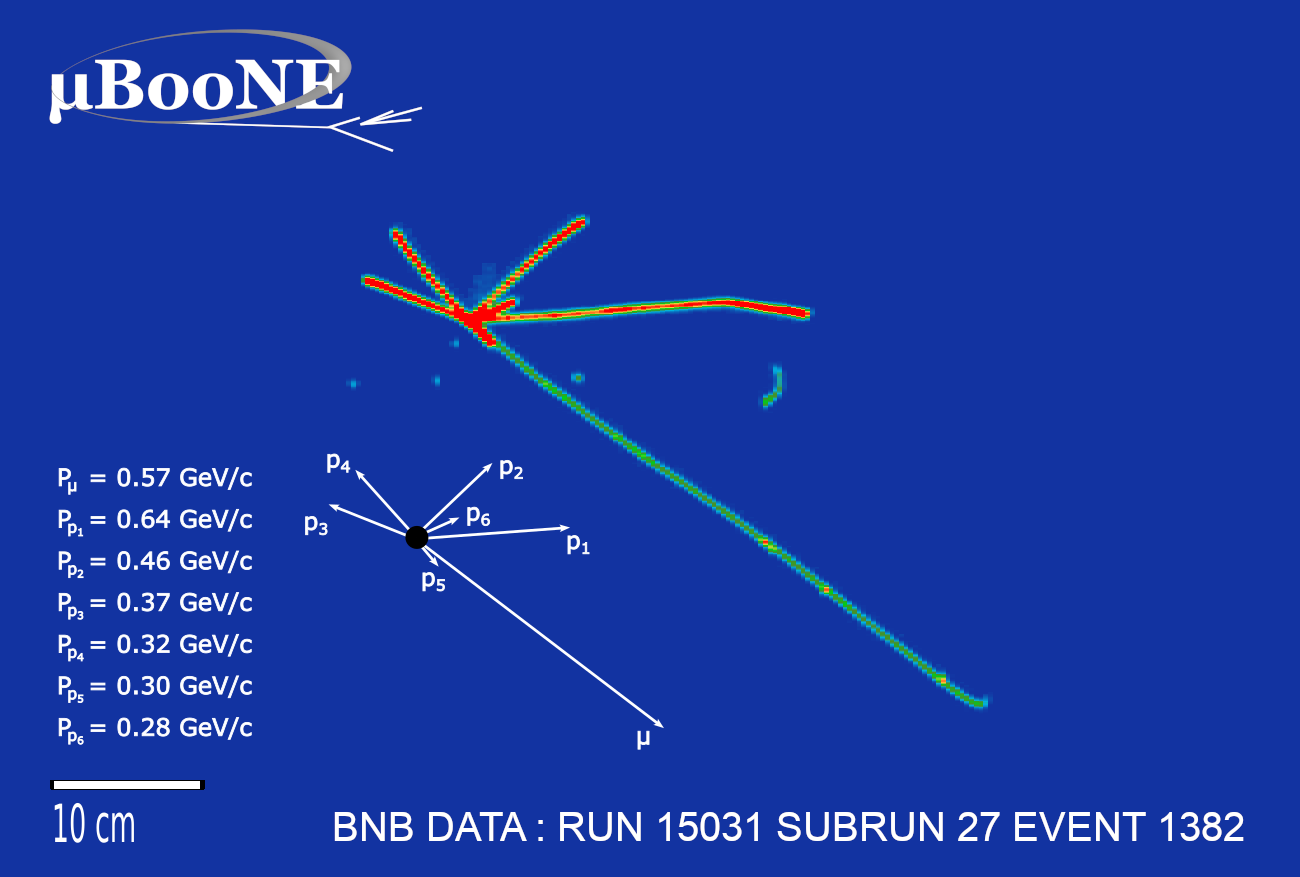}
    \centerline{(b)}
  \end{minipage}
  \caption{Event displays for two selected events: (a) candidate event with 2 reconstructed protons, and (b) candidate event with 6 reconstructed protons.}
  \label{fig:event_displays}
\end{figure*}

The preselection tries to classify valid neutrino events through the identification of a primary muon candidate. First, the reconstructed neutrino vertex is required to be contained within the fiducial volume. The fiducial volume in the analysis is a rectangular prism, chosen in such a way that all of its sides are $21.5 ~ \mathrm{cm}$ inward of the active volume edges, except for the downstream $z$ direction where this distance is increased to $70.0 ~ \mathrm{cm}$ to improve the acceptance of forward-going muons. Next, all the primary neutrino interaction products are required to have their starting points inside a looser containment volume, $10.0 ~ \mathrm{cm}$ inward with respect to the active volume.

Additional cosmic rejection is achieved by applying a cut on the topological score, an event-level variable computed by a support-vector machine trained to classify neutrino interaction candidates as neutrino-like or cosmic-like \cite{VanDePontseele:2020tqz}. Events with a topological score below $0.1$ are rejected.

The final step in the preselection is identifying a viable muon candidate. This is achieved in four steps, starting from all the direct products of the neutrino interaction in the events passing the previous selection cuts. The following particle-level selection requirements are applied sequentially to identify the muon candidates:
\begin{enumerate}
    \item The track score must be higher than $0.8$ to reject shower-like particles.
    \item Discard particles with a distance between the neutrino vertex and their start position greater than $4.0 ~ \mathrm{cm}$.
    \item Require a track length greater than $10.0 ~ \mathrm{cm}$, which excludes short tracks such as low-energy protons and cosmic rays.
    \item Proton- and muon-like particles are separated by requiring the log-likelihood ratio PID score \cite{MicroBooNE:2021ddy} to be greater than 0.2.
\end{enumerate}
For each event, the particle passing all the selections and with the highest PID score is taken to be the muon candidate.

In order to isolate the desired topology, namely one or more protons and no mesons in the final state, a number of additional requirements are applied. To match the phase-space limit imposed on the signal definition, the reconstructed muon candidate momentum must be greater than $0.1 ~ \mathrm{GeV}/c$. The muon momentum is estimated either from range or the MCS method, depending on whether the muon is contained or not.

Events containing shower-like particles are rejected by requiring all reconstructed particles to have a track score greater than $0.5$, suppressing backgrounds from primary electrons and neutral pions. All particles passing this requirement, excluding the muon candidate, are considered proton candidates. At least one proton candidate must be present, and all proton candidate endpoints must lie within the containment volume defined above.

An additional log-likelihood ratio PID score cut is applied to the proton candidates. A value less than $0.2$ allows for good acceptance of low-energy protons, which tend to have PID scores near zero. Finally, the leading proton candidate's momentum, estimated using a range-based kinetic energy reconstruction, must lie within the range $[0.25, \,1.00] ~ \mathrm{GeV}/c$.

Events passing the inclusive $\nu_{\mu}$ CC preselection and the additional selection criteria are taken to be CC$0\pi Np$ candidates, the signal event category for this measurement.

\begin{table}[!h]
	\caption{Evolution of the number of selected events, efficiency, and purity as the selection criteria are applied to the simulation sample.}
	\begin{center}
        \begin{small}
			\begin{tabular}{cccc}
                \parbox{6em}{\centering Selection cut} & \parbox{4em}{\centering Events} & \parbox{4em}{\centering Efficiency} & \parbox{4em}{\centering Purity} \\[1mm] \hline
                \strutlike No cuts                     &                         5365313 &                               1.000 &                           0.026 \\[1mm]
                In FV                                  &                          477172 &                               0.789 &                           0.230 \\[1mm]
                Start points contained                 &                          437360 &                               0.729 &                           0.231 \\[1mm]
                Topological cut                        &                          274544 &                               0.654 &                           0.330 \\[1mm]
                $\nu_{\mu}$ CC incl.                   &                          210832 &                               0.591 &                           0.389 \\[1mm]
                $\mu$ phase-space limits               &                          210793 &                               0.591 &                           0.389 \\[1mm]
                No showers                             &                          157058 &                               0.522 &                           0.461 \\[1mm]
                Has $p$ candidate                      &                          126591 &                               0.473 &                           0.518 \\[1mm]
                $p$ contained                          &                          113051 &                               0.449 &                           0.552 \\[1mm]
                $p$ PID cut                            &                           73149 &                               0.398 &                           0.755 \\[1mm]
                $p$ phase-space limits                 &                           71342 &                               0.393 &                           0.764
                \end{tabular} 
        \end{small}
	\end{center}
	\label{tab:selection_metrics}
\end{table}

\begin{figure*}[!t]
	\centering
	\includegraphics[width=0.99\linewidth]{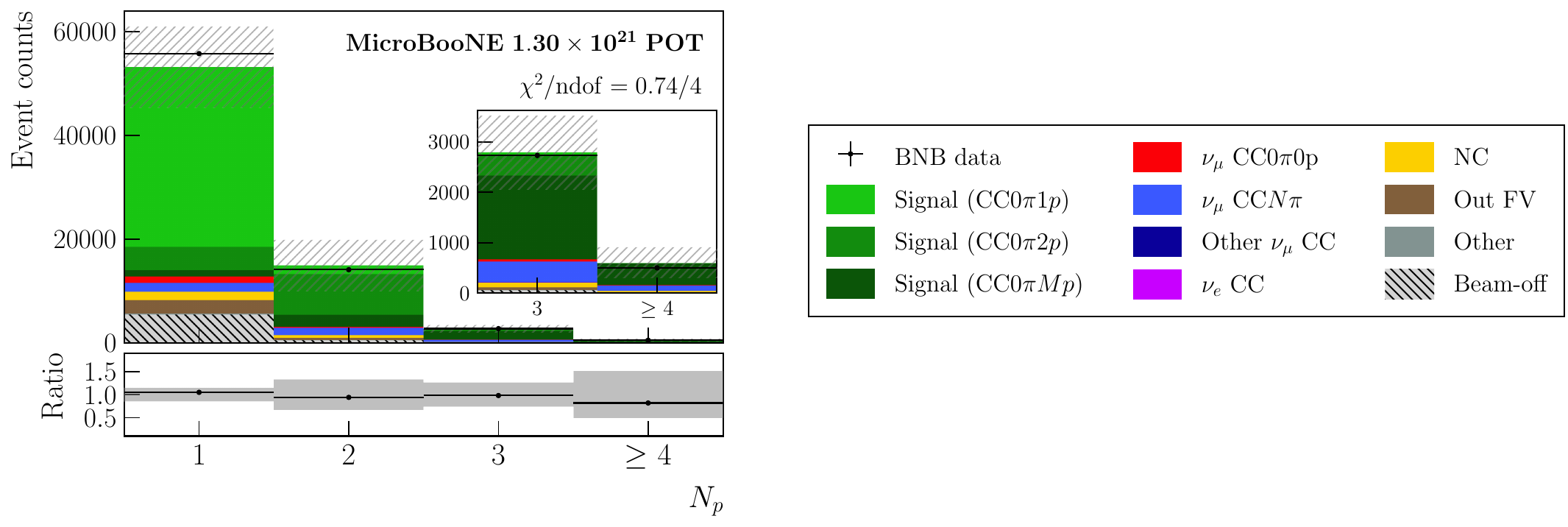}
	\caption{Reconstructed proton multiplicity event rate distribution. The total uncertainty on the prediction is indicated by the hatched boxes. The bottom panel shows the ratio of the observed data to the central-value prediction, with the gray band indicating the fractional uncertainty. The inset shows an enlarged view of the $N_{p} = 3$ and $N_{p} \geq 4$ bins.}
	\label{fig:event_rates_mult_collapsed}
\end{figure*}

\begin{figure*}[!t]
	\centering
	\includegraphics[width=0.91\linewidth]{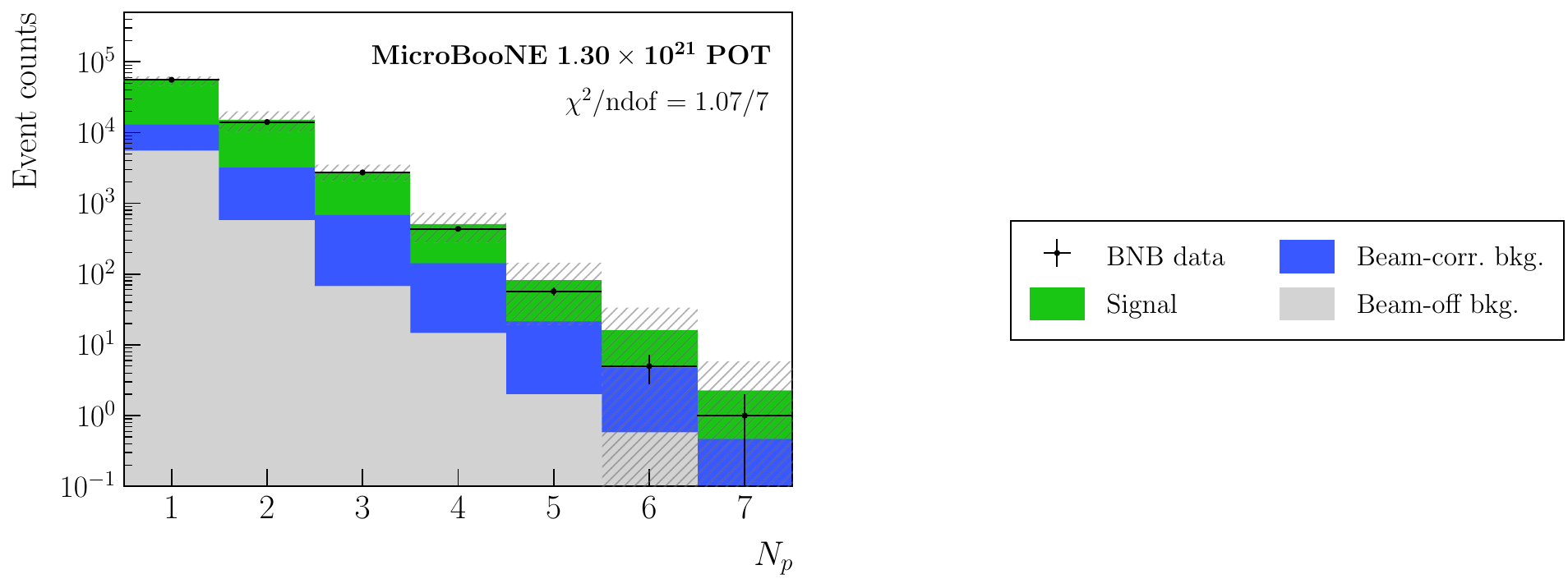}
	\caption{Reconstructed proton multiplicity event rate distribution shown on a logarithmic scale, with every multiplicity bin resolved individually. The signal is shown as a single category and the background is split into beam-correlated and beam-off contributions. The total uncertainty on the prediction is indicated by the hatched boxes.}
	\label{fig:event_rates_mult_log_merged}
\end{figure*}

Using this selection, we obtain a purity of $76.4\%$ and a selection efficiency of $39.3\%$. A step-by-step summary of the number of selected events, the selection efficiency and purity achieved is presented in Table \ref{tab:selection_metrics}. Figure \ref{fig:event_displays} shows two signal candidate events selected from the MicroBooNE BNB dataset, with 2 and 6 reconstructed protons.

Compared to the previous MicroBooNE CC$0\pi Np$ analysis from Ref.~\cite{MicroBooNE:2024yzp} the selection efficiency of $39.3\%$ is significantly higher than the $12.3\%$ obtained in that work, as a result of the more relaxed muon requirements described above. In addition, the present analysis benefits from the full MicroBooNE dataset of $1.30 \times 10^{21}$ POT, compared to the $6.79 \times 10^{20}$ POT used in the previous measurement.

As a first look at the selected sample, we compare the GENIE MicroBooNE tune prediction to the collected BNB data. The selected data sample contains $73058$ events, comprising $55740$ $1p$, $14087$ $2p$, $2732$ $3p$, and $499$ $\geq 4p$ events. Figure \ref{fig:event_rates_mult_collapsed} shows the reconstructed proton multiplicity distribution, and Fig.~\ref{fig:event_rates_mult_log_merged} shows the same distribution on a logarithmic scale with each multiplicity bin resolved separately. The data points include only statistical uncertainties, while the hatched boxes indicate the total uncertainty on the prediction (MC simulation and beam-off statistical plus systematic contributions, described in Sec.~\ref{subsec:uncertainty}). The different event categories shown are described in Sec.~\ref{subsec:signal_definition}.

\subsection{Measurement strategy}
\label{subsec:observables}

\begin{figure*}[htbp]
  \centering

  \begin{minipage}[b]{\linewidth}
    \centering
    \includegraphics[width=\linewidth]{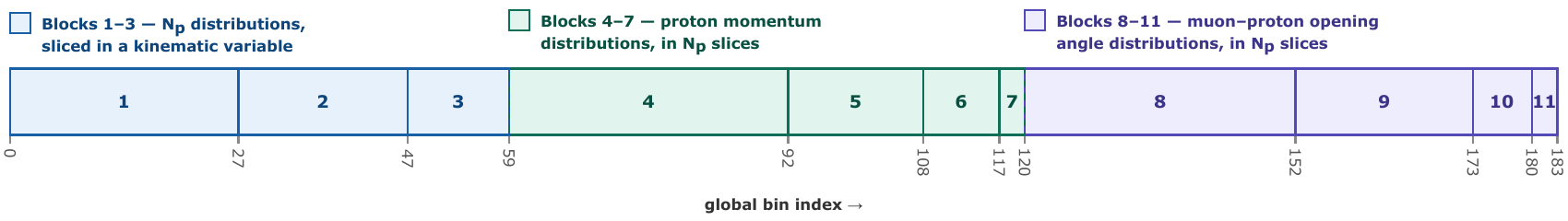}
    \centerline{(a)}
  \end{minipage}

  \vspace{0.5em}

  \begin{minipage}[b]{0.32\linewidth}
    \centering
    \includegraphics[width=\linewidth]{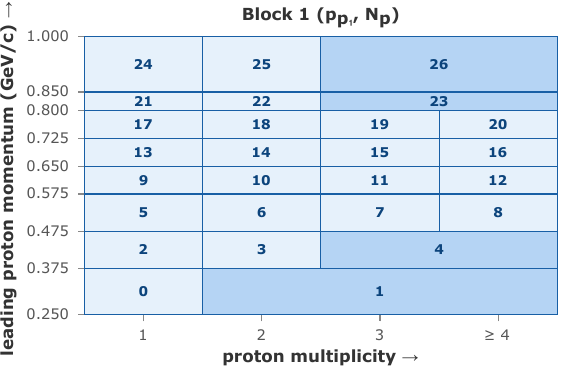}
    \centerline{(b)}
  \end{minipage}
  \hfill
  \begin{minipage}[b]{0.32\linewidth}
    \centering
    \includegraphics[width=\linewidth]{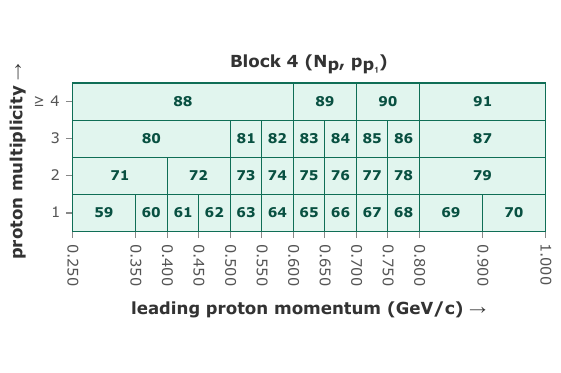}
    \centerline{(c)}
  \end{minipage}
  \hfill
  \begin{minipage}[b]{0.32\linewidth}
    \centering
    \includegraphics[width=\linewidth]{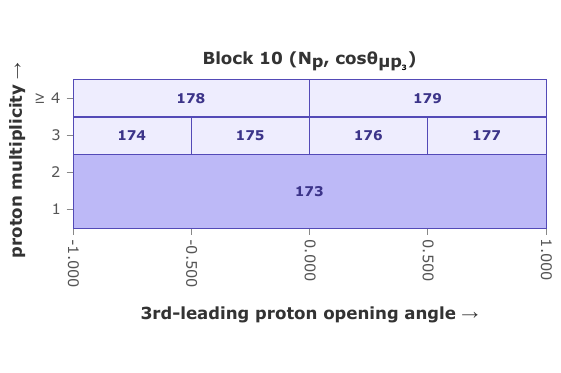}
    \centerline{(d)}
  \end{minipage}

  \caption{Schematic of the bin block structure used in the analysis. (a) The 11 blocks mapped onto the global bin index, color-coded by group: blocks 1--3 bin in proton multiplicity sliced in a kinematic variable (blue), blocks 4--7 bin in proton momentum in multiplicity slices (green), and blocks 8--11 bin in opening angle in multiplicity slices (purple). (b)--(d) Two-dimensional bin structure for three representative blocks. Darker cells indicate bins where adjacent multiplicity values are merged.}
  \label{fig:block_schematic}
\end{figure*}

We focus on observables related to the proton multiplicity and kinematics. The proton multiplicity is defined as the number of protons in the event with momentum above the $0.25 ~ \mathrm{GeV}/c$ threshold, consistent with the lower momentum threshold in the signal definition. For the kinematic variables, we study the magnitude of the momentum for the leading and subleading protons up to the 4th-leading, which we denote $p_{p_{i}} \equiv |\mathbf{p}_{p_{i}}|$ where $i=1,2,3,4$ indicates the momentum-ordered ranking of the proton. In addition, we compute the opening angles between the muon and each proton, $\mathrm{cos} \, \theta_{\mu p_{i}}$, from their three-momenta. We also measure the total hadronic available energy
\begin{equation}
    E_{\mathrm{avail}} = \sum_{j=1}^{N_{p}} T_{p_{j}},
\end{equation}
where $N_{p}$ is the proton multiplicity, and $T_{p_{j}}$ denotes the kinetic energy of the $j$th-leading proton in the event. The combination of these observables with proton multiplicity is directly sensitive to FSI and multi-nucleon processes.

The aim of the analysis is measuring double-differential cross sections in combinations of proton multiplicity, (sub)leading proton momentum, proton-muon opening angles, and available energy. This gives a total of 11 distributions. We adopt the approach proposed in Ref.~\cite{MicroBooNE:2024yzp} to report the correlations between the different measurements. By doing this we exploit the full power of the MicroBooNE multivariable dataset. This approach requires grouping all the bins representing the same kinematic distribution in bin blocks. These blocks are unfolded independently, following the procedure described in Sec.~\ref{subsec:unfolding}, while keeping the off-block-diagonal entries in the covariance for the final measurement. A detailed discussion about this blockwise method can be found in Ref.~\cite{Gardiner:2024gdy}. The first 3 blocks in the analysis represent bins of proton multiplicity sliced in leading proton momentum, opening angle, and available energy. The next 4 bin blocks contain distributions of leading, subleading, 3rd-, and 4th-leading proton momenta in proton multiplicity slices. The remaining 4 blocks contain the corresponding distributions for the proton opening angles. The overall block structure and three representative examples are illustrated in Fig. \ref{fig:block_schematic}.

The binning scheme used in the analysis is tabulated in Appendix~\ref{app:bin_defs}. This binning is used to build histograms for the observables of interest both in reconstructed and true space. The estimated mapping between the reconstructed- and true-level variables is given by the detector response matrix, which accounts for selection efficiency and reconstruction smearing effects. It is defined as
\begin{equation}\label{eq:response_matrix}
    \Delta_{i\mu} \equiv \frac{\varphi_{i\mu}}{\varphi_{\mu}},
\end{equation}
where $\varphi_{i \mu}$ is the number of expected selected signal events simultaneously falling in reconstructed bin $i$ and true bin $\mu$, while $\varphi_{\mu}$ is the total number of signal events in true bin $\mu$. The detector response matrix can be factorized to separate the efficiency and smearing contributions
\begin{equation}
    \Delta_{i\mu} = \epsilon_{\mu} \, M_{i\mu},
\end{equation}
with $\epsilon_{\mu}$ representing the selection efficiency for true bin $\mu$, and $M_{i\mu}$, the migration matrix, describing the probability that an event generated in true bin $\mu$ is reconstructed in bin $i$. Because the protons are ordered by momentum, reconstruction occasionally swaps their ranking; the leading and subleading protons are estimated to exchange rank in approximately $10\%$ of $N_{p} \geq 2$ events, contributing to the off-diagonal structure of the migration matrices. The efficiency histogram and migration matrix for each of the bin blocks are included in Sec.~II of the supplemental materials.

\subsection{Uncertainty estimation}
\label{subsec:uncertainty}

We consider the contributions of multiple sources of systematic uncertainty. Many of these are evaluated by repeatedly estimating the expected number of selected events in each bin using varied simulation parameters, under a multi-universe approach. This leads to a covariance matrix in reconstructed space given by
\begin{equation}
    \text{Cov}(n_{i}, \, n_{j}) = \frac{1}{N_{\mathrm{univ}}} \sum_{u=1}^{N_{\mathrm{univ}}} (n_{i} - n_{i}^{u}) \, (n_{j} - n_{j}^{u}),
\end{equation}
where $n_{i}$ is the number of selected events in bin $i$ according to our nominal simulation, and $n_{i}^{u}$ is the corresponding quantity obtained in the $u$th alternate universe. Here, the number of selected events per bin includes the true signal events and both beam-correlated and beam-off background events. The number of universes $N_{\mathrm{univ}}$ considered depends on the systematic effect.

A reweighting scheme based on the approach developed by the MiniBooNE collaboration \cite{MiniBooNE:2008hfu} is adopted to estimate the systematic uncertainties related to the BNB neutrino flux prediction. The alternate universes can be divided into two groups: single-universe variations affecting the geometry in the beam simulation, and hadron production modeling uncertainties.

The systematic uncertainties related to the neutrino-nucleus interaction modeling are included via a reweighting of our baseline interaction model, as described in Ref.~\cite{MicroBooNE:2021ccs}. Since the GENIE systematic variations do not cover all relevant sources of modeling uncertainty, an extra systematic uncertainty is introduced to span the difference between the MicroBooNE GENIE tune and the NuWro event generator \texttt{v19.02.2} \cite{Golan:2012rfa} in the analysis variables. Other MicroBooNE cross-section analyses \cite{MicroBooNE:2023cmw,MicroBooNE:2023tzj,MicroBooNE:2024yzp} include this additional uncertainty when working with similar mesonless signal definitions.

The contribution of these interaction systematic uncertainties enters through both the signal and background event rate estimations. For signal events, only the response matrix defined in Eq.~\eqref{eq:response_matrix} is varied and not the predicted distribution in true kinematics, as these uncertainties modify the efficiency and smearing estimates used in the final cross-section extraction. The detector response matrix is reevaluated in each alternate universe and applied to the truth-level signal events $\varphi_{\mu}$ in the central-value universe to calculate the varied number of expected selected signal events.

Reinteraction uncertainties account for the effects related to particle propagation in the detector medium. The Geant4Reweight package \cite{Calcutt:2021zck} is used to calculate the corresponding weights from varying the hadronic total cross-section models. This treatment includes proton and pion reinteraction uncertainties.

The systematic uncertainties on the modeling of the detector response are calculated using dedicated MC simulation samples, where different aspects of the LArTPC and light collection system simulations are varied. The LArTPC response variations change the electron--ion recombination model \cite{ArgoNeuT:2013kpa}, the space-charge effects, and the wire response, via data-driven transformations \cite{MicroBooNE:2021roa}. Light system variations account for the uncertainties related to the response of the photomultiplier tubes, including light yield mismodeling and decline over the operational period, and the uncertainty in the Rayleigh scattering length \cite{MicroBooNE:2026kfn}.


To account for the uncertainty on the beam exposure and the number of target nuclei in the fiducial volume, two additional normalization uncertainties are introduced. These are flat, fully-correlated $2\%$ and $1\%$ uncertainties, respectively, affecting only the signal and beam-correlated background events. Beam-off background events, being directly measured from data, are not affected by these normalization uncertainties.

Because we measure different kinematic distributions simultaneously, the statistical covariance is not a combination of independent Poisson distributions resulting in a diagonal matrix. In this case, a single event belongs to more than one reconstructed bin at the same time, as it appears in multiple distributions within the blockwise scheme. The general prescription to calculate the statistical covariance is
\begin{equation}
    \text{Cov}(n_{i}, \, n_{j}) = \sum_{k=1}^{n_{ij}} w_{k}^{2},
\end{equation}
where $n_{ij}$ is the number of selected events which simultaneously belong to the reconstructed bins $i$ and $j$ and $w_{k}$ are the event weights. For beam data and beam-off background events, $w_{k} = 1$, and the covariance reduces to $n_{ij}$. Reference \cite{Gardiner:2024gdy} discusses this approach to the statistical uncertainty estimation.

The covariance matrices are calculated for each individual source of uncertainty, and then summed to obtain the total covariance matrix.

\section{Cross-section measurement}
\label{sec:results}

\subsection{Event rate distributions}
\label{subsec:event_rates}

\begin{figure*}[!htbp]
	\centering
	\includegraphics[width=0.90\linewidth]{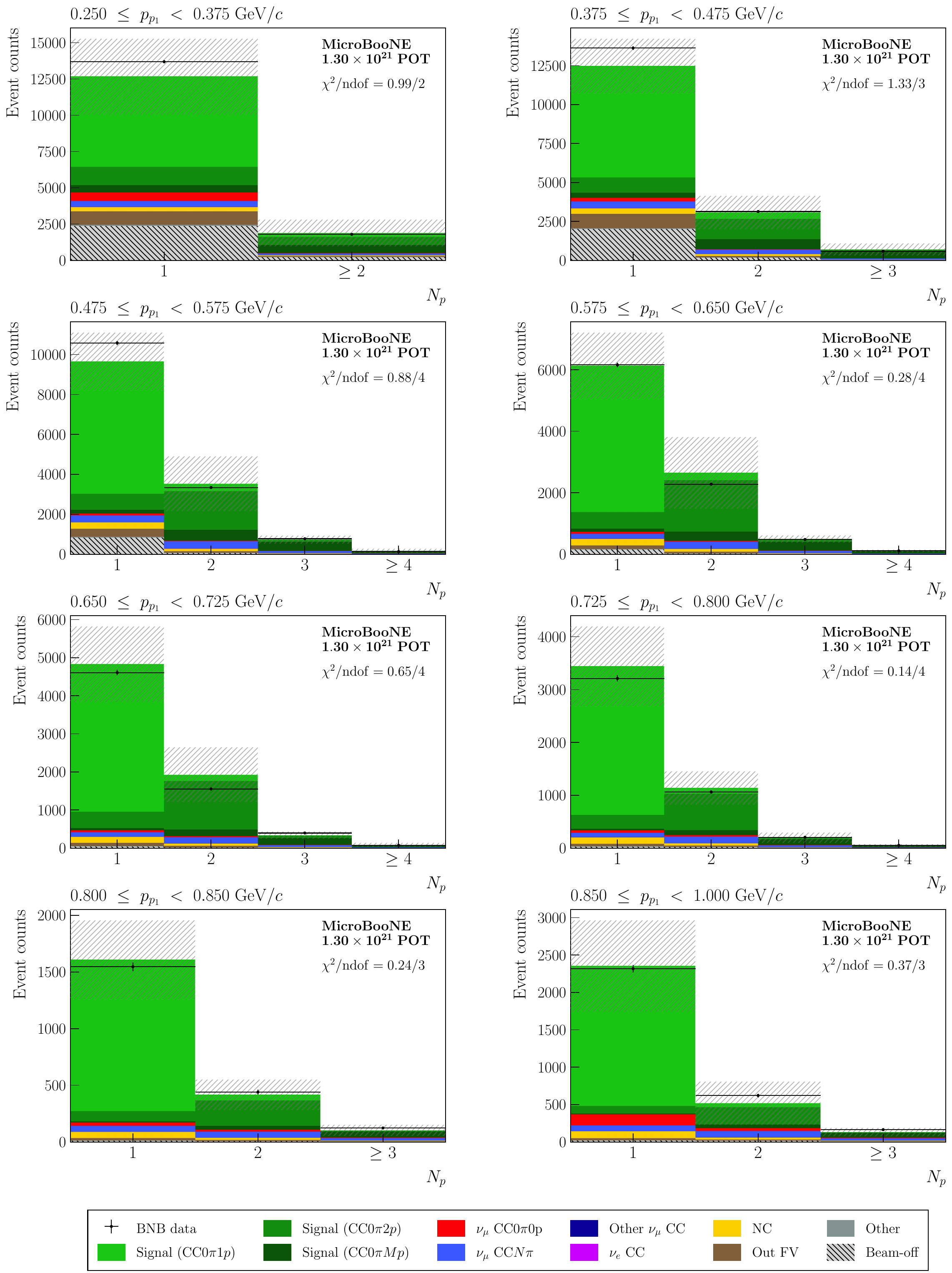}
	\caption{Reconstructed event rate distributions for block 1, corresponding to the double-differential measurement $(p_{p_{1}}, \, N_{p})$. The total uncertainty on the prediction is indicated by the hatched boxes. The per-slice $\chi^{2}/\text{ndof}$ for the central-value prediction is shown in each individual panel.}
	\label{fig:event_rates_block_1}
\end{figure*}

\begin{figure*}[!t]
	\centering
	\includegraphics[width=0.90\linewidth]{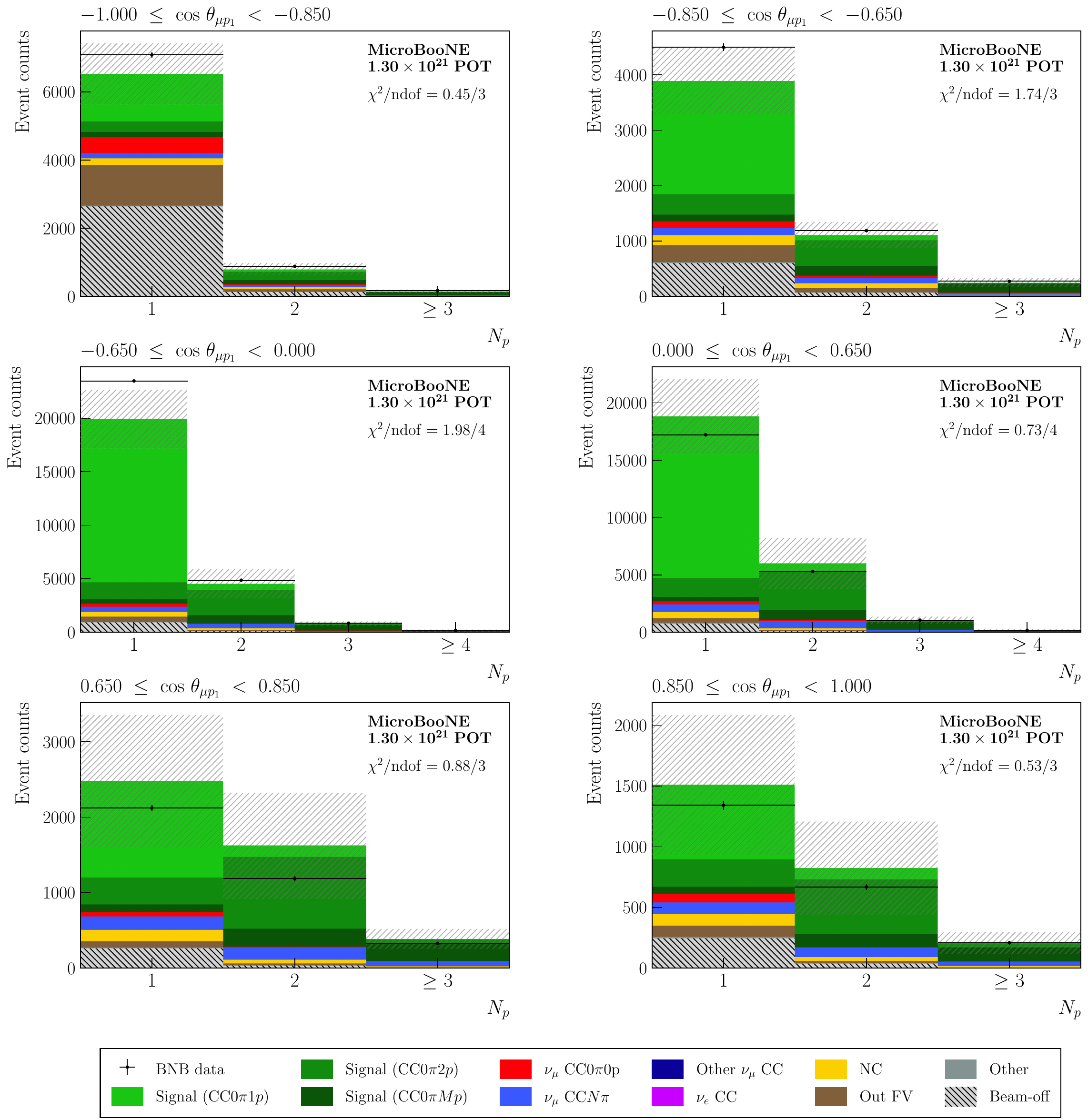}
	\caption{Reconstructed event rate distributions for block 2, corresponding to the double-differential measurement $(\mathrm{cos} \, \theta_{\mu p_{1}}, \, N_{p})$. The total uncertainty on the prediction is indicated by the hatched boxes. The per-slice $\chi^{2}/\text{ndof}$ for the central-value prediction is shown in each individual panel.}
	\label{fig:event_rates_block_2}
\end{figure*}

\begin{figure*}[!htbp]
	\centering
	\includegraphics[width=0.90\linewidth]{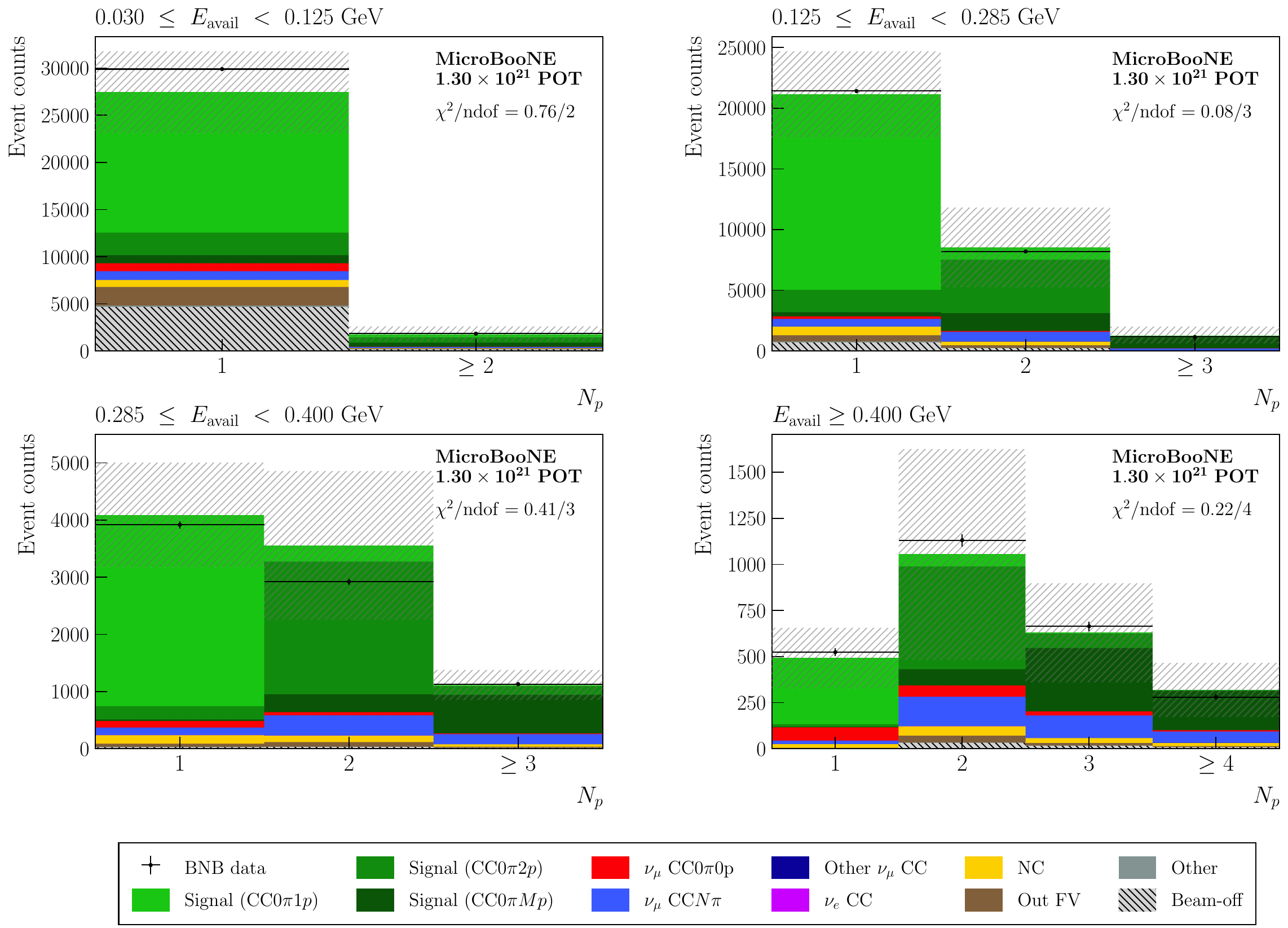}
	\caption{Reconstructed event rate distributions for block 3, corresponding to the double-differential measurement $(E_{\mathrm{avail}}, \, N_{p})$. The total uncertainty on the prediction is indicated by the hatched boxes. The per-slice $\chi^{2}/\text{ndof}$ for the central-value prediction is shown in each individual panel.}
	\label{fig:event_rates_block_3}
\end{figure*}

\begin{figure*}[!htbp]
	\centering
	\includegraphics[width=1.0\linewidth]{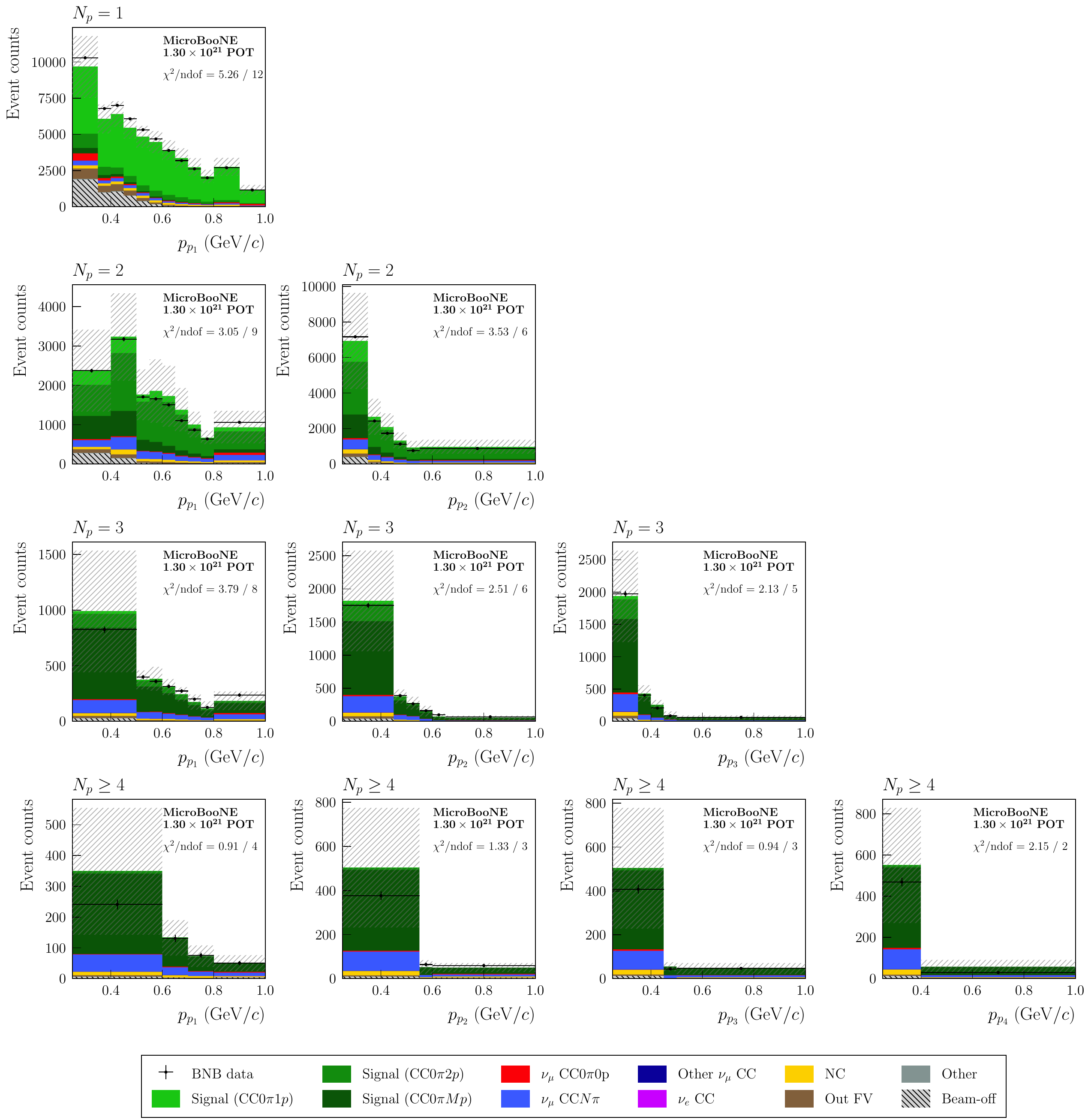}
	\caption{Reconstructed event rate distributions for blocks 4, 5, 6, and 7, corresponding to the double-differential measurements $(N_{p}, \, p_{p_{i}})$. The total uncertainty on the prediction is indicated by the hatched boxes. The per-slice $\chi^{2}/\text{ndof}$ for the central-value prediction is shown in each individual panel.}
	\label{fig:event_rates_blocks_4_5_6_7}
\end{figure*}

\begin{figure*}[!htbp]
	\centering
	\includegraphics[width=1.0\linewidth]{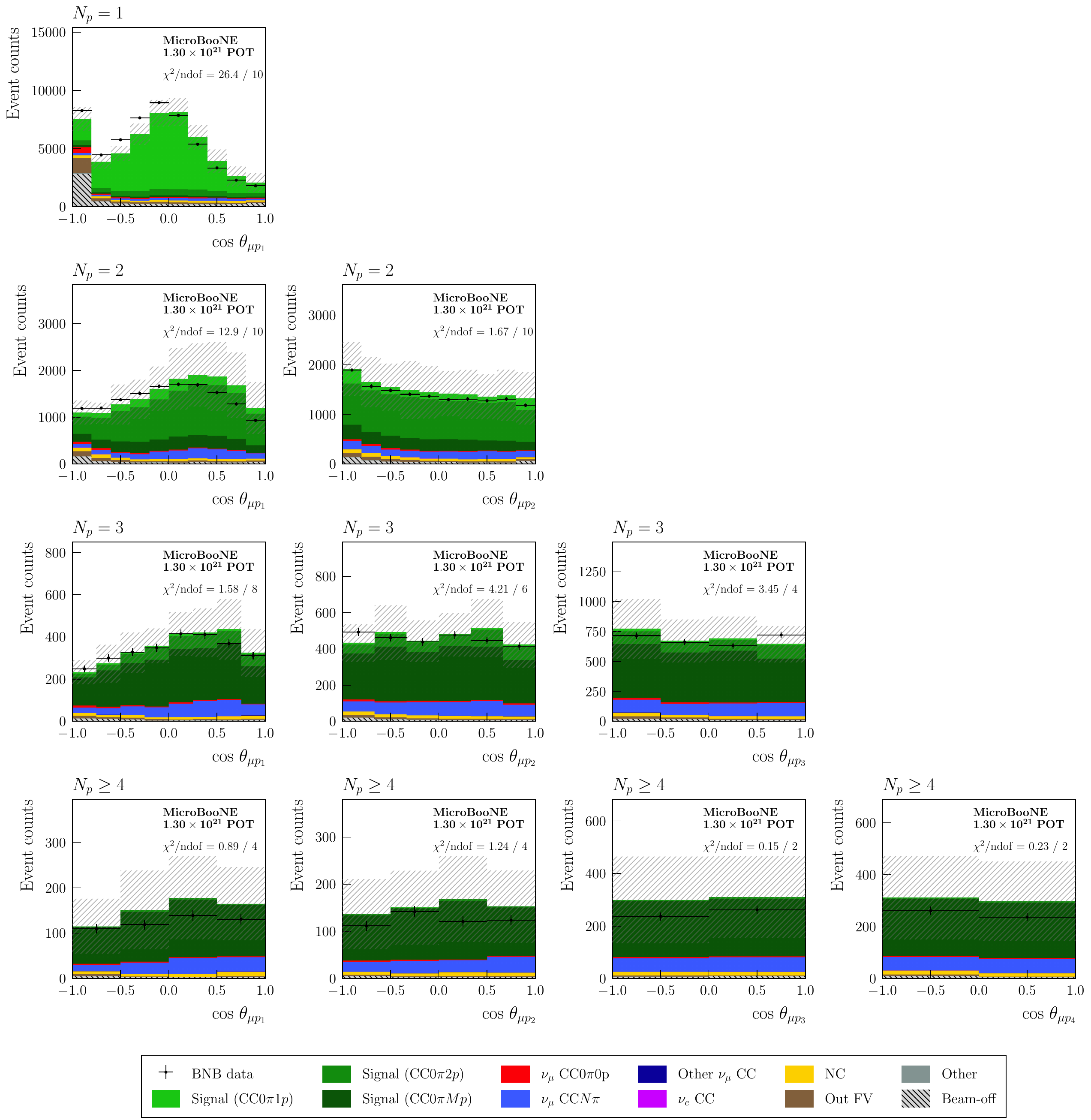}
	\caption{Reconstructed event rate distributions for blocks 8, 9, 10, and 11, corresponding to the double-differential measurements $(N_{p}, \, \mathrm{cos} \, \theta_{\mu p_{i}})$. The total uncertainty on the prediction is indicated by the hatched boxes. The per-slice $\chi^{2}/\text{ndof}$ for the central-value prediction is shown in each individual panel.}
	\label{fig:event_rates_blocks_8_9_10_11}
\end{figure*}

The event rate distributions comparing the central-value predictions to the data for bin blocks 1, 2, and 3 are shown in Figs. \ref{fig:event_rates_block_1}, \ref{fig:event_rates_block_2}, and \ref{fig:event_rates_block_3}. Figure \ref{fig:event_rates_blocks_4_5_6_7} shows blocks 4, 5, 6, and 7, corresponding to the proton momenta distributions in slices of fixed proton multiplicity. Similarly, Fig. \ref{fig:event_rates_blocks_8_9_10_11} contains the distributions for blocks 8, 9, 10, and 11, showing proton opening angles for the different multiplicity slices. The corresponding distributions with the signal component decomposed by primary interaction mode can be found in Sec.~III of the supplemental materials.

\begin{table}[!h]
	\caption{Goodness-of-fit between the observed data and the central-value prediction for the reconstructed event rate distributions in each block computed using the total covariance (statistical plus systematic).}
	\begin{center}
        \begin{small}
			\begin{tabular}{lcc}
                \parbox{6em}{\raggedright Block}                     & \parbox{6em}{\centering $\chi^{2} / \text{ndof}$} & \parbox{6em}{\centering p-value} \\[1mm] \hline
                \strutlike1: $(p_{p_{1}}, \, N_{p})$                 &                                        20.59 / 27 &                             0.81 \\[1mm]
                 2: $(\mathrm{cos} \, \theta_{\mu p_{1}}, \, N_{p})$ &                                        45.39 / 20 &                          $<0.01$ \\[1mm]
                 3: $(E_{\mathrm{avail}}, \, N_{p})$                 &                                         7.30 / 12 &                             0.84 \\[1mm]
                 4: $(N_{p}, \, p_{p_{1}})$                          &                                        28.41 / 33 &                             0.69 \\[1mm]
                 5: $(N_{p}, \, p_{p_{2}})$                          &                                        17.12 / 15 &                             0.31 \\[1mm]
                 6: $(N_{p}, \, p_{p_{3}})$                          &                                         8.14 /  8 &                             0.42 \\[1mm]
                 7: $(N_{p}, \, p_{p_{4}})$                          &                                         2.15 /  2 &                             0.34 \\[1mm]
                 8: $(N_{p}, \, \mathrm{cos} \, \theta_{\mu p_{1}})$ &                                        45.87 / 32 &                             0.05 \\[1mm]
                 9: $(N_{p}, \, \mathrm{cos} \, \theta_{\mu p_{2}})$ &                                         8.74 / 20 &                             0.99 \\[1mm]
                10: $(N_{p}, \, \mathrm{cos} \, \theta_{\mu p_{3}})$ &                                         3.53 /  6 &                             0.74 \\[1mm]
                11: $(N_{p}, \, \mathrm{cos} \, \theta_{\mu p_{4}})$ &                                         0.23 /  2 &                             0.89
            \end{tabular} 
        \end{small}
	\end{center}
	\label{tab:event_rate_chi2}
\end{table}

Good agreement between data and the central-value prediction is observed for most bin blocks. Table \ref{tab:event_rate_chi2} shows a per-block goodness-of-fit summary. The leading proton opening angle distributions (blocks 2 and 8) are a notable exception, showing significant tension between data and MC. Similar discrepancies in this observable have been reported in previous MicroBooNE $\text{CC} 0\pi Np$ measurements \cite{MicroBooNE:2024yzp}.

The adequacy of the background modeling is assessed through a sideband study. A combined sideband selection is formed merging the contributions from the three main background components in the analysis. The background-enhanced selections are defined by inverting some of the selection cuts:
\begin{itemize}
    \item[-] Out FV: events with a reconstructed vertex outside the fiducial volume.
    \item[-] NC: events for which a muon candidate could not be identified.
    \item[-] $\mathrm{CC} N\pi$: events with some proton candidates failing the log-likelihood ratio PID score cut.
\end{itemize}
Events passing any of these three selection criteria are accepted by the combined sideband selection.

The combined sideband study shows overall good agreement between data and prediction for most bin blocks. The event rate distributions for the different blocks can be found in Sec.~IV of the supplemental materials. A dedicated study is performed to investigate the discrepancy in the leading proton opening angle distributions, in which bin-by-bin signal and background correction factors are simultaneously extracted from the selection and sideband distributions. The null hypothesis that the background is correctly modeled cannot be rejected (p-values of $0.98$ and $0.77$ for blocks 2 and 8, respectively), indicating that the observed disagreement is consistent with signal mismodeling. This confirms that the background subtraction is not affected by this discrepancy. The procedure followed is described in Appendix~\ref{app:sideband_background}. The potential impact of this signal mismodeling on the response matrix is examined in Sec.~\ref{subsec:validation}.

\subsection{Cross-section extraction}
\label{subsec:unfolding}

The observed data consists of reconstructed event counts that have been distorted by detector inefficiencies and resolution effects, as discussed in Sec.~\ref{subsec:observables}. These distortions cause some selected events to migrate from the true kinematic bins in which they were generated to different bins in reconstructed space. To recover the underlying true distribution of events, one must correct for these detector effects through a procedure called unfolding. This is typically written in terms of a matrix multiplication
\begin{equation}\label{eq:unfolding}
    \hat{\varphi}_{\mu} = \sum_{i} U_{\mu i} \, d_{i},
\end{equation}
where $\hat{\varphi}_{\mu}$ is an estimator of the signal event counts in true bin $\mu$, $d_{i}$ are the background-subtracted event counts in reconstructed bin $i$, and $U_{\mu i}$ is the unfolding matrix. The background-subtracted event counts are calculated by subtracting the beam-correlated and beam-off expected number of background events from the total number of measured events in each reconstructed bin.

The covariance matrix between the estimated signal event counts can be computed via the unfolding matrix as
\begin{equation}\label{eq:unfold_cov}
    \text{Cov}(\hat{\varphi}_{\mu}, \, \hat{\varphi}_{\nu}) = \sum_{a, b} U_{\mu a} \, \text{Cov}(d_{a}, \, d_{b}) \, U_{b, \nu}^{\mathsf{T}},
\end{equation}
using the covariance matrix between the background-subtracted event counts in reconstructed space. This is estimated from the expected number of selected events
\begin{equation}
    \text{Cov}(d_{a}, \, d_{b}) \approx \text{Cov}(n_{a}, \, n_{b}),
\end{equation}
following the approach described in Sec.~\ref{subsec:uncertainty}.

The unfolding problem is fundamentally ill-posed. Naively, the unfolding matrix could be constructed by directly inverting the response matrix from Eq.~\eqref{eq:response_matrix}. However, in most cases the direct inversion is not viable because the response matrix is ill-conditioned. Small fluctuations in measured data lead to large, unstable fluctuations in the inferred true distribution. To address this, the analysis must introduce additional assumptions or constraints, a process known as regularization. These constraints help stabilize the solution, for example, by favoring distributions that are smooth or consistent with physical expectations.

The effects introduced by the regularization methods can be encapsulated in an additional smearing matrix $A_{C}$, which relates the regularized and direct-inversion unfolding matrices. It is defined as the product of the unfolding matrix $U$ and the detector response matrix $\Delta$:
\begin{equation}\label{eq:addtional_smearing}
    A_{C} \equiv U \, \Delta.
\end{equation}
In the limit of direct matrix inversion, $A_{C}$ reduces to the identity matrix. This formulation was proposed in Ref.~\cite{Tang:2017rob} for an unfolding scheme based on a Wiener filter, although the relation in Eq.~\eqref{eq:addtional_smearing} allows the calculation of the additional smearing matrix for any unfolding method and regularization scheme. In this work, all the theoretical predictions are multiplied by the $A_{C}$ matrix when checking for goodness-of-fit with the unfolded results, to account for the regularization bias. However, the predictions are shown unaltered in the plots. The elements of the additional smearing matrix for each block are included in Sec.~VI of the supplemental materials.

In the present analysis, we use the unfolding method proposed by G. D'Agostini \cite{DAgostini:1994fjx}, which applies Bayes' theorem iteratively. Starting from an arbitrary prior, each iteration refines the estimate in successive steps using the measured data and the detector response matrix. In the many iterations limit, this method approaches the direct inversion result. Regularization is introduced by limiting the number of iterations, avoiding overfitting statistical fluctuations and retaining only the dominant features of the distribution. In the calculation, the selection efficiency $\epsilon_{\mu}$ and response matrix $\Delta_{i\mu}$ are kept constant. Each iteration updates the conditional probability to assign a signal event belonging in true bin $\mu$ to reconstructed bin $i$ according to Bayes' theorem. The unfolding matrix is obtained from this probability, and then used in Eq.~\eqref{eq:unfolding} to produce the updated estimate for the next iteration.

The number of iterations used in the unfolding is fixed using the average fractional difference between iterations
\begin{equation}
    \mathfrak{F} \equiv \frac{1}{N_{\text{true}}} \, \sum_{\mu} \frac{|\hat{\varphi}^{(n+1)}_{\mu} - \hat{\varphi}^{(n)}_{\mu}|}{\hat{\varphi}^{(n+1)}_{\mu}},
\end{equation}
where $N_{\text{true}}$ is the total number of true bins. We stop iterating when this fractional difference between consecutive iterations falls below $2.5\%$.

The unfolding procedure described above is applied to each of the bin blocks in the analysis independently. In the blockwise approach adopted, the total unfolding matrix is constructed as the direct sum of all the block-level unfolding matrices, $U = \bigoplus_{b} U_{b}$, resulting in a block diagonal matrix. This total unfolding matrix is used to compute the unfolded covariance matrix according to Eq.~\eqref{eq:unfold_cov}.

The nominal-flux-integrated \cite{Butkevich:2010cr} differential cross section is derived from the unfolded event counts as
\begin{equation}\label{eq:xsec_units}
    \sigma_{\mu} \equiv \left< \frac{\mathrm{d}^{n} \sigma}{\mathrm{d} \mathbf{x}} \right>_{\mu} = \frac{\hat{\varphi}_{\mu}}{\Phi ~ \mathcal{N}_{\mathrm{Ar}} ~ \Delta \mathbf{x}_{\mu}},
\end{equation}
where $\Phi$ is the integrated $\nu_{\mu}$ flux over all neutrino energies for the corresponding exposure, $\mathcal{N}_{\mathrm{Ar}}$ is the number of argon nuclei in the FV, and $\Delta \mathbf{x}_{\mu}$ is the product of the bin widths in each differential variable for true bin $\mu$. Similarly, the covariance in cross-section units is given by
\begin{equation}\label{eq:xsec_cov}
    \text{Cov}(\sigma_{\mu}, \sigma_{\nu}) = \frac{\text{Cov}(\hat{\varphi}_{\mu}, \hat{\varphi}_{\nu})}{\Phi^{2} ~ \mathcal{N}_{\mathrm{Ar}}^{2} ~ \Delta \mathbf{x}_{\mu} ~ \Delta \mathbf{x}_{\nu}}.
\end{equation}
The integrated muon neutrino flux is obtained from the BNB flux prediction \cite{MiniBooNE:2008hfu} yielding $\Phi = 9.596 \times 10^{11} ~ \nu_{\mu} / \mathrm{cm}^{2}$. The number of argon nuclei is computed to be $\mathcal{N}_{\mathrm{Ar}} = 7.992 \times 10^{29}$, using the pure argon density and the FV defined in Sec.~\ref{subsec:event_selection}.

\subsection{Validation studies}
\label{subsec:validation}

The outcome of the unfolding is sensitive to the choice of prior and to the modeling of the detector response. For this reason, it is essential to validate the cross-section extraction method using closure tests and fake data studies, which check whether the unfolding procedure can recover a known input distribution from simulated data.

We perform a fake data study in which the beam-on data is replaced with the NuWro single-universe variation samples described in Sec.~\ref{subsec:uncertainty}. Since the underlying true distribution differs from the GENIE MicroBooNE tune prediction used to build the detector response matrix, this test verifies that the unfolding procedure can recover a signal shape different from the prior. MC statistical and neutrino interaction model uncertainties are considered. The unfolded fake data distributions are shown in Sec.~V A of the supplemental materials. For all blocks, the distributions show better agreement with the NuWro prediction than with the MicroBooNE tune prediction.

\begin{table}[!h]
	\caption{Goodness-of-fit for the comparison between fake data unfolded with the data- and MC-derived resimulated response matrices for each block, computed using the response matrix statistical covariance.}
	\begin{center}
        \begin{small}
			\begin{tabular}{lcc}
                \parbox{6em}{\raggedright Block}                     & \parbox{6em}{\centering $\chi^{2} / \text{ndof}$} & \parbox{6em}{\centering p-value} \\[1mm] \hline
                \strutlike1: $(p_{p_{1}}, \, N_{p})$                 &                                         1.92 / 27 &                             1.00 \\[1mm]
                 2: $(\mathrm{cos} \, \theta_{\mu p_{1}}, \, N_{p})$ &                                         0.88 / 20 &                             1.00 \\[1mm]
                 3: $(E_{\mathrm{avail}}, \, N_{p})$                 &                                         0.49 / 12 &                             1.00 \\[1mm]
                 4: $(N_{p}, \, p_{p_{1}})$                          &                                         1.33 / 33 &                             1.00 \\[1mm]
                 5: $(N_{p}, \, p_{p_{2}})$                          &                                         0.74 / 15 &                             1.00 \\[1mm]
                 6: $(N_{p}, \, p_{p_{3}})$                          &                                         0.40 /  8 &                             1.00 \\[1mm]
                 7: $(N_{p}, \, p_{p_{4}})$                          &                                         0.31 /  2 &                             0.86 \\[1mm]
                 8: $(N_{p}, \, \mathrm{cos} \, \theta_{\mu p_{1}})$ &                                         2.34 / 32 &                             1.00 \\[1mm]
                 9: $(N_{p}, \, \mathrm{cos} \, \theta_{\mu p_{2}})$ &                                         0.83 / 20 &                             1.00 \\[1mm]
                10: $(N_{p}, \, \mathrm{cos} \, \theta_{\mu p_{3}})$ &                                         2.09 /  6 &                             0.91 \\[1mm]
                11: $(N_{p}, \, \mathrm{cos} \, \theta_{\mu p_{4}})$ &                                         1.98 /  2 &                             0.37
            \end{tabular} 
        \end{small}
	\end{center}
	\label{tab:resim_chi2}
\end{table}

A separate data-driven validation study is performed to assess the sensitivity of the response matrix to potential mismodeling of the subleading proton kinematics near the reconstruction threshold. Selected data and MC events are resimulated using their reconstructed momenta as inputs to the full simulation and reconstruction chain, and the resulting samples are used to construct alternative response matrices. These are then used to unfold central-value fake data, and the two sets of unfolded results are compared. Table \ref{tab:resim_chi2} summarizes the $\chi^{2}$ values for the comparison between the data- and MC-derived results. The two are compatible across all blocks, indicating no evidence of bias from this source. Good closure is also achieved individually for both response matrices. The individual distributions can be found in Sec.~V B of the supplemental materials, and the full procedure is described in Appendix~\ref{app:resim_validation}.

This study also addresses the concern raised by the disagreement between data and prediction in the leading proton opening angle event rate distributions in Sec.~\ref{subsec:event_rates}. The resimulated data and MC response matrices produce compatible unfolded results for these blocks, showing that the response matrix is not significantly biased by the observed differences in proton kinematics between data and simulation. Together with the sideband study, this indicates that neither the background modeling nor the response matrix is affected by that discrepancy.

\subsection{Generator configurations}
\label{subsec:generators}

The nominal generator prediction used in this analysis is the GENIE MicroBooNE tune. It uses the local Fermi gas (LFG) model \cite{Chiang:1989nh} as a description of the nucleus. The Valencia model describes the QE and MEC CC interactions \cite{Nieves:2004wx,Nieves:2011pp}. The Kuzmin-Lyubushkin-Naumov and Berger-Sehgal (KLN-BS) model \cite{Kuzmin:2003ji,Berger:2007rq,Berger:2008xs} is used to simulate the resonant pion production. DIS interactions are described by the Bodek-Yang (BY) model \cite{Bodek:2002vp,Bodek:2004pc}, together with PYTHIA \cite{Sjostrand:2006za} for the hadronization. The FSI are simulated using the effective intranuclear transport model (hA2018) included in GENIE \cite{Dytman:2021ohr}. Additionally, this prediction includes some MicroBooNE-specific tuning of the model parameters \cite{MicroBooNE:2021ccs}.

\begin{figure*}[!htbp]
	\centering
	\includegraphics[width=0.88\linewidth]{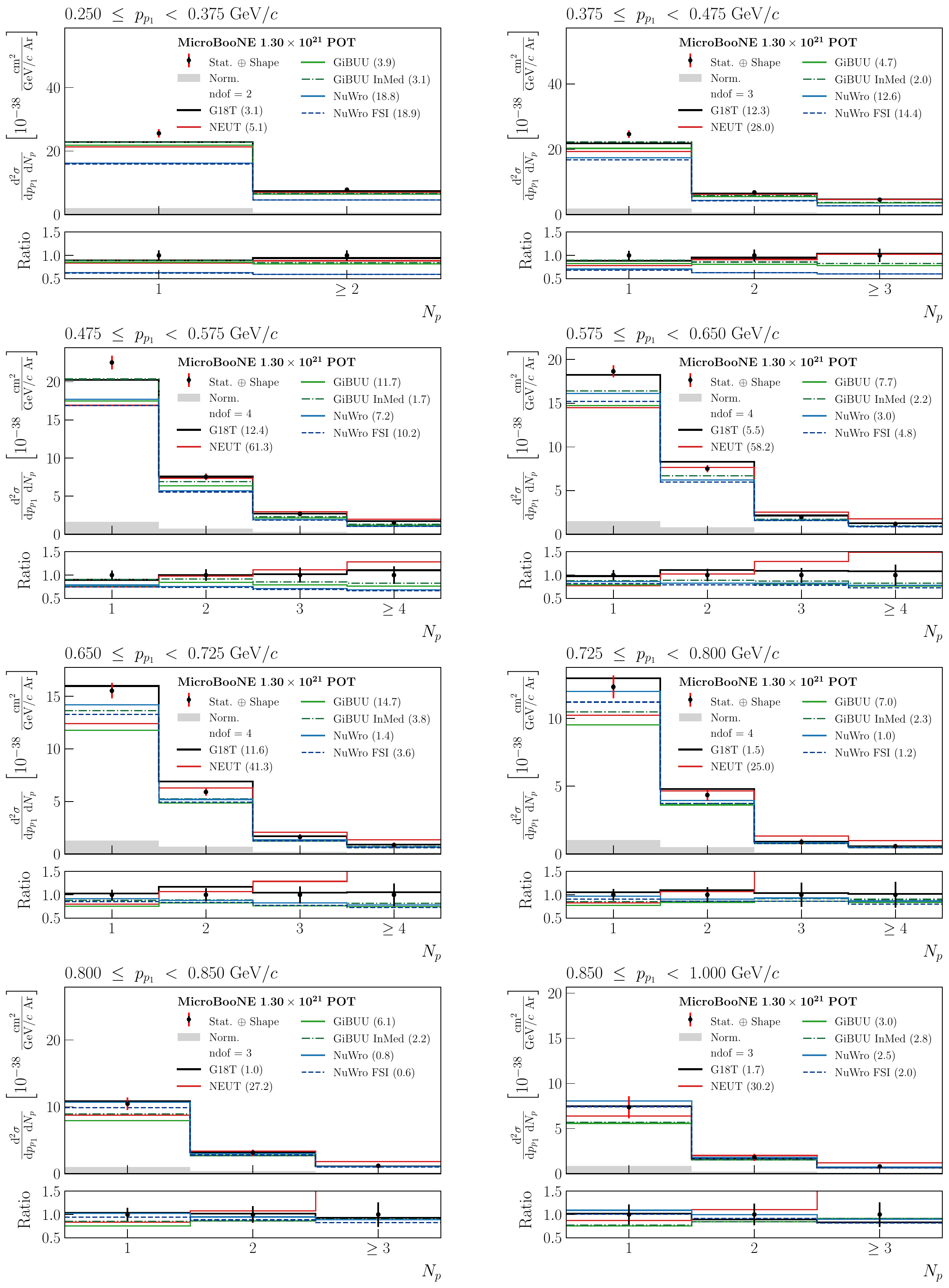}
	\caption{Flux-integrated double-differential cross sections for block 1 $(p_{p_{1}}, \, N_{p})$, extracted from the full MicroBooNE BNB dataset (black data points). Error bars indicate statistical and shape-only uncertainties, while the gray band indicates the normalization uncertainty. Predictions from the GENIE MicroBooNE tune, NuWro, NEUT, and GiBUU are overlaid as indicated in the legend. The $\chi^{2}$ for each prediction is indicated in parentheses. The bottom panels show the ratio of the predictions to the unfolded data.}
	\label{fig:unfolded_xsec_block_1}
\end{figure*}

\begin{figure*}[!htbp]
	\centering
	\includegraphics[width=0.88\linewidth]{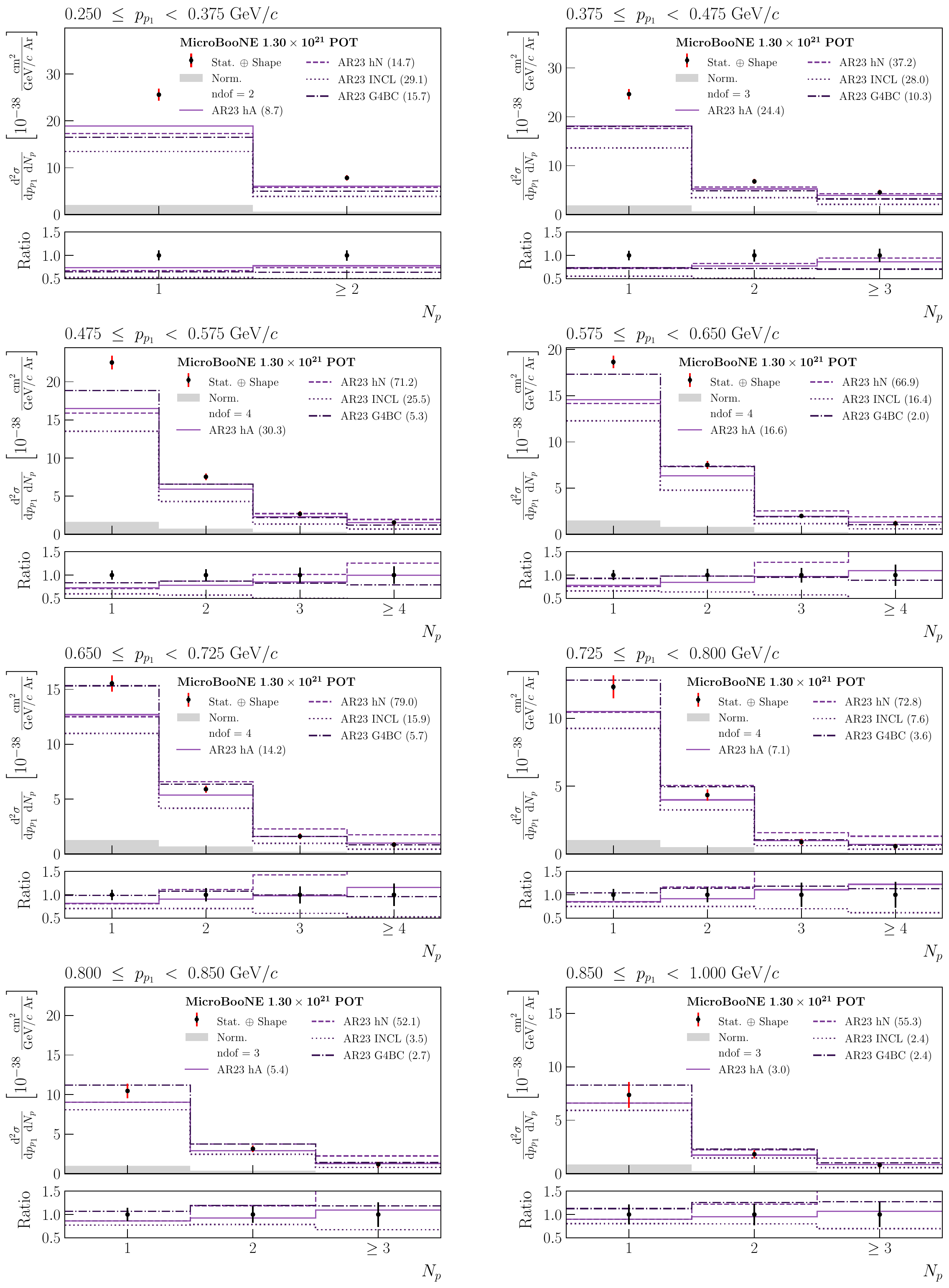}
	\caption{Flux-integrated double-differential cross sections for block 1 $(p_{p_{1}}, \, N_{p})$, extracted from the full MicroBooNE BNB dataset (black data points). Error bars indicate statistical and shape-only uncertainties, while the gray band indicates the normalization uncertainty. Predictions from the GENIE \texttt{AR23} configuration with the hA, hN, INCL, and G4BC FSI models are overlaid as indicated in the legend. The $\chi^{2}$ for each prediction is indicated in parentheses. The bottom panels show the ratio of the predictions to the unfolded data.}
	\label{fig:unfolded_xsec_ar23_block_1}
\end{figure*}

\begin{figure*}[!htbp]
	\centering
	\includegraphics[width=0.88\linewidth]{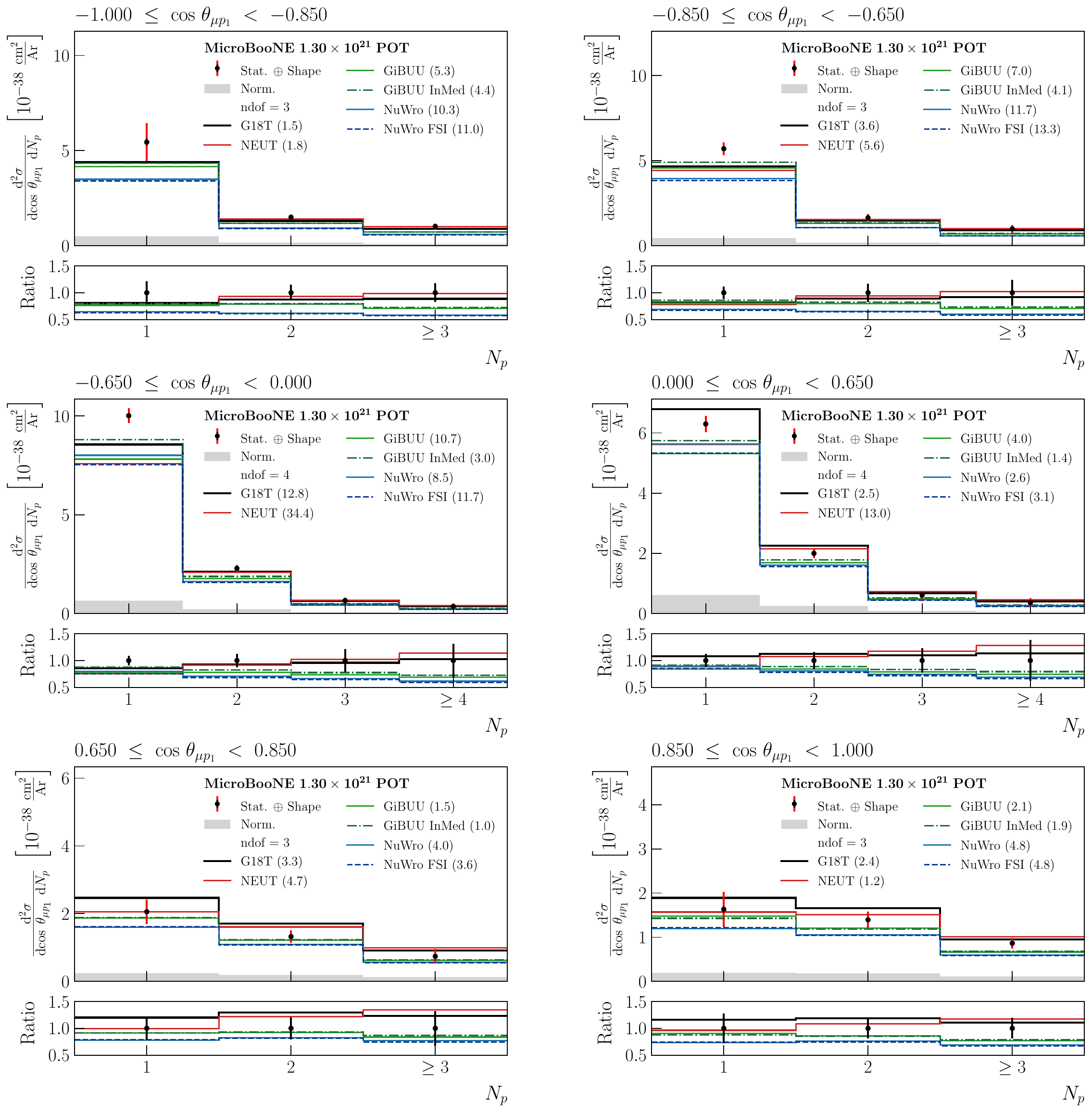}
	\caption{Flux-integrated double-differential cross sections for block 2 $(\mathrm{cos} \, \theta_{\mu p_{1}}, \, N_{p})$, extracted from the full MicroBooNE BNB dataset (black data points). Error bars indicate statistical and shape-only uncertainties, while the gray band indicates the normalization uncertainty. Predictions from the GENIE MicroBooNE tune, NuWro, NEUT, and GiBUU are overlaid as indicated in the legend. The $\chi^{2}$ for each prediction is indicated in parentheses. The bottom panels show the ratio of the predictions to the unfolded data.}
	\label{fig:unfolded_xsec_block_2}
\end{figure*}

\begin{figure*}[!htbp]
	\centering
	\includegraphics[width=0.88\linewidth]{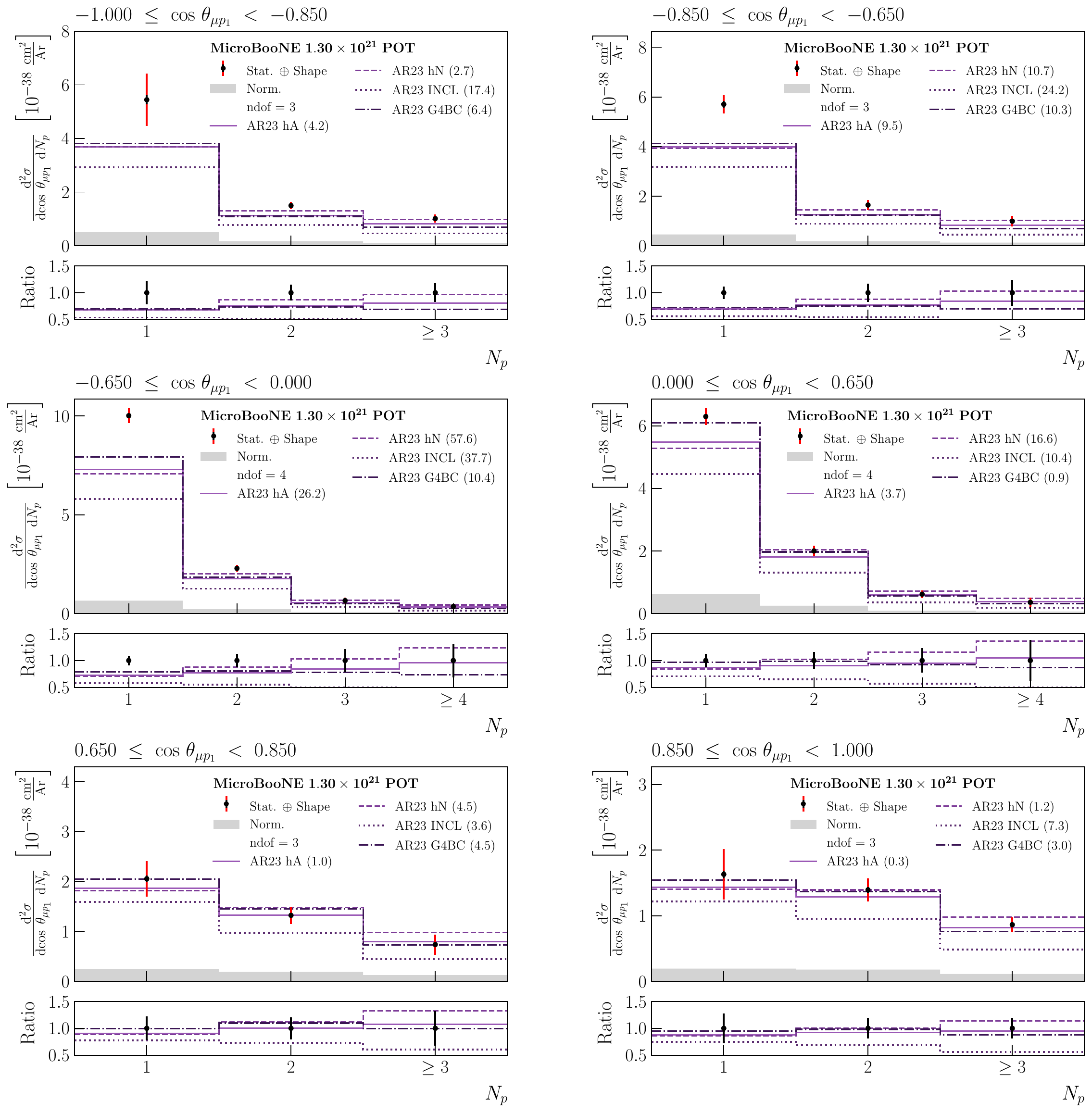}
	\caption{Flux-integrated double-differential cross sections for block 2 $(\mathrm{cos} \, \theta_{\mu p_{1}}, \, N_{p})$, extracted from the full MicroBooNE BNB dataset (black data points). Error bars indicate statistical and shape-only uncertainties, while the gray band indicates the normalization uncertainty. Predictions from the GENIE \texttt{AR23} configuration with the hA, hN, INCL, and G4BC FSI models are overlaid as indicated in the legend. The $\chi^{2}$ for each prediction is indicated in parentheses. The bottom panels show the ratio of the predictions to the unfolded data.}
	\label{fig:unfolded_xsec_ar23_block_2}
\end{figure*}

\begin{figure*}[!htbp]
	\centering
	\includegraphics[width=0.88\linewidth]{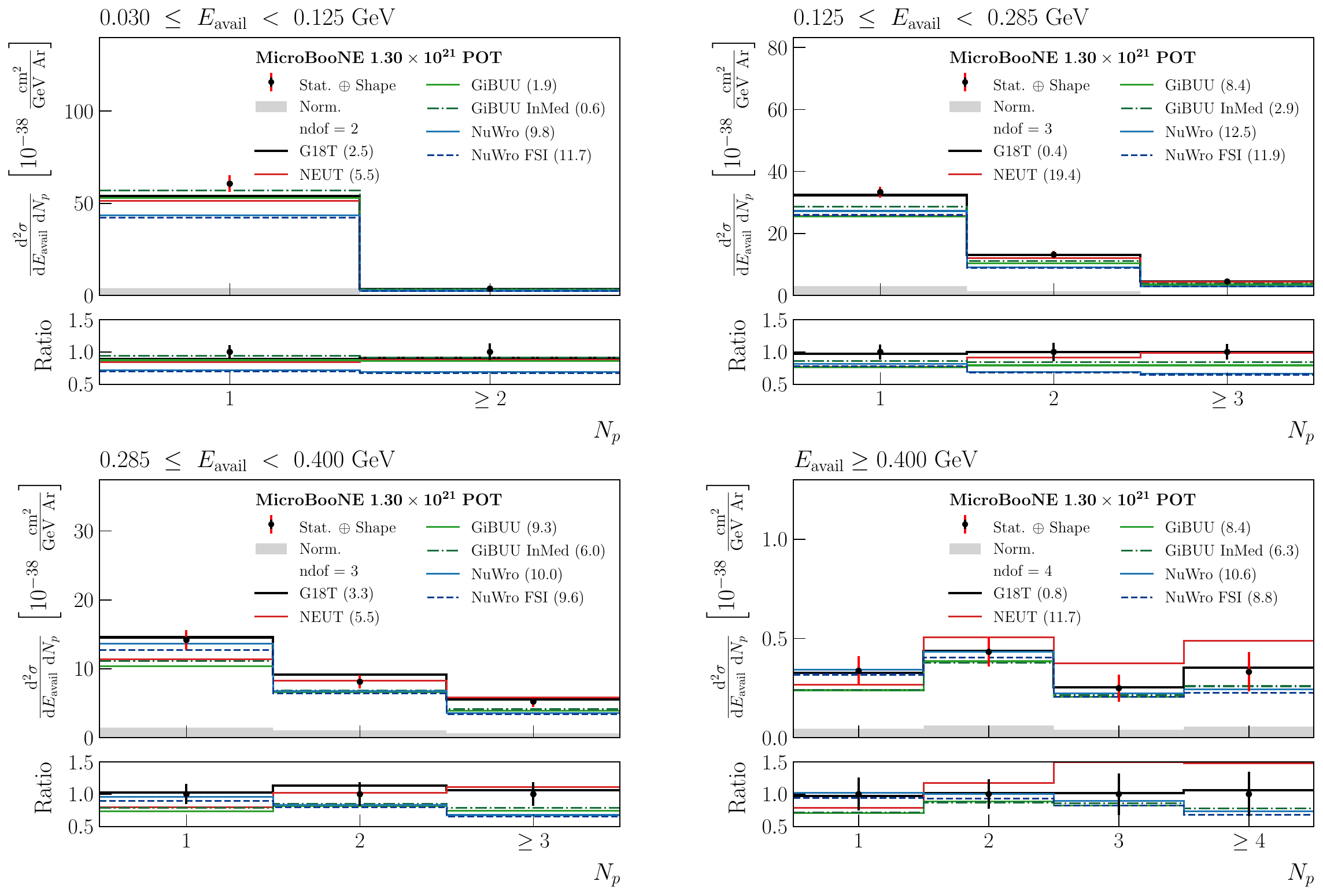}
	\caption{Flux-integrated double-differential cross sections for block 3 $(E_{\mathrm{avail}}, \, N_{p})$, extracted from the full MicroBooNE BNB dataset (black data points). Error bars indicate statistical and shape-only uncertainties, while the gray band indicates the normalization uncertainty. Predictions from the GENIE MicroBooNE tune, NuWro, NEUT, and GiBUU are overlaid as indicated in the legend. The $\chi^{2}$ for each prediction is indicated in parentheses. The bottom panels show the ratio of the predictions to the unfolded data.}
	\label{fig:unfolded_xsec_block_3}
\end{figure*}

\begin{figure*}[!htbp]
	\centering
	\includegraphics[width=0.88\linewidth]{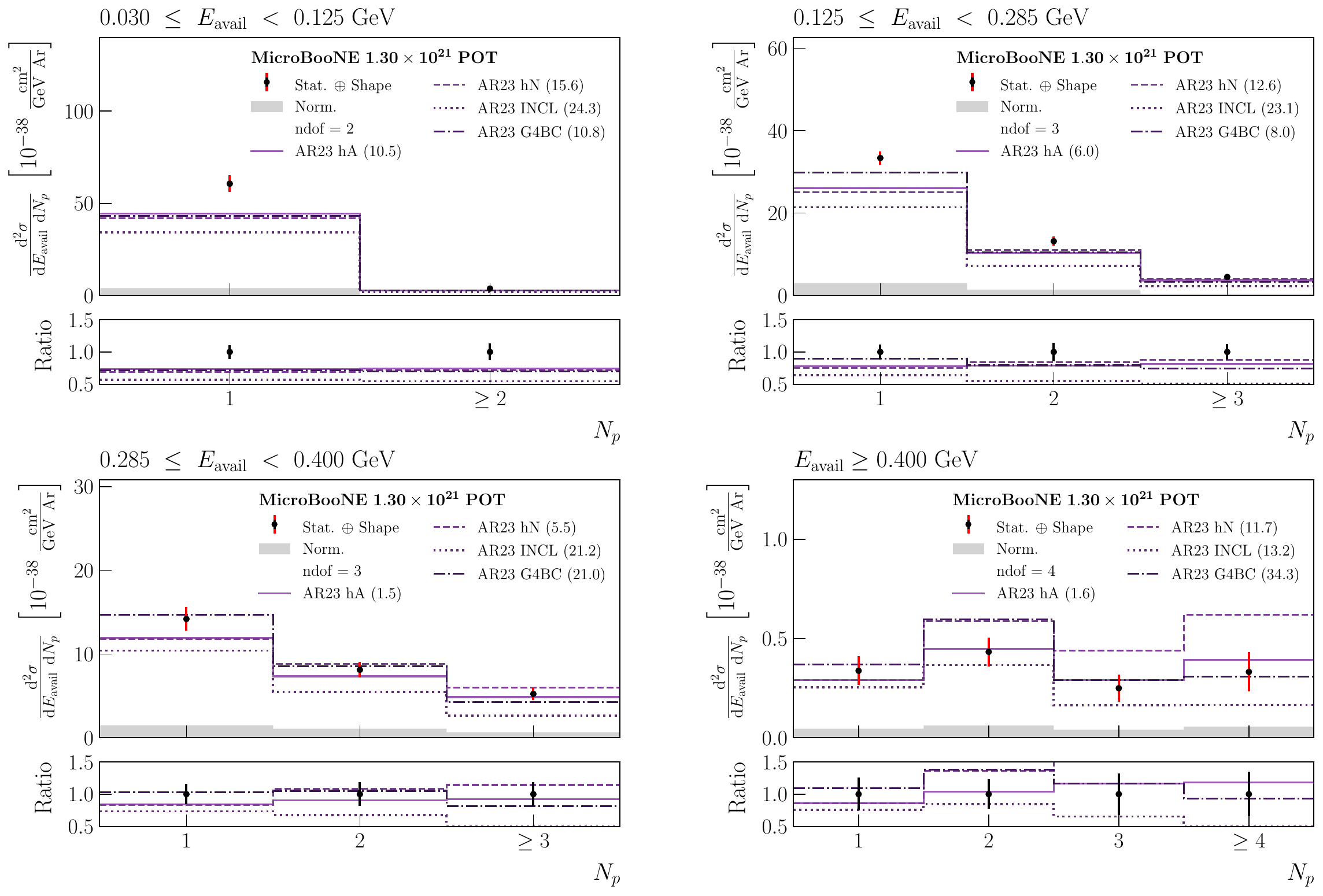}
	\caption{Flux-integrated double-differential cross sections for block 3 $(E_{\mathrm{avail}}, \, N_{p})$, extracted from the full MicroBooNE BNB dataset (black data points). Error bars indicate statistical and shape-only uncertainties, while the gray band indicates the normalization uncertainty. Predictions from the GENIE \texttt{AR23} configuration with the hA, hN, INCL, and G4BC FSI models are overlaid as indicated in the legend. The $\chi^{2}$ for each prediction is indicated in parentheses. The bottom panels show the ratio of the predictions to the unfolded data.}
	\label{fig:unfolded_xsec_ar23_block_3}
\end{figure*}

\begin{figure*}[!htbp]
	\centering
	\includegraphics[width=1.00\linewidth]{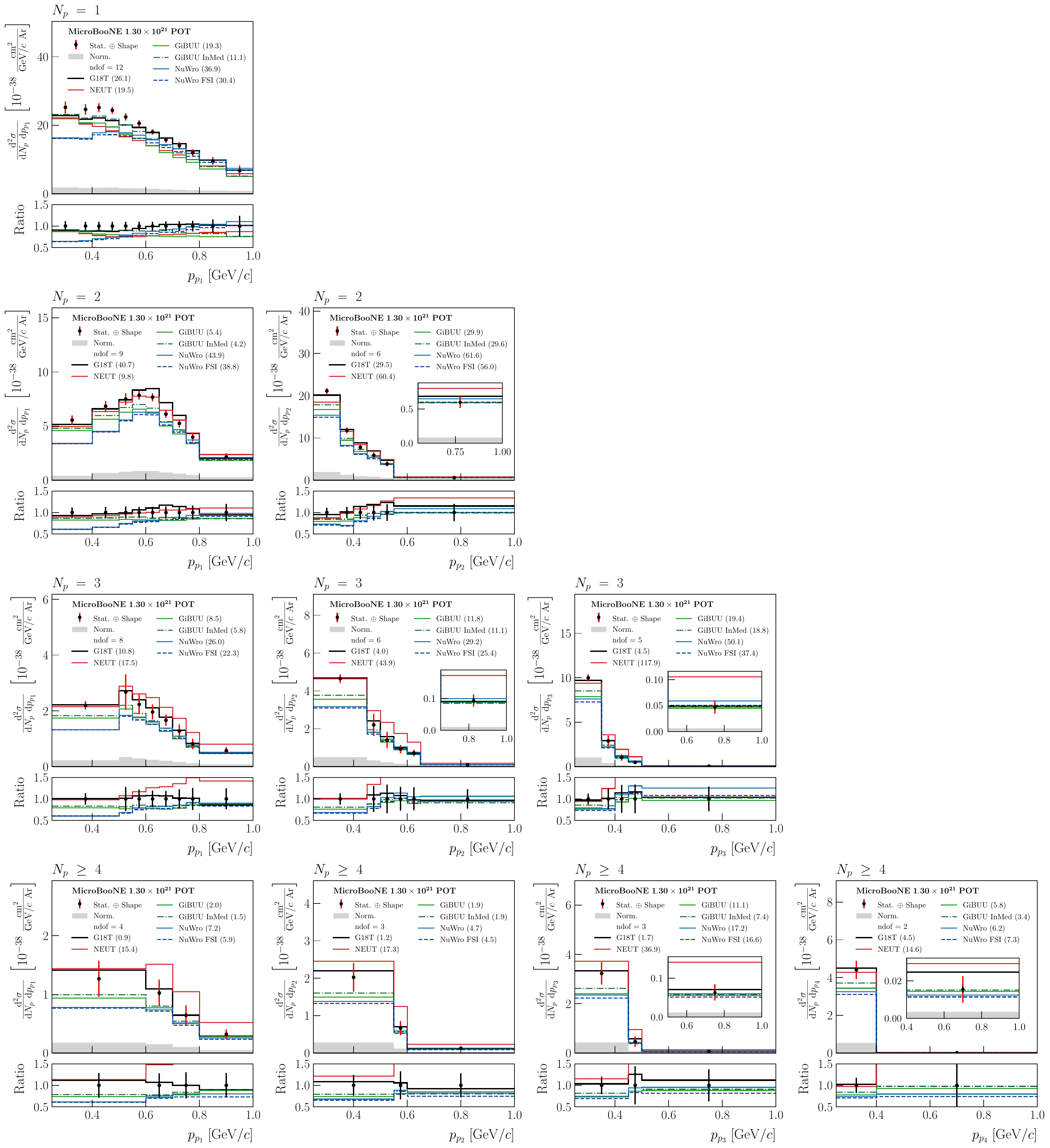}
	\caption{Flux-integrated double-differential cross sections for blocks 4, 5, 6, and 7 $(N_{p}, \, p_{p_{i}})$, extracted from the full MicroBooNE BNB dataset (black data points). Error bars indicate statistical and shape-only uncertainties, while the gray band indicates the normalization uncertainty. Predictions from the GENIE MicroBooNE tune, NuWro, NEUT, and GiBUU are overlaid as indicated in the legend. The $\chi^{2}$ for each prediction is indicated in parentheses. The bottom panels show the ratio of the predictions to the unfolded data.}
	\label{fig:unfolded_xsec_blocks_4_5_6_7}
\end{figure*}

\begin{figure*}[!htbp]
	\centering
	\includegraphics[width=1.00\linewidth]{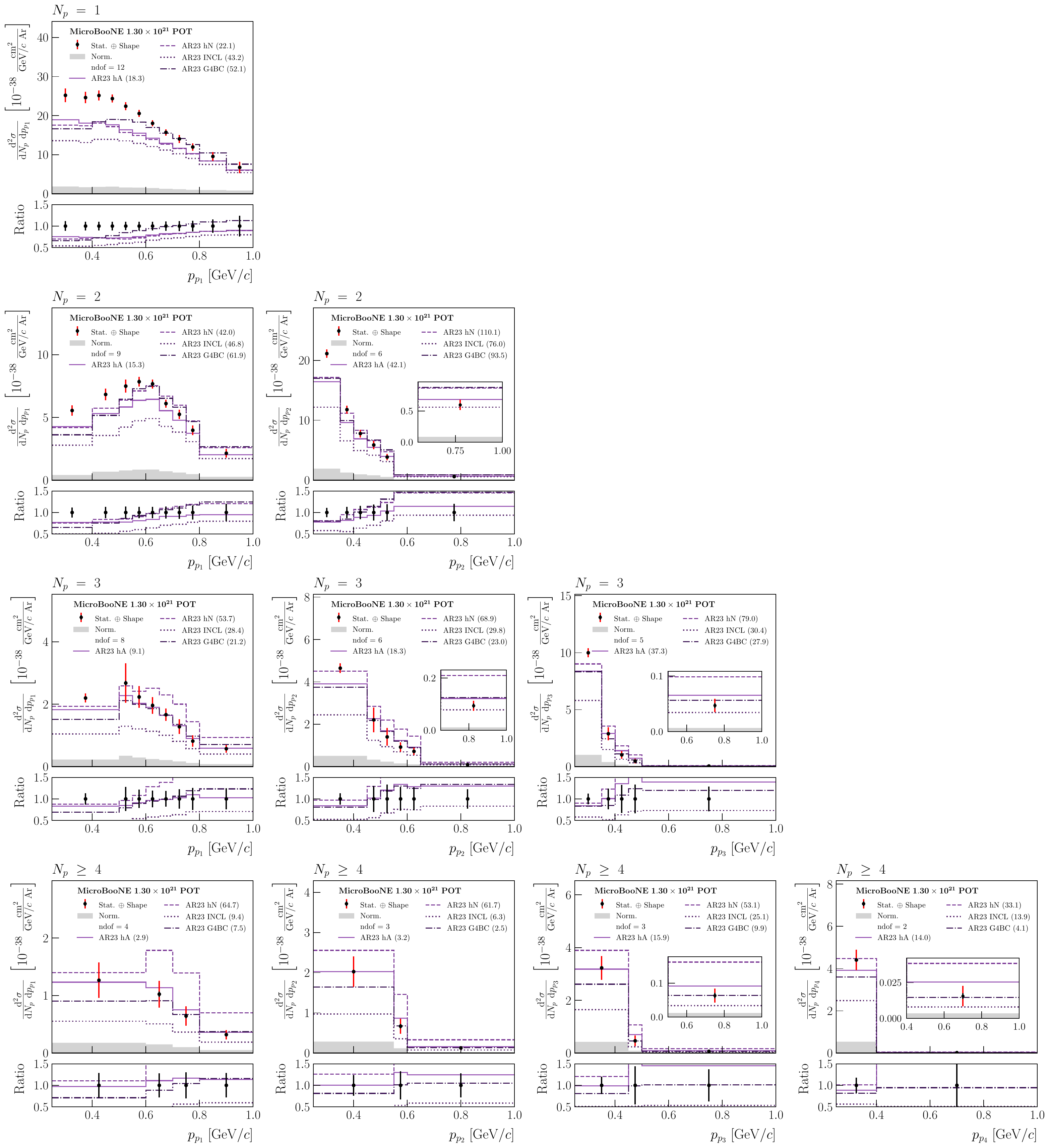}
	\caption{Flux-integrated double-differential cross sections for blocks 4, 5, 6, and 7 $(N_{p}, \, p_{p_{i}})$, extracted from the full MicroBooNE BNB dataset (black data points). Error bars indicate statistical and shape-only uncertainties, while the gray band indicates the normalization uncertainty. Predictions from the GENIE \texttt{AR23} configuration with the hA, hN, INCL, and G4BC FSI models are overlaid as indicated in the legend. The $\chi^{2}$ for each prediction is indicated in parentheses. The bottom panels show the ratio of the predictions to the unfolded data.}
	\label{fig:unfolded_xsec_ar23_blocks_4_5_6_7}
\end{figure*}

\begin{figure*}[!htbp]
	\centering
	\includegraphics[width=1.00\linewidth]{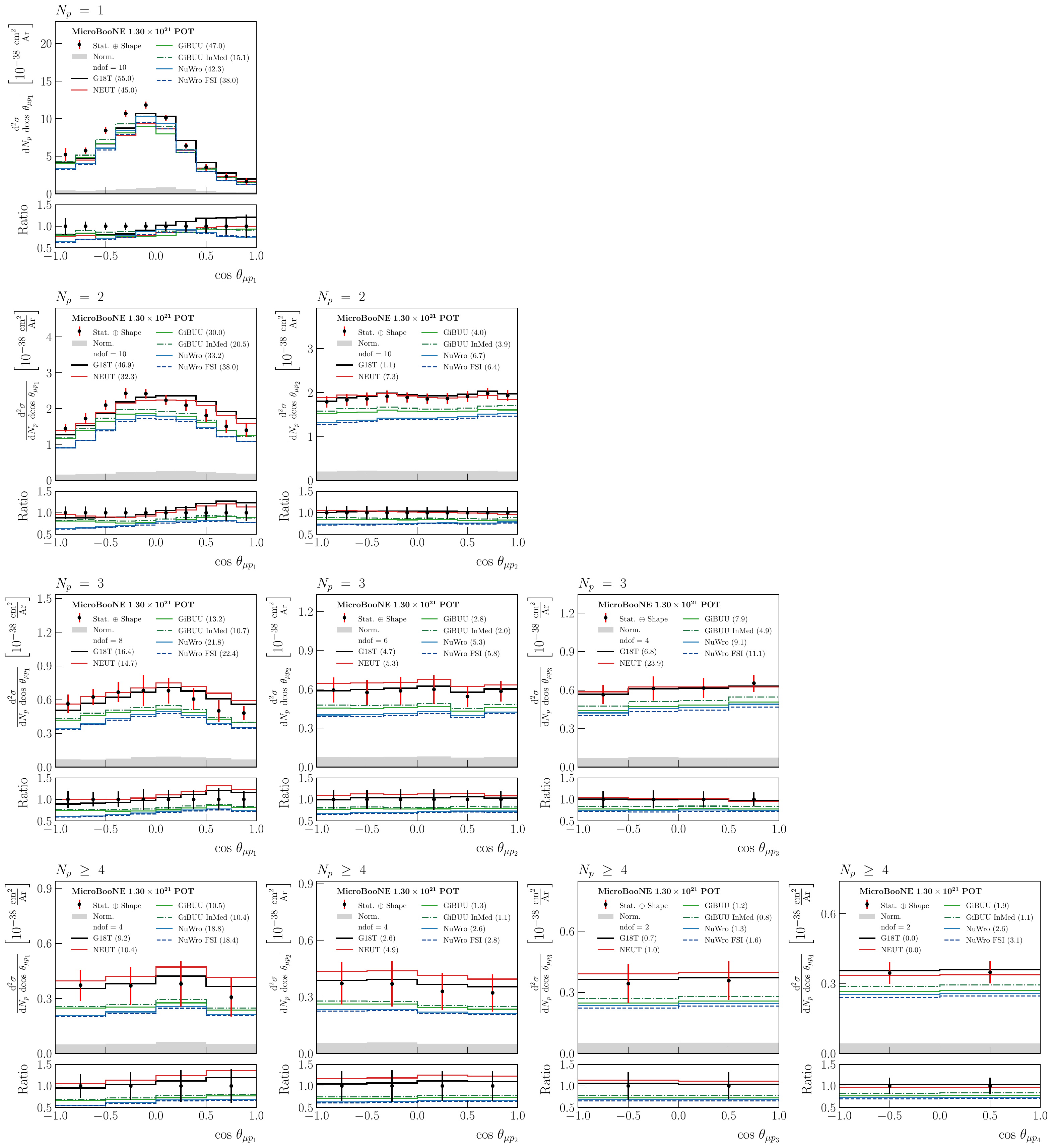}
	\caption{Flux-integrated double-differential cross sections for blocks 8, 9, 10, and 11 $(N_{p}, \, \mathrm{cos} \, \theta_{\mu p_{i}})$, extracted from the full MicroBooNE BNB dataset (black data points). Error bars indicate statistical and shape-only uncertainties, while the gray band indicates the normalization uncertainty. Predictions from the GENIE MicroBooNE tune, NuWro, NEUT, and GiBUU are overlaid as indicated in the legend. The $\chi^{2}$ for each prediction is indicated in parentheses. The bottom panels show the ratio of the predictions to the unfolded data.}
	\label{fig:unfolded_xsec_blocks_8_9_10_11}
\end{figure*}

\begin{figure*}[!htbp]
	\centering
	\includegraphics[width=1.00\linewidth]{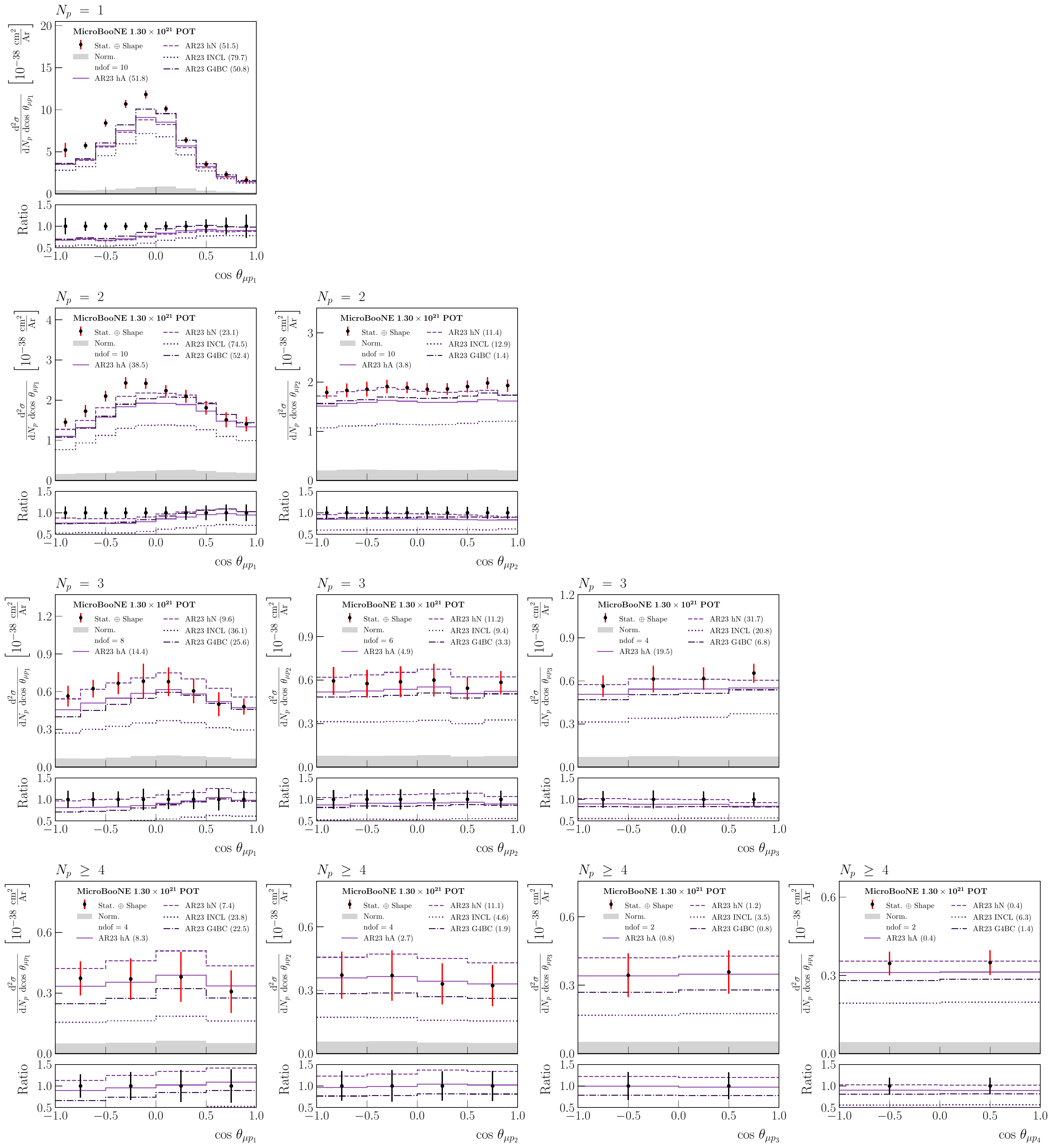}
	\caption{Flux-integrated double-differential cross sections for blocks 8, 9, 10, and 11 $(N_{p}, \, \mathrm{cos} \, \theta_{\mu p_{i}})$, extracted from the full MicroBooNE BNB dataset (black data points). Error bars indicate statistical and shape-only uncertainties, while the gray band indicates the normalization uncertainty. Predictions from the GENIE \texttt{AR23} configuration with the hA, hN, INCL, and G4BC FSI models are overlaid as indicated in the legend. The $\chi^{2}$ for each prediction is indicated in parentheses. The bottom panels show the ratio of the predictions to the unfolded data.}
	\label{fig:unfolded_xsec_ar23_blocks_8_9_10_11}
\end{figure*}

The unfolded cross sections are compared to predictions from several alternative event generators widely used in the neutrino scattering community, including models currently adopted by the SBN and DUNE programs. These generators differ in their treatment of the nuclear structure, interaction dynamics, and final-state interactions. The comparisons are performed using the NUISANCE framework \cite{Stowell:2016jfr}. Additional generator configurations exploring further model variations are described and compared in Sec.~VII of the supplemental materials.

NuWro \texttt{v25.11.1} uses a spectral function (SF) approach to model the ground state of the argon nucleus in the QE channel \cite{Banerjee:2023hub}, while a LFG model is used for all other dynamics. It uses the Llewellyn Smith (LS) model \cite{LlewellynSmith:1971uhs} for QE events, and the Valencia 2020 model \cite{Sobczyk:2020dkn} for MEC. Single pion production is described by the Ghent hybrid model \cite{Gonzalez-Jimenez:2016qqq}, while DIS uses the BY model. For QE events, NuWro applies a convolution scheme which broadens the cross section via a folding function and shifts the energy of the struck nucleon before propagating nucleons through a standard intranuclear cascade \cite{Ankowski:2025umq}, while non-QE channels are treated with the cascade alone. An additional NuWro sample with fine-tuned FSI is also included, in which the effective strength of the nucleon cascade is increased by approximately $24\%$ based on fits to MINERvA transverse kinematics data across multiple nuclear targets \cite{Prasad:2025hty}.

NEUT \texttt{v5.6.2} uses a LFG model to describe the nuclear state. The Nieves model is used for CCQE interactions, and the Valencia 2p2h model describes the MEC contribution. RES events use the BS model, and the BY model describes DIS. FSI are modeled by a semi-classical cascade model with medium corrections for pions \cite{Salcedo:1987md}.

GiBUU \cite{Buss:2011mx} \texttt{2025} also includes the LFG model for the initial state of the nucleus. GiBUU uses its own parametrizations for the QE and RES events \cite{Leitner:2008ue}. MEC interactions are modeled using semi-inclusive electron scattering data \cite{Bosted:1994tm}. PYTHIA \cite{Sjostrand:2006za} is used for DIS events. Hadrons are propagated through the remnant nucleus by numerically solving the Boltzmann-Uehling-Uhlenbeck (BUU) transport equation, with a nuclear potential consistent with the initial state. An additional GiBUU \texttt{2025} sample with in-medium modifications to the nucleon-nucleon cross sections, including a density-dependent lowering of the $NN$ cross section \cite{Bogart:2024gmb}, is also studied.

The GENIE \texttt{AR23\_20i\_00\_000} configuration \cite{GENIE:2021npt}, generated with \texttt{v3.6.2}, was developed for use in the DUNE and SBN programs. It uses a modified LFG model that populates a broader region of missing energy and momentum space compared to the standard LFG. Like the \texttt{G18\_10a} configurations, it uses the Nieves QE model and the Valencia 2p2h MEC, but employs the $z$-expansion parameterization of the axial form factor rather than the dipole approximation \cite{Meyer:2016oeg}. Three additional \texttt{AR23} variants are produced with alternative FSI models, with all other settings unchanged: \texttt{AR23\_20j\_00\_000} uses the hN intranuclear cascade, \texttt{AR23\_20k\_00\_000} uses the INCL (Li\`ege) cascade \cite{Boudard:2012wc,Mancusi:2014fba}, and \texttt{AR23\_20l\_00\_000} uses the Geant4 Bertini cascade \cite{Heikkinen:2003sc}. The configurations for these FSI variants were prepared for this analysis.

\subsection{Unfolded cross-section results}
\label{subsec:cross_section_results}

The unfolded differential cross-section results for all bin blocks are shown in Figs. \ref{fig:unfolded_xsec_block_1} to \ref{fig:unfolded_xsec_ar23_blocks_8_9_10_11}, presented in pairs for each block. The first figure in each pair compares the data to generator predictions extracted from the GENIE MicroBooNE tune (G18T), NuWro \texttt{v25.11.1} (NuWro), NuWro \texttt{v25.11.1} with fine-tuned FSI \cite{Prasad:2025hty} (NuWro FSI), NEUT \texttt{v5.6.2} (NEUT), GiBUU \texttt{2025} (GiBUU), and GiBUU \texttt{2025} with in-medium corrections \cite{Bogart:2024gmb} (GiBUU InMed), described above. The second figure in the pair compares the results to the GENIE \texttt{AR23} configuration predictions using the hA, hN, INCL, and Geant4 Bertini cascade (G4BC) FSI models. Comparisons of the unfolded cross-section results to additional generator configurations are provided in Sec.~VII of the supplemental materials. For each individual slice and prediction, a $\chi^{2}$ metric describing its agreement with the data is reported in the legend. We use the covariance matrix decomposition described in Appendix~\ref{app:covariance_decomp} to separate the uncertainties into normalization, shape, and mixed contributions, following the convention of previous MicroBooNE analyses \cite{MicroBooNE:2023cmw,MicroBooNE:2023tzj,MicroBooNE:2023krv,MicroBooNE:2024yzp,MicroBooNE:2025phj,MicroBooNE:2025ooi}. The inner (black) error bars in the data points represent statistical uncertainties, while the outer (red) error bars include shape-only systematic contributions. Normalization and mixed uncertainties are indicated by the gray histograms. The bottom panels contain the ratio of the different predictions to the data. In this case, the data error bars represent the total uncertainty in the measurement. A detailed breakdown of the fractional uncertainty from each systematic source is provided in Sec.~VIII of the supplemental materials. Tables \ref{tab:unfold_data_summary} and \ref{tab:unfold_data_summary_variations} summarize the per-block $\chi^{2}$ values for all generator predictions.

Figures \ref{fig:unfolded_xsec_block_1} and \ref{fig:unfolded_xsec_ar23_block_1} show the double-differential cross-section measurement in leading proton momentum and proton multiplicity. In this case, GiBUU InMed provides the best description of the data across all slices ($\chi^{2} / \text{ndof} = 25.3 / 27$ for the full block), being the only generator consistent with the observed distributions at the $1\sigma$ level. The default GiBUU prediction is also in good agreement with the data, suggesting that the BUU transport framework captures the essential features of proton production. NEUT exhibits the largest disagreement with the measurement, systematically overpredicting the $N_{p} \geq 2$ events relative to the $N_{p} = 1$ cross section. The \texttt{AR23} FSI variations span a wide range of $\chi^{2}$ from $77.2/27$ (INCL) to $370.8/27$ (hN), underlining the sensitivity of multi-proton observables to the cascade model. NuWro tends to underpredict the cross section across all multiplicities, particularly at low leading proton momentum. The FSI tuning marginally improves the block $\chi^{2}$; however, it does not resolve the deficit. The MicroBooNE tune shows some discrepancy in the mid-momentum range, slightly overpredicting the $N_{p}=2$ events relative to $N_{p}=1$ and higher multiplicity events.

The extracted double-differential cross-section values in slices of leading proton opening angle and bins of proton multiplicity are reported in Figs. \ref{fig:unfolded_xsec_block_2} and \ref{fig:unfolded_xsec_ar23_block_2}. The GiBUU predictions again provide the best overall agreement with the data. The disagreement between data and models is most pronounced for backward angles, where most generators underpredict the $N_{p} = 1$ cross section, particularly NuWro. NEUT performs comparatively well in the very backward and forward regions, although it accumulates tension at intermediate angles where it overestimates the multi-proton contribution. The \texttt{AR23} hA and G4BC variations describe the measurement well at forward angles, but show increasing disagreement at backward angles where $N_{p}=1$ dominates.

Figures \ref{fig:unfolded_xsec_block_3} and \ref{fig:unfolded_xsec_ar23_block_3} present the double-differential cross sections measured in total available energy and proton multiplicity. GiBUU InMed gives the best overall description of the measured data, followed by the NuWro prediction with fine-tuned FSI. However, the two generators perform best in complementary kinematic regimes: GiBUU InMed at low available energies and NuWro FSI in the high-energy region. The MicroBooNE tune achieves good per-slice agreement with the data, yet its block-global $\chi^{2}$ of $42.2 / 12$ indicates a poor overall fit introduced by cross-slice correlations. Both NuWro configurations underpredict the cross section at low and intermediate energies, consistent with the results from blocks 1 and 2. NEUT again shows a characteristic shape distortion: it overestimates the high-multiplicity events at high energy, similar to what is observed in the leading proton momentum distributions. A similar NEUT overprediction of the hadronic system at high available energy was observed in previous MicroBooNE inclusive $\nu_{\mu}$ CC measurements \cite{MicroBooNE:2024zwf}. Among the \texttt{AR23} FSI variations, the hN cascade yields the largest disagreement, underpredicting the $N_{p} = 1$ cross section while overpredicting higher multiplicities, a pattern consistent with excessive rescattering of primary protons through the cascade. INCL and hA are the best-performing variations, although both underpredict the $N_{p} = 1$ cross section at low energies. Although INCL achieves a comparable $\chi^{2}$ to hA in this block, this is partly driven by the normalization component of the covariance absorbing its uniform deficit, rather than reflecting a genuine shape agreement.

The measured double-differential cross sections in the leading and subleading proton momenta and slices of fixed proton multiplicity are displayed in Figs. \ref{fig:unfolded_xsec_blocks_4_5_6_7} and \ref{fig:unfolded_xsec_ar23_blocks_4_5_6_7}. The GiBUU InMed prediction gives the best agreement for the leading proton momentum block, accurately reproducing the spectral shape across all multiplicity slices. The standard GiBUU also provides an acceptable description of the data. A hierarchy emerges as one moves to the subleading proton distributions: for the 2nd-leading proton all generators yield $\chi^{2}/\text{ndof}$ values well above 1. The MicroBooNE tune and GiBUU (both configurations) perform comparably at the block-level, with significant tension concentrated in the $N_{p} = 2$ slice. The MicroBooNE tune prediction outperforms the rest of the generators in the 3rd-leading proton block. For the 4th-leading proton, GiBUU InMed, \texttt{AR23} G4BC, and the MicroBooNE tune all describe the measurement adequately, although the binning limits the discriminating power of this distribution. GiBUU InMed excels for the leading proton but shows reduced agreement at higher proton ranks, suggesting an underestimation of the cross section for the softer, subleading protons produced through secondary rescattering. Conversely, NEUT systematically produces a harder momentum spectrum for the subleading protons, overpredicting the high-momentum tails, particularly for $N_{p} \geq 3$. NuWro exhibits a persistent deficit at low momenta, indicating an underprediction of low-momentum secondary protons. The NuWro FSI configuration only partially improves the agreement. The \texttt{AR23} INCL configuration consistently underpredicts the cross section across all slices, continuing the pattern observed in the previous blocks. The \texttt{AR23} hN prediction shares the underprediction of the leading proton distributions common to all \texttt{AR23} configurations at low multiplicities, but additionally overestimates the high-momentum region at higher multiplicities, consistent with the excessive rescattering pattern observed in block 1.

\begin{table*}[!htbp]
    \caption{Summary of the $\chi^{2}$ values obtained for each bin block and globally for the GENIE MicroBooNE tune, NuWro (without and with fine-tuned FSI), NEUT, and GiBUU (without and with in-medium corrections) predictions compared to the unfolded cross-section data.}
    \begin{center}
        \begin{small}
            \begin{tabular}{lccccccc}
                \multirow{2}{9em}{Block}                             & \multirow{2}{3em}{\centering ndof} & \multirow{2}{9em}{\centering GENIE uB Tune} &    \multicolumn{2}{c}{\parbox{9em}{\centering NuWro \texttt{v25.11.1}}} & \multirow{2}{9em}{\centering NEUT \texttt{v5.6.2}} &         \multicolumn{2}{c}{\parbox{9em}{\centering GiBUU \texttt{2025}}} \\
                                                                     &                                    &                                             &   \parbox{4.5em}{\centering Default} &   \parbox{4.5em}{\centering FSI} &                                                    &  \parbox{4.5em}{\centering Default} &  \parbox{4.5em}{\centering InMed}  \\[1mm] \hline
                \strutlike 1: $(p_{p_{1}}, \, N_{p})$                &                                 27 &                                      121.30 &                                80.82 &                            64.35 &                                             206.18 &                               45.88 &                              25.33 \\[1.5mm]
                2: $(\mathrm{cos} \, \theta_{\mu p_{1}}, \, N_{p})$  &                                 20 &                                       80.20 &                                58.35 &                            59.67 &                                              78.41 &                               40.26 &                              27.73 \\[1.5mm]
                3: $(E_{\mathrm{avail}}, \, N_{p})$                  &                                 12 &                                       42.21 &                                34.92 &                            31.74 &                                             133.01 &                               33.51 &                              25.20 \\[1.5mm]
                4: $(N_{p}, \, p_{p_{1}})$                           &                                 33 &                                      136.90 &                                71.85 &                            59.03 &                                             221.54 &                               48.95 &                              29.96 \\[1.5mm]
                5: $(N_{p}, \, p_{p_{2}})$                           &                                 15 &                                       47.62 &                                99.70 &                            86.09 &                                             143.05 &                               46.09 &                              47.39 \\[1.5mm]
                6: $(N_{p}, \, p_{p_{3}})$                           &                                  8 &                                        7.56 &                                58.76 &                            43.18 &                                             120.97 &                               23.06 &                              23.09 \\[1.5mm]
                7: $(N_{p}, \, p_{p_{4}})$                           &                                  2 &                                        4.54 &                                 7.19 &                             8.90 &                                              14.64 &                                5.77 &                               3.38 \\[1.5mm]
                8: $(N_{p}, \, \mathrm{cos} \, \theta_{\mu p_{1}})$  &                                 32 &                                       87.20 &                                90.00 &                            74.89 &                                             102.99 &                               68.23 &                              42.66 \\[1.5mm]
                9: $(N_{p}, \, \mathrm{cos} \, \theta_{\mu p_{2}})$  &                                 20 &                                        9.90 &                                14.34 &                            16.49 &                                              29.39 &                                9.51 &                               9.29 \\[1.5mm]
                10: $(N_{p}, \, \mathrm{cos} \, \theta_{\mu p_{3}})$ &                                  6 &                                        7.59 &                                 9.41 &                            11.50 &                                              25.12 &                                8.15 &                               5.36 \\[1.5mm]
                11: $(N_{p}, \, \mathrm{cos} \, \theta_{\mu p_{4}})$ &                                  2 &                                        0.03 &                                 4.63 &                             6.24 &                                               0.05 &                                1.86 &                               1.14 \\[1.5mm] \hline
                \strutlike Global                                    &                                145 &                                       569.9 &                                678.7 &                            574.8 &                                             1132.0 &                               480.9 &                              428.8
            \end{tabular} 
        \end{small}
    \end{center}
    \label{tab:unfold_data_summary}
\end{table*}

\begin{table*}[!t]
    \caption{Summary of the $\chi^{2}$ values obtained for each bin block and globally for the GENIE \texttt{AR23} configuration predictions using the hA, hN, INCL, and G4BC FSI models compared to the unfolded cross-section data.}
	\begin{center}
        \begin{small}
			\begin{tabular}{lccccc}
                 \parbox{9em}{\raggedright Block}                     &                 \parbox{3em}{ndof} &   \parbox{9em}{\texttt{AR23} hA} &   \parbox{9em}{\texttt{AR23} hN} & \parbox{9em}{\texttt{AR23} INCL} & \parbox{9em}{\texttt{AR23} G4BC} \\[1mm] \hline
                 \strutlike 1: $(p_{p_{1}}, \, N_{p})$                &                                 27 &                            97.77 &                           370.76 &                            77.18 &                           142.73 \\[1.5mm]
                 2: $(\mathrm{cos} \, \theta_{\mu p_{1}}, \, N_{p})$  &                                 20 &                            77.87 &                           103.68 &                           117.52 &                            71.08 \\[1.5mm]
                 3: $(E_{\mathrm{avail}}, \, N_{p})$                  &                                 12 &                            62.15 &                           243.69 &                            60.78 &                           114.63 \\[1.5mm]
                 4: $(N_{p}, \, p_{p_{1}})$                           &                                 33 &                            90.45 &                           388.50 &                            91.07 &                           142.94 \\[1.5mm]
                 5: $(N_{p}, \, p_{p_{2}})$                           &                                 15 &                            60.67 &                           236.52 &                           102.95 &                           139.05 \\[1.5mm]
                 6: $(N_{p}, \, p_{p_{3}})$                           &                                  8 &                            39.13 &                            91.29 &                            33.59 &                            33.27 \\[1.5mm]
                 7: $(N_{p}, \, p_{p_{4}})$                           &                                  2 &                            14.00 &                            33.07 &                            13.88 &                             4.11 \\[1.5mm]
                 8: $(N_{p}, \, \mathrm{cos} \, \theta_{\mu p_{1}})$  &                                 32 &                            88.16 &                           113.59 &                           125.82 &                            93.06 \\[1.5mm]
                 9: $(N_{p}, \, \mathrm{cos} \, \theta_{\mu p_{2}})$  &                                 20 &                            18.85 &                            41.16 &                            18.47 &                             8.49 \\[1.5mm]
                 10: $(N_{p}, \, \mathrm{cos} \, \theta_{\mu p_{3}})$ &                                  6 &                            21.89 &                            36.91 &                            22.33 &                             7.20 \\[1.5mm]
                 11: $(N_{p}, \, \mathrm{cos} \, \theta_{\mu p_{4}})$ &                                  2 &                             0.36 &                             0.37 &                             6.32 &                             1.37 \\[1.5mm] \hline
				 \strutlike Global                                    &                                145 &                            488.1 &                           1200.6 &                            685.3 &                            822.0
			\end{tabular} 
        \end{small}
	\end{center}
	\label{tab:unfold_data_summary_variations}
\end{table*}

\begin{table*}[!t]
    \caption{Summary of the fractional uncertainties on the extracted cross section for each bin block, in percent. The values are the cross-section-weighted root mean square of the fractional uncertainty over the bins of each block, as defined in Eq.~\eqref{eq:frac_unc}.}
	\begin{center}
        \begin{small}
			\begin{tabular}{lccccccc}
                 \parbox{6em}{\raggedright Block}                     & \parbox{6em}{Flux} & \parbox{6em}{Interaction} & \parbox{6em}{Reinteraction} & \parbox{6em}{Detector} & \parbox{6em}{Normalization} & \parbox{6em}{Statistical} & \parbox{6em}{Total} \\[1mm] \hline
                 \strutlike 1: $(p_{p_{1}}, \, N_{p})$                &               7.58 &                      5.10 &                        3.65 &                   5.12 &                        2.61 &                      1.70 &               11.52 \\[1.5mm]
                 2: $(\mathrm{cos} \, \theta_{\mu p_{1}}, \, N_{p})$  &               7.81 &                      9.54 &                        0.87 &                   4.13 &                        2.65 &                      1.59 &               13.39 \\[1.5mm]
                 3: $(E_{\mathrm{avail}}, \, N_{p})$                  &               7.57 &                      8.25 &                        3.09 &                   4.85 &                        2.60 &                      1.24 &               12.91 \\[1.5mm]
                 4: $(N_{p}, \, p_{p_{1}})$                           &               7.61 &                      6.96 &                        3.71 &                   5.20 &                        2.62 &                      1.79 &               12.54 \\[1.5mm]
                 5: $(N_{p}, \, p_{p_{2}})$                           &               7.54 &                      6.48 &                        1.06 &                   4.15 &                        2.60 &                      1.20 &               11.20 \\[1.5mm]
                 6: $(N_{p}, \, p_{p_{3}})$                           &               7.51 &                      5.93 &                        0.80 &                   3.87 &                        2.60 &                      0.92 &               10.71 \\[1.5mm]
                 7: $(N_{p}, \, p_{p_{4}})$                           &               7.47 &                      5.11 &                        0.76 &                   3.72 &                        2.60 &                      0.68 &               10.17 \\[1.5mm]
                 8: $(N_{p}, \, \mathrm{cos} \, \theta_{\mu p_{1}})$  &               7.86 &                      9.83 &                        0.89 &                   4.15 &                        2.66 &                      1.76 &               13.66 \\[1.5mm]
                 9: $(N_{p}, \, \mathrm{cos} \, \theta_{\mu p_{2}})$  &               7.55 &                      8.69 &                        0.80 &                   3.86 &                        2.61 &                      1.09 &               12.49 \\[1.5mm]
                 10: $(N_{p}, \, \mathrm{cos} \, \theta_{\mu p_{3}})$ &               7.50 &                      7.59 &                        0.77 &                   3.75 &                        2.60 &                      0.73 &               11.65 \\[1.5mm]
                 11: $(N_{p}, \, \mathrm{cos} \, \theta_{\mu p_{4}})$ &               7.46 &                      5.26 &                        0.76 &                   3.69 &                        2.60 &                      0.64 &               10.23 \\[1.5mm]
			\end{tabular} 
        \end{small}
	\end{center}
	\label{tab:unfold_data_uncertainty}
\end{table*}

Figures \ref{fig:unfolded_xsec_blocks_8_9_10_11} and \ref{fig:unfolded_xsec_ar23_blocks_8_9_10_11} report the double-differential cross sections measured in the leading and subleading proton opening angles and proton multiplicity. For the leading proton opening angle block, GiBUU InMed is the only generator with a p-value above $0.05$ ($\chi^{2}/\text{ndof} = 42.7/32$). This is primarily driven by its agreement with the data in the $N_{p} = 1$ slice, where all other models yield $\chi^{2} = 38 \text{--} 80$ for the $10$ bins. The tension in this distribution is concentrated at backward angles, $\mathrm{cos} \, \theta_{\mu p_{1}} < 0$, where data significantly exceed all predictions. This suggests that single-proton events populate wider opening angles than most generators predict, a feature only partially recovered by the in-medium corrections in GiBUU; the same excess at wide lepton-proton opening angles in recent MicroBooNE $\nu_{e}$ measurements using the BNB and NuMI beams \cite{MicroBooNE:2026ifl,MicroBooNE:2025aiw} points to a common, flavor-independent feature of the hadronic final state. In the $N_{p} = 2$ slices an informative contrast emerges between the leading and subleading proton angular distributions: the leading proton retains a strongly peaked structure, reflecting its correlation with the momentum-transfer direction, while the 2nd-leading proton distribution is nearly isotropic across the full angular range. The peaked $\mathrm{cos} \, \theta_{\mu p_{1}}$ shape in the $N_{p} = 2$ slice proves difficult to model, with GiBUU InMed and \texttt{AR23} hN providing the closest agreement. The isotropy of the subleading protons is well reproduced by all generators regardless of their FSI implementation, suggesting that this is a generic feature of intranuclear rescattering rather than a detail sensitive to a particular cascade model. The 3rd- and 4th-leading proton opening angle blocks offer limited discriminating power due to the reduced number of bins, with most generators providing an adequate description. The \texttt{AR23} INCL prediction continues to systematically underpredict the cross section across all proton ranks and angular bins. Together with the results from the previous blocks, this points to a systematic underprediction of the overall proton yield in the INCL cascade model on argon.

Table~\ref{tab:unfold_data_uncertainty} summarizes the contribution of each source of uncertainty to the extracted cross section in every bin block. For a source $s$, with associated covariance matrix $\text{Cov}_{s}(\sigma_{\mu}, \, \sigma_{\nu})$, the quoted value for block $b$ is the root mean square of the fractional uncertainty over its bins, weighted by the extracted cross section
\begin{equation}\label{eq:frac_unc}
    f_{s}^{(b)} = \sqrt{\sum_{\mu \in \mathrm{bins}(b)} w_{\mu} \, \frac{\text{Cov}_{s}(\sigma_{\mu}, \, \sigma_{\mu})}{\sigma_{\mu}^{2}}},
\end{equation}
where the weights are the fraction of the cross section integrated over the block that is contained in each bin
\begin{equation}\label{eq:frac_unc_weights}
    w_{\mu} = \frac{\sigma_{\mu} \, \Delta \mathbf{x}_{\mu}} {\sum_{\nu \in \mathrm{bins}(b)} \sigma_{\nu} \, \Delta \mathbf{x}_{\nu}}.
\end{equation}
Because the weights are common to all sources, the individual contributions add in quadrature to the total. These values indicate the relative importance of each source across the measured phase space and do not account for the bin-to-bin correlations, so they differ from the breakdown obtained for a cross section integrated over all the bins in the block.

The BNB flux prediction is the dominant source of uncertainty in most blocks, contributing approximately $7.5\%$ almost uniformly, as it acts predominantly as a normalization uncertainty. The neutrino interaction modeling contribution is largest in the blocks measuring the leading proton opening angle, reaching $9.8\%$ in block 8. The hadron reinteraction uncertainty is only significant in the blocks binned in leading proton momentum and available energy, reflecting its impact on the proton kinematics. The statistical uncertainty is subleading in all blocks.

Finally, we assess the global agreement between the data and each generator prediction, combining all blocks while accounting for the correlations between them. These blockwise correlated $\chi^{2}$ values are reported in Tables \ref{tab:unfold_data_summary} and \ref{tab:unfold_data_summary_variations} alongside the per-block results. When the reconstructed and true binning are identical, as in this analysis, event sharing between blocks introduces structural constraints that reduce the number of independent degrees of freedom in the final measurement. From the bin construction, $38$ constrained degrees of freedom are identified, reducing the effective $\mathrm{ndof}$ for data-model comparisons from $183$ to $145$. The range-projected $\chi^{2}$ test statistic of Ref.~\cite{MartinezLopez:2026asn}, described in Appendix~\ref{app:global_chi2}, accounts for these constraints when computing global goodness-of-fit across blocks. GiBUU InMed achieves the best global agreement ($\chi^{2} / \text{ndof} = 428.8/145$), followed by the default GiBUU ($480.9/145$), and the GENIE \texttt{AR23} hA configuration ($488.1/145$). The global ranking is consistent with the block-by-block observations. The predictions using the BUU transport framework provide the most complete description of the data, while NEUT and the hN cascade show the largest overall tension. The consistent advantage of the BUU transport treatment over the semi-classical cascades in these FSI-sensitive observables mirrors the hierarchy seen in previous MicroBooNE measurements of mesonless final states \cite{MicroBooNE:2024yzp} and of transverse and generalized kinematic imbalance variables \cite{MicroBooNE:2023tzj,MicroBooNE:2023krv}. Among the FSI variants, the global $\chi^{2}$ spans from $488.1$ (hA) to $1200.6$ (hN), quantifying the strong FSI model sensitivity observed throughout the individual blocks. The GiBUU and NuWro predictions provide two further comparisons of the same kind, in which the only change is to the hadronic transport model. The in-medium corrections to the nucleon-nucleon cross section in GiBUU and the increased strength of the intranuclear cascade in NuWro each improve the global agreement relative to their default configurations. Across all three generators, varying the FSI treatment alone affects the global $\chi^{2}$ substantially, isolating final-state interactions as a leading driver of the observed data-model discrepancies. Even so, no generator achieves a statistically acceptable description of the full dataset, underscoring the need for continued refinement of hadronic transport models on argon.

\section{Summary and conclusions}
\label{sec:conclusions}

This analysis presents the first measurement of proton multiplicity and leading and subleading proton kinematics in charged-current $\nu_{\mu}$ interactions on argon producing no mesons and one or more protons in the final state. We report flux-integrated double-differential cross-section measurements in proton multiplicity, momenta and opening angles for the first four protons, ordered by momentum, and available energy. The analysis uses the full MicroBooNE Booster Neutrino Beam dataset corresponding to $1.30 \times 10^{21}$ POT. The blockwise approach from Ref.~\cite{Gardiner:2024gdy} is adopted, allowing the simultaneous reporting of all distributions with their full cross-block correlations.

The extracted cross sections are compared to predictions from several neutrino event generators. Overall, GiBUU \texttt{2025} with in-medium corrections to the nucleon-nucleon cross sections \cite{Bogart:2024gmb} provides the best description of the data at the block level and globally. However, no generator achieves an adequate description of the full dataset, indicating that significant modeling deficiencies remain. The measurement reveals strong sensitivity to the treatment of intra-nuclear hadron propagation, as demonstrated by the wide spread in $\chi^{2}$ values across the GENIE \texttt{AR23} configuration FSI variants and by the improved agreement with the data of both GiBUU and NuWro when their FSI treatment alone is modified. A backward-angle $N_{p} = 1$ deficit in the predictions appears to be a persistent feature that challenges all generators, suggesting a common deficiency in modeling single-proton final states at wide muon-proton opening angles. The systematic underprediction at low momenta for NuWro and the overprediction of multi-proton events at high energies seen in NEUT point to complementary deficiencies in the treatment of low-energy proton production and cascade strength.

The results presented in this paper are collected in a comprehensive data release. It includes the measured cross sections per bin, total statistical and systematic uncertainties, and predictions from the generators considered in the analysis. The covariance matrices describing the correlated uncertainties across the 183 bins are also included, separated by uncertainty source. The structure of the data release is described in detail in Sec.~I of the supplemental materials.

Using the full MicroBooNE statistics, this study complements and updates the previous $\text{CC} 0\pi Np$ measurement \cite{MicroBooNE:2024yzp} focused on muon-proton correlations. Together, they offer a more complete picture of the hadronic final state in neutrino-argon scattering. The reported measurements provide direct input for the interaction modeling needs of the SBN and DUNE programs, and the comprehensive data release enables the generator community to use them for tuning and validation.

\section{Acknowledgments}

This document was prepared by the MicroBooNE collaboration using the resources of the Fermi National Accelerator Laboratory (Fermilab), a U.S. Department of Energy, Office of Science, Office of High Energy Physics HEP User Facility. Fermilab is managed by Fermi Forward Discovery Group, LLC, acting under Contract No. 89243024CSC000002. MicroBooNE is supported by the following: the U.S. Department of Energy, Office of Science, Offices of High Energy Physics and Nuclear Physics; the U.S. National Science Foundation; the Swiss National Science Foundation; the Science and Technology Facilities Council (STFC), part of United Kingdom Research and Innovation (UKRI); the Royal Society (United Kingdom); the UKRI Future Leaders Fellowship; the NSF AI Institute for Artificial Intelligence and Fundamental Interactions; and the European Union's Horizon 2020 research and innovation programme under the Marie Sk\l{}odowska-Curie grant agreement No. 101003460 (PROBES). Additional support for the laser calibration system and cosmic ray tagger was provided by the Albert Einstein Center for Fundamental Physics, Bern, Switzerland. We also acknowledge the contributions of technical and scientific staff to the design, construction, and operation of the MicroBooNE detector as well as the contributions of past collaborators to the development of MicroBooNE analyses, without whom this work would not have been possible. For the purpose of open access, the authors have applied a Creative Commons Attribution (CC BY) public copyright license to any Author Accepted Manuscript version arising from this submission.


\appendix

\section{Bin definitions}
\label{app:bin_defs}

Table \ref{tab:block_bin_definitions} presents the binning scheme used in the analysis, organized in blocks representing the same kinematic distributions. For each bin, the corresponding selection efficiency and diagonal entry in the migration matrix are also reported.

\phantom{AAAA} 

\begin{longtable*}{
    w{c}{2.5cm}
    w{c}{2.5cm}
    w{c}{2.5cm}
    w{c}{2.5cm}
    w{c}{2.5cm}
    w{c}{2.5cm}
}
    \noalign{\vskip 1cm}
	\caption{Bin definitions used in the analysis.}
	\label{tab:block_bin_definitions}\\[5mm]

	\multicolumn{6}{c}{Block 1: $(p_{p_{1}}, \, N_{p})$} \\[3mm]
	Global bin number & $p_{p_{1}}^{\text{low}} ~ (\mathrm{GeV}/c)$ & $p_{p_{1}}^{\text{high}} ~ (\mathrm{GeV}/c)$ & $N_{p}$ & Efficiency & Migration diagonal \\[2mm] \hline \strutlike
	\endfirsthead
 
	\endlastfoot
 
	  0 &  0.250 &  0.375 &        1 & 0.747 & 0.870 \\[1mm]
	  1 &        &        & $\geq 2$ & 0.617 & 0.301 \\[2.5mm]
	  2 &  0.375 &  0.475 &        1 & 0.678 & 0.720 \\[1mm]
	  3 &        &        &        2 & 0.495 & 0.407 \\[1mm]
	  4 &        &        & $\geq 3$ & 0.408 & 0.229 \\[2.5mm]
	  5 &  0.475 &  0.575 &        1 & 0.712 & 0.729 \\[1mm]
	  6 &        &        &        2 & 0.534 & 0.502 \\[1mm]
	  7 &        &        &        3 & 0.393 & 0.251 \\[1mm]
	  8 &        &        & $\geq 4$ & 0.324 & 0.174 \\[2.5mm]
	  9 &  0.575 &  0.650 &        1 & 0.728 & 0.708 \\[1mm]
	 10 &        &        &        2 & 0.562 & 0.521 \\[1mm]
	 11 &        &        &        3 & 0.372 & 0.262 \\[1mm]
	 12 &        &        & $\geq 4$ & 0.275 & 0.195 \\[2.5mm]
	 13 &  0.650 &  0.725 &        1 & 0.699 & 0.672 \\[1mm]
	 14 &        &        &        2 & 0.522 & 0.489 \\[1mm]
	 15 &        &        &        3 & 0.329 & 0.255 \\[1mm]
	 16 &        &        & $\geq 4$ & 0.245 & 0.230 \\[2.5mm]
	 17 &  0.725 &  0.800 &        1 & 0.677 & 0.627 \\[1mm]
	 18 &        &        &        2 & 0.447 & 0.441 \\[1mm]
	 19 &        &        &        3 & 0.263 & 0.291 \\[1mm]
	 20 &        &        & $\geq 4$ & 0.192 & 0.197 \\[2.5mm]
	 21 &  0.800 &  0.850 &        1 & 0.659 & 0.575 \\[1mm]
	 22 &        &        &        2 & 0.353 & 0.382 \\[1mm]
	 23 &        &        & $\geq 3$ & 0.173 & 0.304 \\[2.5mm]
	 24 &  0.850 &  1.000 &        1 & 0.646 & 0.688 \\[1mm]
	 25 &        &        &        2 & 0.246 & 0.449 \\[1mm]
	 26 &        &        & $\geq 3$ & 0.096 & 0.358 \\[10mm]
 
	\multicolumn{6}{c}{Block 2: $(\mathrm{cos} \, \theta_{\mu p_{1}}, \, N_{p})$} \\[3mm]
	Global bin number & $\mathrm{cos} \, \theta_{\mu p_{1}}^{\text{low}}$ & $\mathrm{cos} \, \theta_{\mu p_{1}}^{\text{high}}$ & $N_{p}$ & Efficiency & Migration diagonal \\[2mm] \hline \strutlike
 
	 27 & -1.000 & -0.850 &        1 & 0.326 & 0.714 \\[1mm]
	 28 &        &        &        2 & 0.263 & 0.459 \\[1mm]
	 29 &        &        & $\geq 3$ & 0.174 & 0.251 \\[2.5mm]
	 30 & -0.850 & -0.650 &        1 & 0.628 & 0.706 \\[1mm]
	 31 &        &        &        2 & 0.425 & 0.423 \\[1mm]
	 32 &        &        & $\geq 3$ & 0.249 & 0.253 \\[2.5mm]
	 33 & -0.650 &  0.000 &        1 & 0.736 & 0.834 \\[1mm]
	 34 &        &        &        2 & 0.432 & 0.485 \\[1mm]
	 35 &        &        &        3 & 0.268 & 0.232 \\[1mm]
	 36 &        &        & $\geq 4$ & 0.159 & 0.169 \\[2.5mm]
	 37 &  0.000 &  0.650 &        1 & 0.773 & 0.808 \\[1mm]
	 38 &        &        &        2 & 0.453 & 0.545 \\[1mm]
	 39 &        &        &        3 & 0.245 & 0.254 \\[1mm]
	 40 &        &        & $\geq 4$ & 0.133 & 0.165 \\[2.5mm]
	 41 &  0.650 &  0.850 &        1 & 0.666 & 0.592 \\[1mm]
	 42 &        &        &        2 & 0.487 & 0.483 \\[1mm]
	 43 &        &        & $\geq 3$ & 0.219 & 0.284 \\[2.5mm]
	 44 &  0.850 &  1.000 &        1 & 0.560 & 0.534 \\[1mm]
	 45 &        &        &        2 & 0.389 & 0.412 \\[1mm]
	 46 &        &        & $\geq 3$ & 0.168 & 0.231 \\[10mm]

	\multicolumn{6}{c}{Block 3: $(E_{\mathrm{avail}}, \, N_{p})$} \\[3mm]
	Global bin number & $E_{\mathrm{avail}}^{\text{low}} ~ (\mathrm{GeV})$ & $E_{\mathrm{avail}}^{\text{high}} ~ (\mathrm{GeV})$ & $N_{p}$ & Efficiency & Migration diagonal \\[2mm] \hline \strutlike
 
     47 &  0.030 &  0.125 &        1 & 0.711 & 0.939 \\[1mm]
     48 &        &        & $\geq 2$ & 0.632 & 0.188 \\[2.5mm]
     49 &  0.125 &  0.285 &        1 & 0.708 & 0.783 \\[1mm]
     50 &        &        &        2 & 0.533 & 0.482 \\[1mm] 
     51 &        &        & $\geq 3$ & 0.465 & 0.150 \\[2.5mm]
     52 &  0.285 &  0.400 &        1 & 0.656 & 0.623 \\[1mm]
     53 &        &        &        2 & 0.501 & 0.473 \\[1mm]
     54 &        &        & $\geq 3$ & 0.355 & 0.150 \\[2.5mm]
     55 &  0.400 &  10.00 &        1 & 0.272 & 0.546 \\[1mm]
     56 &        &        &        2 & 0.162 & 0.457 \\[1mm]
     57 &        &        &        3 & 0.125 & 0.328 \\[1mm]
     58 &        &        & $\geq 4$ & 0.105 & 0.215 \\[10mm]

	\multicolumn{6}{c}{Block 4: $(N_{p}, \, p_{p_{1}})$} \\[3mm]
	Global bin number & $N_{p}$ & $p_{p_{1}}^{\text{low}} ~ (\mathrm{GeV}/c)$ & $p_{p_{1}}^{\text{high}} ~ (\mathrm{GeV}/c)$ & Efficiency & Migration diagonal \\[2mm] \hline \strutlike

     59 &        1 &  0.250 &  0.350 & 0.760 & 0.819 \\[1mm]
     60 &          &  0.350 &  0.400 & 0.701 & 0.513 \\[1mm]
     61 &          &  0.400 &  0.450 & 0.677 & 0.575 \\[1mm]
     62 &          &  0.450 &  0.500 & 0.679 & 0.598 \\[1mm] 
     63 &          &  0.500 &  0.550 & 0.714 & 0.661 \\[1mm]
     64 &          &  0.550 &  0.600 & 0.731 & 0.666 \\[1mm]
     65 &          &  0.600 &  0.650 & 0.724 & 0.659 \\[1mm]
     66 &          &  0.650 &  0.700 & 0.706 & 0.634 \\[1mm]
     67 &          &  0.700 &  0.750 & 0.686 & 0.605 \\[1mm]
     68 &          &  0.750 &  0.800 & 0.669 & 0.578 \\[1mm]
     69 &          &  0.800 &  0.900 & 0.662 & 0.658 \\[1mm]
     70 &          &  0.900 &  1.000 & 0.627 & 0.620 \\[2.5mm]
     71 &        2 &  0.250 &  0.400 & 0.599 & 0.282 \\[1mm]
     72 &          &  0.400 &  0.500 & 0.491 & 0.438 \\[1mm]
     73 &          &  0.500 &  0.550 & 0.535 & 0.436 \\[1mm]
     74 &          &  0.550 &  0.600 & 0.558 & 0.463 \\[1mm] 
     75 &          &  0.600 &  0.650 & 0.561 & 0.477 \\[1mm]
     76 &          &  0.650 &  0.700 & 0.528 & 0.458 \\[1mm]
     77 &          &  0.700 &  0.750 & 0.491 & 0.433 \\[1mm]
     78 &          &  0.750 &  0.800 & 0.431 & 0.404 \\[1mm]
     79 &          &  0.800 &  1.000 & 0.287 & 0.465 \\[2.5mm]
     80 &        3 &  0.250 &  0.500 & 0.447 & 0.174 \\[1mm]
     81 &          &  0.500 &  0.550 & 0.386 & 0.227 \\[1mm]
     82 &          &  0.550 &  0.600 & 0.412 & 0.224 \\[1mm]
     83 &          &  0.600 &  0.650 & 0.343 & 0.235 \\[1mm] 
     84 &          &  0.650 &  0.700 & 0.333 & 0.234 \\[1mm]
     85 &          &  0.700 &  0.750 & 0.309 & 0.270 \\[1mm]
     86 &          &  0.750 &  0.800 & 0.242 & 0.262 \\[1mm]
     87 &          &  0.800 &  1.000 & 0.135 & 0.301 \\[2.5mm]
     88 & $\geq 4$ &  0.250 &  0.600 & 0.323 & 0.150 \\[1mm]
     89 &          &  0.600 &  0.700 & 0.262 & 0.215 \\[1mm]
     90 &          &  0.700 &  0.800 & 0.200 & 0.210 \\[1mm]
     91 &          &  0.800 &  1.000 & 0.085 & 0.234 \\[10mm]

	\multicolumn{6}{c}{Block 5: $(N_{p}, \, p_{p_{2}})$} \\[3mm]
	Global bin number & $N_{p}$ & $p_{p_{2}}^{\text{low}} ~ (\mathrm{GeV}/c)$ & $p_{p_{2}}^{\text{high}} ~ (\mathrm{GeV}/c)$ & Efficiency & Migration diagonal \\[2mm] \hline \strutlike

     92 &    $< 2$ &                      &                      & 0.693 & 0.952 \\[2.5mm]
     93 &        2 &                0.250 &                0.350 & 0.560 & 0.343 \\[1mm]
     94 &          &                0.350 &                0.400 & 0.446 & 0.336 \\[1mm]
     95 &          &                0.400 &                0.450 & 0.387 & 0.390 \\[1mm] 
     96 &          &                0.450 &                0.500 & 0.382 & 0.414 \\[1mm]
     97 &          &                0.500 &                0.550 & 0.394 & 0.463 \\[1mm]
     98 &          &                0.550 &                1.000 & 0.203 & 0.545 \\[2.5mm]
     99 &        3 &                0.250 &                0.450 & 0.369 & 0.202 \\[1mm]
    100 &          &                0.450 &                0.500 & 0.240 & 0.227 \\[1mm]
    101 &          &                0.500 &                0.550 & 0.236 & 0.280 \\[1mm] 
    102 &          &                0.550 &                0.600 & 0.179 & 0.328 \\[1mm]
    103 &          &                0.600 &                0.650 & 0.121 & 0.287 \\[1mm]
    104 &          &                0.650 &                1.000 & 0.033 & 0.325 \\[2.5mm]
    105 & $\geq 4$ &                0.250 &                0.550 & 0.238 & 0.174 \\[1mm]
    106 &          &                0.550 &                0.600 & 0.134 & 0.215 \\[1mm]
    107 &          &                0.600 &                1.000 & 0.030 & 0.275 \\[10mm]

	\multicolumn{6}{c}{Block 6: $(N_{p}, \, p_{p_{3}})$} \\[3mm]
	Global bin number & $N_{p}$ & $p_{p_{3}}^{\text{low}} ~ (\mathrm{GeV}/c)$ & $p_{p_{3}}^{\text{high}} ~ (\mathrm{GeV}/c)$ & Efficiency & Migration diagonal \\[2mm] \hline \strutlike

    108 &    $< 3$ &                      &                      & 0.610 & 0.990 \\[2.5mm]
    109 &        3 &                0.250 &                0.350 & 0.357 & 0.173 \\[1mm]
    110 &          &                0.350 &                0.400 & 0.207 & 0.205 \\[1mm]
    111 &          &                0.400 &                0.450 & 0.162 & 0.301 \\[1mm] 
    112 &          &                0.450 &                0.500 & 0.101 & 0.229 \\[1mm]
    113 &          &                0.500 &                1.000 & 0.044 & 0.326 \\[2.5mm]
    114 & $\geq 4$ &                0.250 &                0.450 & 0.214 & 0.160 \\[1mm]
    115 &          &                0.450 &                0.500 & 0.100 & 0.171 \\[1mm]
    116 &          &                0.500 &                1.000 & 0.030 & 0.311 \\[10mm]

	\multicolumn{6}{c}{Block 7: $(N_{p}, \, p_{p_{4}})$} \\[3mm]
	Global bin number & $N_{p}$ & $p_{p_{4}}^{\text{low}} ~ (\mathrm{GeV}/c)$ & $p_{p_{4}}^{\text{high}} ~ (\mathrm{GeV}/c)$ & Efficiency & Migration diagonal \\[2mm] \hline \strutlike

    117 &    $< 4$ &                      &                      & 0.578 & 0.999 \\[2.5mm]
    118 & $\geq 4$ &                0.250 &                0.400 & 0.190 & 0.165 \\[1mm]
    119 &          &                0.400 &                1.000 & 0.037 & 0.241 \\[10mm]

	\multicolumn{6}{c}{Block 8: $(N_{p}, \, \mathrm{cos} \, \theta_{\mu p_{1}})$} \\[3mm]
	Global bin number & $N_{p}$ & $\mathrm{cos} \, \theta_{\mu p_{1}}^{\text{low}}$ & $\mathrm{cos} \, \theta_{\mu p_{1}}^{\text{high}}$ & Efficiency & Migration diagonal \\[2mm] \hline \strutlike

    120 &        1 & -1.000 & -0.800 & 0.365 & 0.764 \\[1mm]
    121 &          & -0.800 & -0.600 & 0.641 & 0.702 \\[1mm]
    122 &          & -0.600 & -0.400 & 0.682 & 0.658 \\[1mm]
    123 &          & -0.400 & -0.200 & 0.738 & 0.659 \\[1mm] 
    124 &          & -0.200 &  0.000 & 0.776 & 0.661 \\[1mm]
    125 &          &  0.000 &  0.200 & 0.788 & 0.657 \\[1mm]
    126 &          &  0.200 &  0.400 & 0.784 & 0.631 \\[1mm]
    127 &          &  0.400 &  0.600 & 0.737 & 0.612 \\[1mm]
    128 &          &  0.600 &  0.800 & 0.678 & 0.595 \\[1mm]
    129 &          &  0.800 &  1.000 & 0.587 & 0.583 \\[2.5mm]
    130 &        2 & -1.000 & -0.800 & 0.300 & 0.483 \\[1mm]
    131 &          & -0.800 & -0.600 & 0.427 & 0.388 \\[1mm]
    132 &          & -0.600 & -0.400 & 0.450 & 0.400 \\[1mm]
    133 &          & -0.400 & -0.200 & 0.430 & 0.368 \\[1mm] 
    134 &          & -0.200 &  0.000 & 0.418 & 0.370 \\[1mm]
    135 &          &  0.000 &  0.200 & 0.422 & 0.399 \\[1mm]
    136 &          &  0.200 &  0.400 & 0.453 & 0.427 \\[1mm]
    137 &          &  0.400 &  0.600 & 0.481 & 0.457 \\[1mm]
    138 &          &  0.600 &  0.800 & 0.487 & 0.472 \\[1mm]
    139 &          &  0.800 &  1.000 & 0.417 & 0.440 \\[2.5mm]
    140 &        3 & -1.000 & -0.750 & 0.226 & 0.209 \\[1mm]
    141 &          & -0.750 & -0.500 & 0.283 & 0.200 \\[1mm]
    142 &          & -0.500 & -0.250 & 0.265 & 0.173 \\[1mm]
    143 &          & -0.250 &  0.000 & 0.268 & 0.196 \\[1mm] 
    144 &          &  0.000 &  0.250 & 0.248 & 0.204 \\[1mm]
    145 &          &  0.250 &  0.500 & 0.238 & 0.194 \\[1mm]
    146 &          &  0.500 &  0.750 & 0.254 & 0.236 \\[1mm]
    147 &          &  0.750 &  1.000 & 0.223 & 0.225 \\[2.5mm]
    148 & $\geq 4$ & -1.000 & -0.500 & 0.146 & 0.160 \\[1mm]
    149 &          & -0.500 &  0.000 & 0.156 & 0.152 \\[1mm]
    150 &          &  0.000 &  0.500 & 0.135 & 0.156 \\[1mm]
    151 &          &  0.500 &  1.000 & 0.119 & 0.169 \\[10mm]

	\multicolumn{6}{c}{Block 9: $(N_{p}, \, \mathrm{cos} \, \theta_{\mu p_{2}})$} \\[3mm]
	Global bin number & $N_{p}$ & $\mathrm{cos} \, \theta_{\mu p_{2}}^{\text{low}}$ & $\mathrm{cos} \, \theta_{\mu p_{2}}^{\text{high}}$ & Efficiency & Migration diagonal \\[2mm] \hline \strutlike

    152 &    $< 2$ &                      &                      & 0.693 & 0.952 \\[2.5mm]
    153 &        2 &               -1.000 &               -0.800 & 0.434 & 0.484 \\[1mm]
    154 &          &               -0.800 &               -0.600 & 0.533 & 0.391 \\[1mm]
    155 &          &               -0.600 &               -0.400 & 0.511 & 0.346 \\[1mm]
    156 &          &               -0.400 &               -0.200 & 0.468 & 0.337 \\[1mm] 
    157 &          &               -0.200 &                0.000 & 0.426 & 0.315 \\[1mm]
    158 &          &                0.000 &                0.200 & 0.393 & 0.290 \\[1mm]
    159 &          &                0.200 &                0.400 & 0.395 & 0.315 \\[1mm]
    160 &          &                0.400 &                0.600 & 0.400 & 0.324 \\[1mm]
    161 &          &                0.600 &                0.800 & 0.420 & 0.344 \\[1mm]
    162 &          &                0.800 &                1.000 & 0.379 & 0.289 \\[2.5mm]
    163 &        3 &               -1.000 &               -0.666 & 0.257 & 0.167 \\[1mm]
    164 &          &               -0.666 &               -0.333 & 0.327 & 0.187 \\[1mm]
    165 &          &               -0.333 &                0.000 & 0.253 & 0.164 \\[1mm]
    166 &          &                0.000 &                0.333 & 0.237 & 0.179 \\[1mm]
    167 &          &                0.333 &                0.666 & 0.237 & 0.171 \\[1mm]
    168 &          &                0.666 &                1.000 & 0.213 & 0.166 \\[2.5mm]
    169 & $\geq 4$ &               -1.000 &               -0.500 & 0.153 & 0.124 \\[1mm]
    170 &          &               -0.500 &                0.000 & 0.155 & 0.120 \\[1mm]
    171 &          &                0.000 &                0.500 & 0.141 & 0.136 \\[1mm]
    172 &          &                0.500 &                1.000 & 0.113 & 0.130 \\[10mm]

	\multicolumn{6}{c}{Block 10: $(N_{p}, \, \mathrm{cos} \, \theta_{\mu p_{3}})$} \\[3mm]
	Global bin number & $N_{p}$ & $\mathrm{cos} \, \theta_{\mu p_{3}}^{\text{low}}$ & $\mathrm{cos} \, \theta_{\mu p_{3}}^{\text{high}}$ & Efficiency & Migration diagonal \\[2mm] \hline \strutlike

    173 &    $< 3$ &                      &                      & 0.610 & 0.990 \\[2.5mm]
    174 &        3 &               -1.000 &               -0.500 & 0.266 & 0.226 \\[1mm]
    175 &          &               -0.500 &                0.000 & 0.268 & 0.166 \\[1mm]
    176 &          &                0.000 &                0.500 & 0.251 & 0.168 \\[1mm]
    177 &          &                0.500 &                1.000 & 0.217 & 0.139 \\[2.5mm]
    178 & $\geq 4$ &               -1.000 &                0.000 & 0.150 & 0.145 \\[1mm]
    179 &          &                0.000 &                1.000 & 0.126 & 0.140 \\[10mm]

	\multicolumn{6}{c}{Block 11: $(N_{p}, \, \mathrm{cos} \, \theta_{\mu p_{4}})$} \\[3mm]
	Global bin number & $N_{p}$ & $\mathrm{cos} \, \theta_{\mu p_{4}}^{\text{low}}$ & $\mathrm{cos} \, \theta_{\mu p_{4}}^{\text{high}}$ & Efficiency & Migration diagonal \\[2mm] \hline \strutlike

    180 &    $< 4$ &                      &                      & 0.578 & 0.999 \\[2.5mm]
    181 & $\geq 4$ &               -1.000 &                0.000 & 0.149 & 0.141 \\[1mm]
    182 &          &                0.000 &                1.000 & 0.126 & 0.130

\end{longtable*}

\section{Data-driven response matrix validation}
\label{app:resim_validation}

The detector response matrix used in the cross-section extraction depends on the simulated proton kinematics, including those of the subleading protons near the reconstruction threshold. If our simulation mismodels the proton kinematics, the response matrix may be incorrect, potentially biasing the unfolded results. To assess the sensitivity to this effect, a data-driven validation is performed in which alternative response matrices are constructed from resimulated data and MC events.

Selected data and MC events are resimulated using the reconstructed muon and proton momenta as inputs to the full simulation and reconstruction chain. Each event is reprocessed multiple times, with the vertex position randomized within the fiducial volume and the three-momenta rotated about the beam axis to sample the detector response more broadly. Since the resimulated samples are constructed from events that already passed the selection, they are biased towards the reconstructable region of phase space and therefore cannot be used to estimate the selection efficiency. Thus, the resimulated response matrices are built by combining the migration matrices from the resimulation with the efficiency from the central-value MC
\begin{equation}
    \Delta_{i\mu}^{\text{resim}} = \epsilon_{\mu}^{\text{CV}} \, M_{i\mu}^{\text{resim}},
\end{equation}
where the resimulated migration matrix is given by
\begin{equation}
    M_{i\mu}^{\text{resim}} = \frac{\varphi_{i\mu}^{\text{resim}}}{\sum_{j} \varphi_{j\mu}^{\text{resim}}},
\end{equation}
with $\varphi_{i\mu}^{\text{resim}}$ being the number of selected events after resimulation simultaneously falling in resimulated reconstructed bin $i$ and original reconstructed bin $\mu$.

The resimulated data and MC response matrices are each used to unfold central-value fake data. The statistical uncertainty on the unfolded result arising from the finite resimulation statistics is propagated through the D'Agostini unfolding via the 3-dimensional Jacobian tensor
\begin{equation}
	j^{\mu \, (n+1)}_{i\nu} = \frac{\partial \hat{\varphi}^{(n+1)}_{\mu}}{\partial \Delta_{i\nu}^{\text{resim}} } = \sum_{j} \frac{\partial U^{(n)}_{j \mu}}{\partial \Delta_{i\nu}^{\text{resim}} } \, d_{j}.
\end{equation}
Since the resimulated response matrix entries follow multinomial statistics, their covariance is
\begin{equation}
    \text{Cov}(\Delta_{i\mu}^{\text{resim}}, \, \Delta_{j\mu}^{\text{resim}}) = \frac{\delta_{ij} \, \Delta_{i\mu}^{\text{resim}} - \Delta_{i\mu}^{\text{resim}} \, \Delta_{j\mu}^{\text{resim}}}{\varphi_{\mu}^{\text{resim}}}.
\end{equation}

The final comparison tests whether the fake data unfolded with the data- and MC-derived response matrices are compatible. The contribution of the response matrix to the statistical covariance between the unfolded event counts in the $\mu$-th and $\nu$-th true bins is given by
\begin{equation}
	V^{\text{respmat}}_{\mu\nu} = \sum_{\alpha} \sum_{i,j} j^{\mu}_{i\alpha} \, \text{Cov}(\Delta_{i\alpha}^{\text{resim}}, \,\Delta_{j\alpha}^{\text{resim}}) \, j^{\nu}_{j\alpha}.
\end{equation}
The covariance of the difference is the sum of the individual response matrix uncertainties, $C_{\Delta} = V_{\text{MC}}^{\text{respmat}} + V_{\text{data}}^{\text{respmat}}$. To account for differences in regularization between the two unfoldings, the residuals are corrected using the respective additional smearing matrices
\begin{equation}
    \delta_{\text{bias}} = (A_{C}^{\text{MC}} - A_{C}^{\text{data}}) \, \varphi^{\text{CV}}.
\end{equation}
The $\chi^{2}$ for the comparison is then
\begin{equation}
    \chi^{2} = (\Delta \hat{\varphi}^{\text{resim}} - \delta_{\text{bias}})^{\mathsf{T}} \, C_{\Delta}^{-1} \, (\Delta \hat{\varphi}^{\text{resim}} - \delta_{\text{bias}}).
\end{equation}

It should be noted that the resimulated response matrices represent a mapping between original reconstructed and resimulated reconstructed bins, rather than the usual truth-to-reconstructed transformation. This approximation affects both the data and MC resimulated response matrices equally, so the comparison between the two remains a valid test of sensitivity to differences in proton kinematics between data and simulation.

\section{Sideband-based background validation}
\label{app:sideband_background}

Cross-section extraction methods that rely on MC predictions for background subtraction are sensitive to the accuracy of the background model. When a discrepancy between data and prediction is observed in the selected sample, it is important to determine whether it originates from mismodeling of the signal, the background, or both. Here, we outline a general procedure to disentangle these two contributions using a background-enhanced sideband selection. In this analysis, this method is applied to the leading proton opening angle distributions (bin blocks 2 and 8), which show significant tension between data and MC.

In each bin, the data in the selection and sideband regions can be modeled as
\begin{equation}\label{eq:sb}
	\begin{split}
		D^{\text{sel}}_{i} &= \alpha_{i} \, \varphi^{\text{sel}}_{i} + \beta_{i} \, b^{\text{sel}}_{i}, \\
    	D^{\text{sb}}_{i} &= \alpha_{i} \, \varphi^{\text{sb}}_{i} + \beta_{i} \, b^{\text{sb}}_{i},
	\end{split}
\end{equation}
where $\alpha_{i}$ and $\beta_{i}$ are bin-dependent correction factors for signal and background events respectively, the superscripts denote the region (selection or sideband), $D_{i}$ is the total number of measured events, $\varphi_{i}$ is the expected number of selected signal events, and $b_{i}$ is the total number of expected background events (both beam-off and beam-correlated backgrounds) for bin $i$. The key assumption is that the correction factors are properties of the underlying physics modeling and are therefore the same in both regions.

The solution in each bin can be written in matrix form as
\begin{equation}\label{eq:correction_matrix_form}
	\begin{pmatrix} \alpha_{i} - 1 \\ \beta_{i} - 1 \end{pmatrix} = \begin{pmatrix} \varphi^{\text{sel}}_{i} & b^{\text{sel}}_{i} \\ \varphi^{\text{sb}}_{i} & b^{\text{sb}}_{i} \end{pmatrix}^{-1} \, \begin{pmatrix} D^{\text{sel}}_{i} - n^{\text{sel}}_{i} \\ D^{\text{sb}}_{i} - n^{\text{sb}}_{i} \end{pmatrix},
\end{equation}
where $n_{i} = \varphi_{i} + b_{i}$ denotes the total number of predicted events in bin $i$. The covariance matrices for $\alpha_{i}$ and $\beta_{i}$ are obtained from the total covariance matrix (data plus prediction) and the Jacobian elements determined by the inverse matrix in the linear system.

The null hypothesis is that the background is correctly modeled, i.e.\ $\beta_{i} = 1$ for all bins, and that all observed discrepancy between data and MC is due to signal mismodeling alone. This is tested with a $\chi^2$ statistic
\begin{equation}\label{eq:chi2}
    \chi^2 = \sum_{i,\,j} \left(\beta_{i} - 1\right) \, \mathrm{Cov}(\beta_{i}, \, \beta_{j})^{-1} \, \left(\beta_{j} - 1\right).
\end{equation}
Under the null hypothesis, this statistic follows a $\chi^2$ distribution with $N_{\text{bins}}$ degrees of freedom, where $N_{\text{bins}}$ is the number of bins in the block.

\section{Covariance matrix decomposition}
\label{app:covariance_decomp}

Given a histogram $\phi$ with an associated covariance matrix $C$ not including statistical uncertainties, this matrix can be decomposed into separate normalization, shape, and mixed contributions following the decomposition \cite{MicroBooNE:2024yzp}:
\begin{equation}
    C_{ij} = C_{ij}^{\text{norm}} + C_{ij}^{\text{shape}} + C_{ij}^{\text{mixed}},
\end{equation}
\begin{equation}
    C_{ij}^{\text{norm}} = \frac{\phi_{i} \, \phi_{j}}{\phi_{\text{tot}}^2} \sum_{k,l} C_{kl},
\end{equation}
\begin{equation}
    C_{ij}^{\text{shape}} = C_{ij} - \frac{\phi_{j}}{\phi_{\text{tot}}} \sum_{k} C_{ik} - \frac{\phi_{i}}{\phi_{\text{tot}}} \sum_{k} C_{kj} + C_{ij}^{\text{norm}},
\end{equation}
\begin{equation}
    C_{ij}^{\text{mixed}} = \frac{\phi_{j}}{\phi_{\text{tot}}} \sum_{k} C_{ik} + \frac{\phi_{i}}{\phi_{\text{tot}}} \sum_{k} C_{kj} - 2 \, C_{ij}^{\text{norm}},
\end{equation}
where $\phi_{\text{tot}} = \sum_{i} \phi_{i}$ is the sum of all the entries in the histogram.

The normalization covariance matrix is related to uncertainties which change the histogram by the same overall factor. The shape matrix is associated with effects that keep the total number of entries constant. The mixed matrix captures the correlation between the overall normalization and the bin-dependent shape variations.

\phantom{AAAA}

\section{Range-projected \boldmath{$\chi^{2}$} test statistic}
\label{app:global_chi2}

The combined measurement approach proposed in Ref.~\cite{Gardiner:2024gdy} introduces event sharing between bins across different distribution blocks. This sharing structure can be encoded in a combination matrix $\Omega$ that represents all bin configurations allowed by the measurement structure. The null space of $\Omega$ defines $N_{\text{null}}$ exact linear constraints on the bin counts. These constraints render the statistical covariance matrix rank-deficient. A standard $\chi^{2}$ test cannot reliably be applied to the full bin vector without addressing these degenerate directions.

The range-projected $\chi^{2}$ test statistic of Ref.~\cite{MartinezLopez:2026asn} solves this by restricting the test to the subspace where independent statistical information exists. The construction proceeds as follows. The combination matrix $\Omega$ is built in reconstructed space from the bin scheme used in the measurement. The eigendecomposition of $\Omega^{\mathsf{T}} \Omega$ yields $r = \mathrm{rank}(\Omega)$ nonzero eigenvalues, whose associated eigenvectors $V_{r}$ span the non-null subspace. These are the directions along which the measurement has independent statistical sensitivity. The projection is performed in reconstructed space, before unfolding, to avoid contamination from null-space components that would otherwise leak into the range through the non-orthogonal unfolding matrix. Given a reconstructed-space residual vector $\mathbf{r} = \mathbf{d} - \Delta \, \boldsymbol{\varphi}_{\text{pred}}$ with an associated total covariance matrix $C_{\text{total}}$, the projected $\chi^{2}$ is computed as
\begin{equation}\label{eq:range_proj_chi2}
    \chi^{2}_{\text{proj}} = \tilde{\mathbf{r}}^{\,\mathsf{T}} \, \tilde{C}_{\text{total}}^{-1} \, \tilde{\mathbf{r}},
\end{equation}
where $\tilde{\mathbf{r}} = V_r^{\mathsf{T}} \, \mathbf{r}$ is the projected residual vector and
$\tilde{C}_{\text{total}} = V_r^{\mathsf{T}} \, C_{\text{total}} \, V_r$ is the $r \times r$ reduced covariance matrix. The matrix $\tilde{C}_{\text{total}}$ is full-rank and can be inverted with standard methods. Under the null hypothesis, this test statistic follows a $\chi^2$ distribution with $r$ degrees of freedom. This test statistic is invariant under unfolding. The reduced system is full-rank, so the standard $\chi^{2}$ invariance under invertible linear transformations applies.

\FloatBarrier

\bibliography{refs}

\end{document}


\title{First double-differential cross-section measurements in proton multiplicity and kinematics for mesonless $\nu_{\mu}$ charged-current interactions on argon using the MicroBooNE detector}

\maketitle

\section{Data release}

The file \texttt{data\_release.tar.gz} includes all information required to reproduce the results and perform new model comparisons using the measurements from this analysis. On Unix-like systems, the files can be extracted by running:
\begin{center}
	\texttt{tar -xvzf data\_release.tar.gz}
\end{center}
The archive contains the following structure:
\begin{center}
	\begin{minipage}{0.25\textwidth}
		\begin{verbatim}
		data_release/
		|-- DataRelease.root
		|-- README.md
		|-- scripts/
			|-- chi2_example.C
			|-- chi2_example.py
		\end{verbatim}
	\end{minipage}
\end{center}

The main data products are stored in the \texttt{DataRelease.root} file as ROOT \texttt{TH1} and \texttt{TH2} objects. The released cross sections are flux-integrated total cross sections (not divided by bin widths), given in units of $10^{-38} ~ \mathrm{cm}^{2}$ per argon nucleus. They are obtained from the unfolded event counts as:
\begin{equation}
\left< \sigma \right>_{\mu} = \frac{\hat{\varphi}_{\mu}}{\Phi \, \mathcal{N}_{\text{Ar}}},
\end{equation}
\begin{equation}
\text{Cov}(\left< \sigma \right>_{\mu}, \left< \sigma \right>_{\nu}) = \frac{\text{Cov}(\hat{\varphi}_{\mu}, \hat{\varphi}_{\nu})}{\Phi^{2} \, \mathcal{N}_{\mathrm{Ar}}^{2}}.
\end{equation}
To convert to differential cross sections, the cross-section histograms and covariance matrices should be divided by the appropriate product of bin widths. Pre-made unit-transformation histograms are provided for this purpose (\texttt{Hist\_Unit\_Transform} for cross sections, \texttt{Matr\_Unit\_Transform} for covariance matrices).

The file includes the extracted cross sections from the full BNB dataset (\texttt{Hist\_Data}), the total covariance matrix (\texttt{Cov\_Total}), and the data statistical covariance (\texttt{Cov\_Stats\_BNB}). Systematic uncertainties are broken down by source (\texttt{Cov\_*}), including individual cross-section (\texttt{Cov\_Xsec\_*}) and detector variation (\texttt{Cov\_DetVar\_*}) contributions. Predicted cross sections from the neutrino event generators used in the comparisons are also included (\texttt{Hist\_*}).

The unfolding matrix (\texttt{Unfolding\_Matrix}) and the additional smearing matrix (\texttt{Additional\_Smearing\_Matrix}) are provided. The additional smearing matrix must be applied to any model prediction before computing a goodness-of-fit metric against the extracted data, as described in Sec. IV B of the main text.

The file also contains the reduced combination matrix $\Omega^{\mathsf{T}}\Omega$ (\texttt{Combination\_Matrix}) and the projection matrix $P$ (\texttt{Projection\_Matrix}) needed to compute the range-projected $\chi^{2}$ described in Appendix E of the main text.

Example C++ and Python scripts demonstrating the computation of per-block and global $\chi^{2}$ values are provided in the \texttt{scripts/} directory. A \texttt{README.md} file describes the usage and expected outputs.

\section{Efficiency and migration matrices}

\begin{figure}[!h]
	\centering
	\includegraphics[width=.80\linewidth]{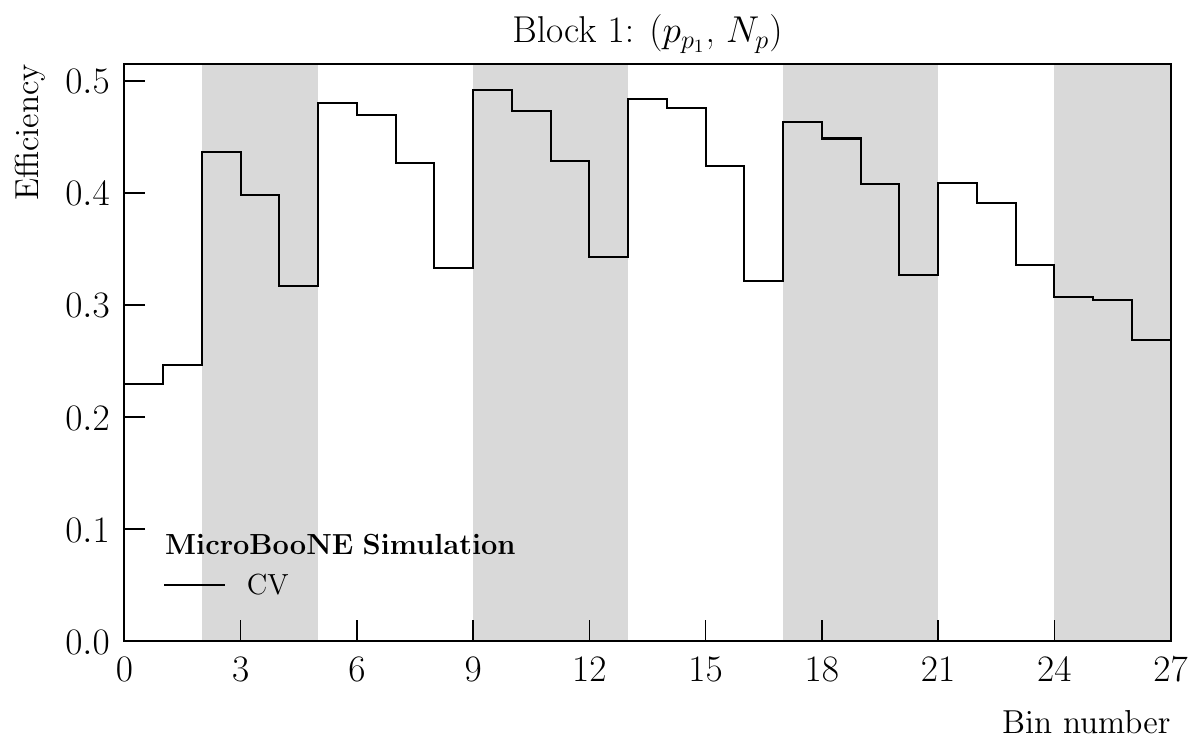}
	\caption{Selection efficiency for the true bins in block 1, used for the double-differential $(p_{p_{1}}, \, N_{p})$ measurement.}
	\label{fig:block_1_efficiency}
\end{figure}

\begin{figure}[!h]
	\centering
	\includegraphics[width=.80\linewidth]{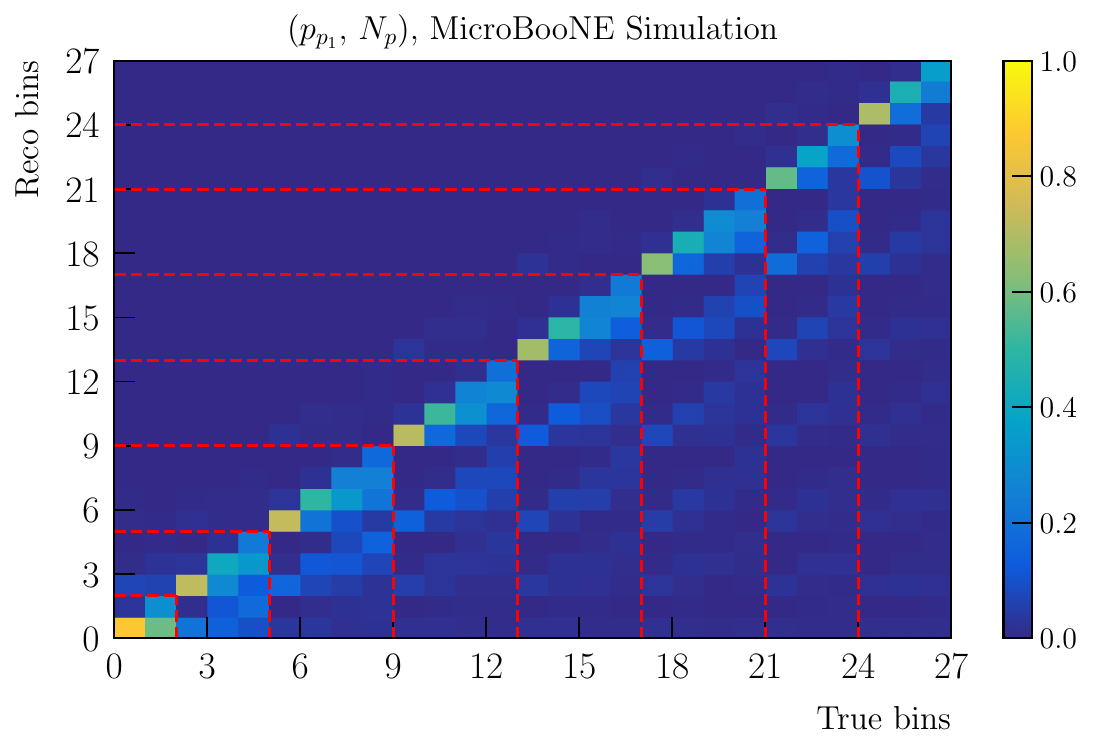}
	\caption{Migration matrix for block 1, used for the double-differential $(p_{p_{1}}, \, N_{p})$ measurement. Dashed red lines indicate the slice boundaries.}
	\label{fig:block_1_migration_matrix}
\end{figure}

\begin{figure}[!h]
	\centering
	\includegraphics[width=.80\linewidth]{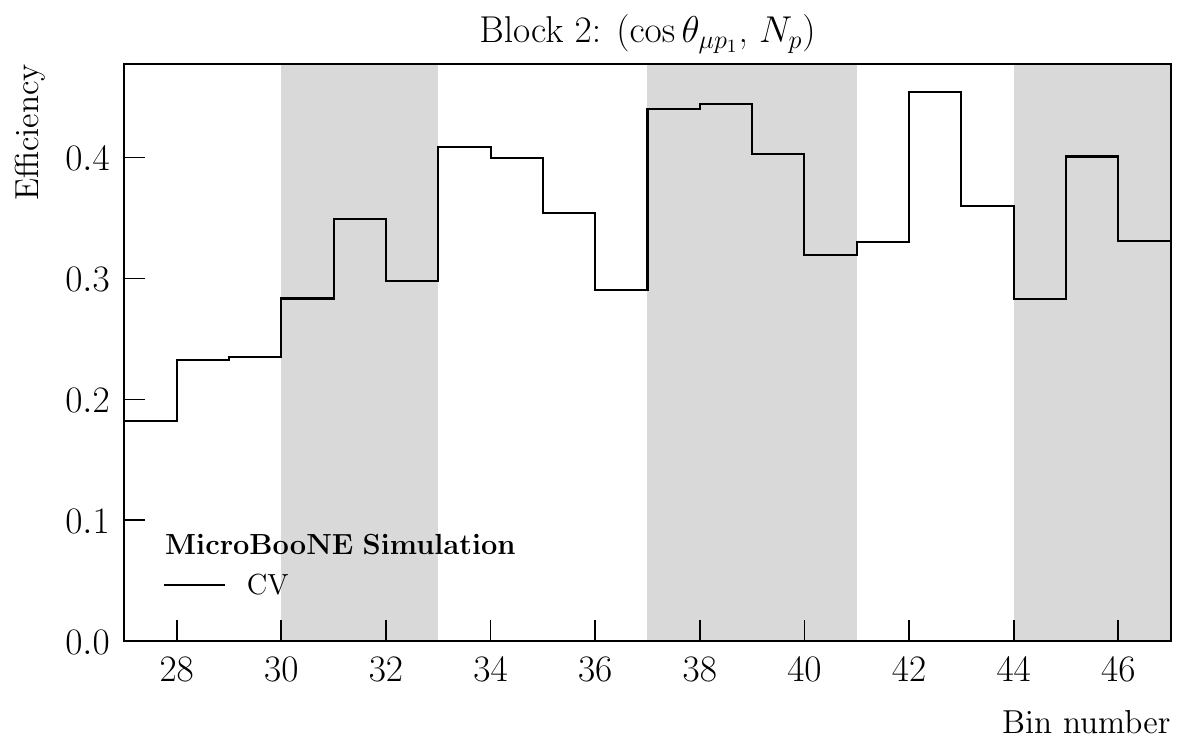}
	\caption{Selection efficiency for the true bins in block 2, used for the double-differential $(\mathrm{cos} \, \theta_{\mu p_{1}}, \, N_{p})$ measurement.}
	\label{fig:block_2_efficiency}
\end{figure}

\begin{figure}[!h]
	\centering
	\includegraphics[width=.80\linewidth]{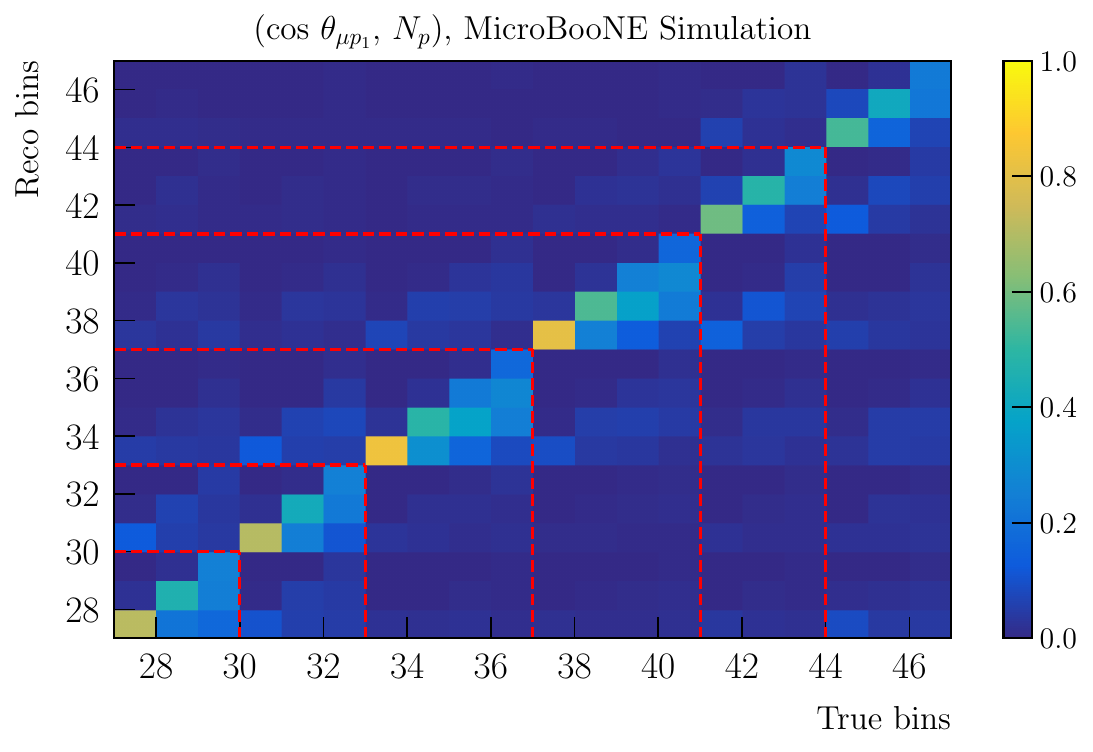}
	\caption{Migration matrix for block 2, used for the double-differential $(\mathrm{cos} \, \theta_{\mu p_{1}}, \, N_{p})$ measurement. Dashed red lines indicate the slice boundaries.}
	\label{fig:block_2_migration_matrix}
\end{figure}

\begin{figure}[!h]
	\centering
	\includegraphics[width=.80\linewidth]{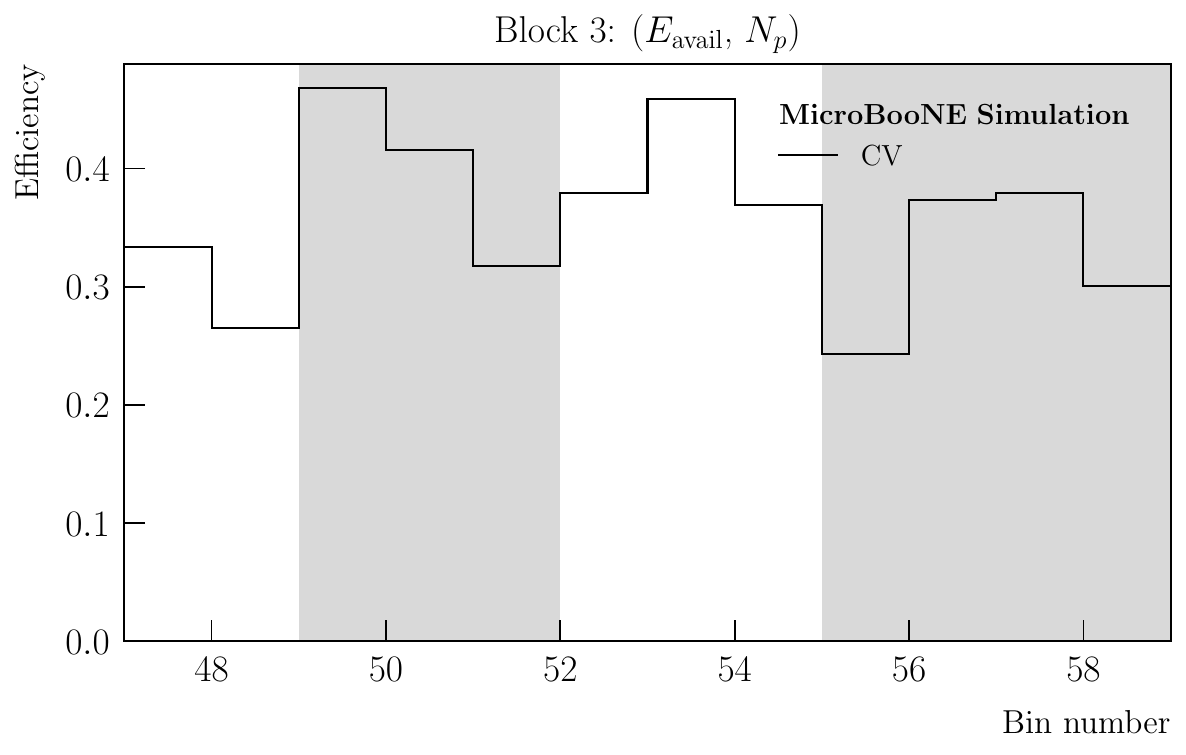}
	\caption{Selection efficiency for the true bins in block 3, used for the double-differential $(E_{\mathrm{avail}}, \, N_{p})$ measurement.}
	\label{fig:block_3_efficiency}
\end{figure}

\begin{figure}[!h]
	\centering
	\includegraphics[width=.80\linewidth]{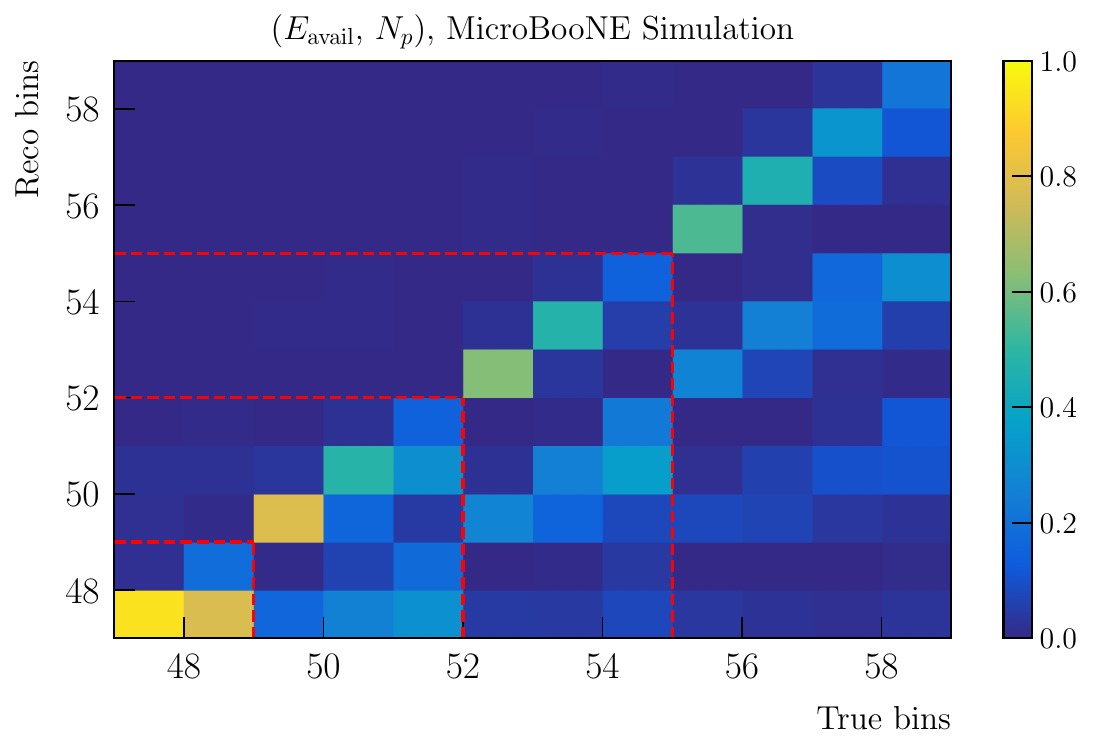}
	\caption{Migration matrix for block 4, used for the double-differential $(E_{\mathrm{avail}}, \, N_{p})$ measurement. Dashed red lines indicate the slice boundaries.}
	\label{fig:block_3_migration_matrix}
\end{figure}

\begin{figure}[!h]
	\centering
	\includegraphics[width=.80\linewidth]{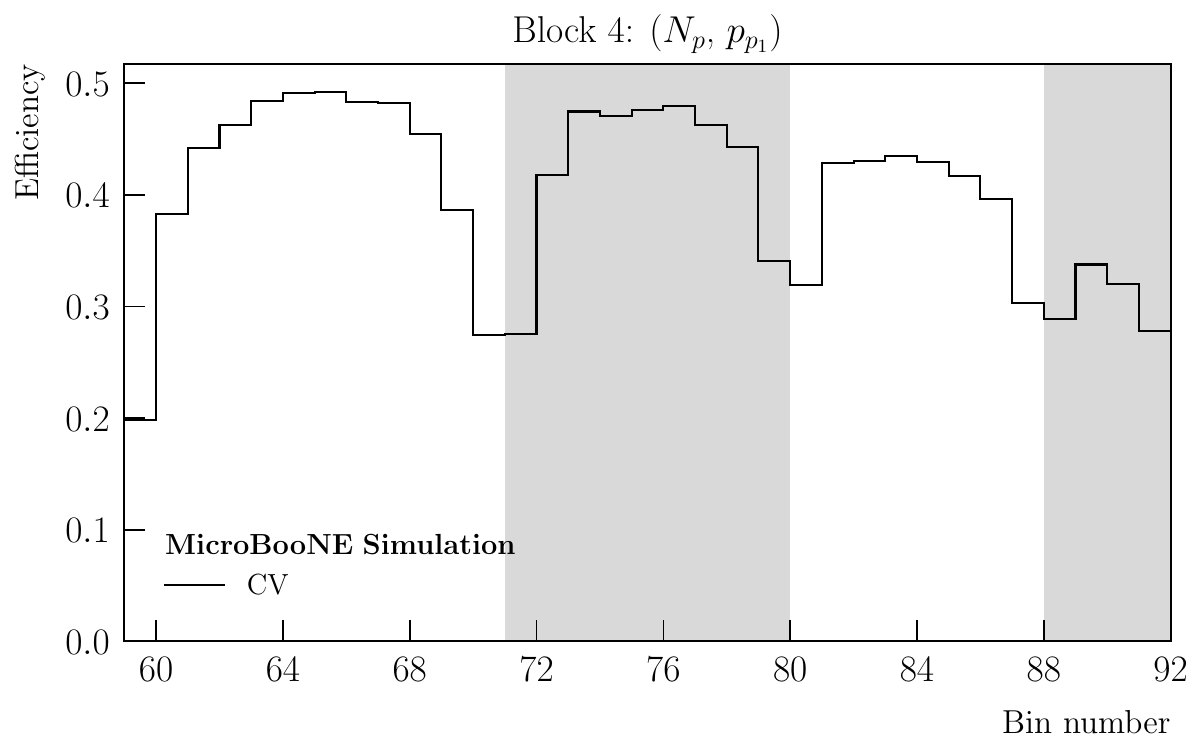}
	\caption{Selection efficiency for the true bins in block 4, used for the double-differential $(N_{p}, \, p_{p_{1}})$ measurement.}
	\label{fig:block_4_efficiency}
\end{figure}

\begin{figure}[!h]
	\centering
	\includegraphics[width=.80\linewidth]{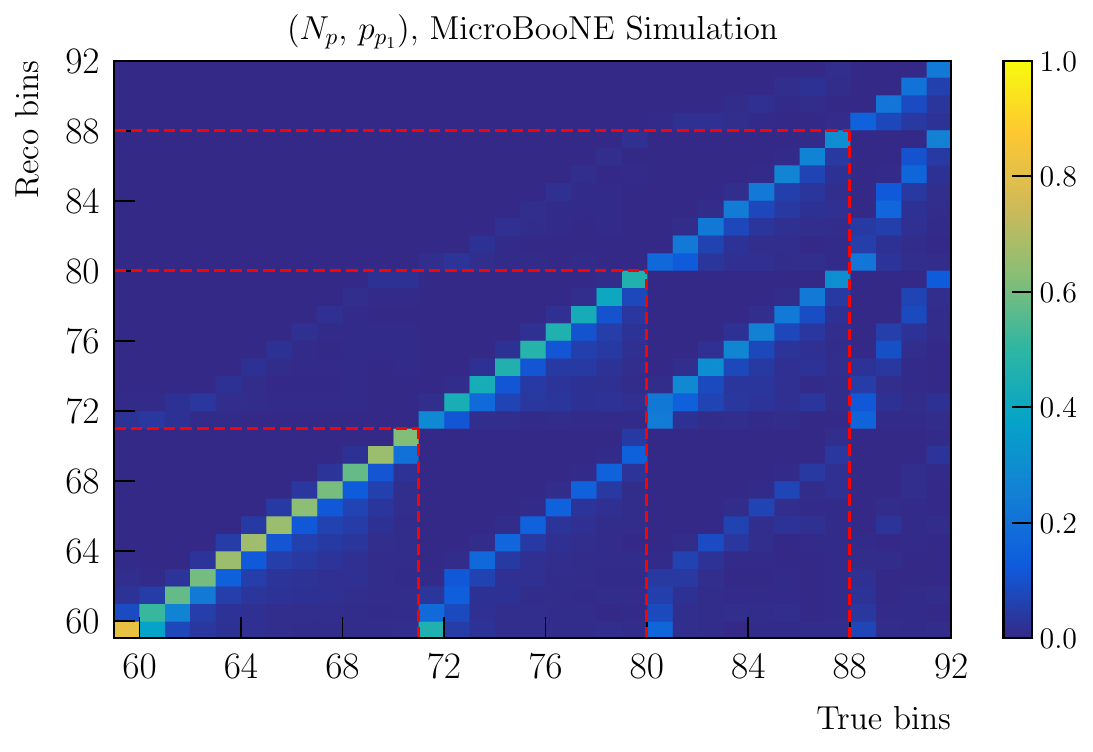}
	\caption{Migration matrix for block 4, used for the double-differential $(N_{p}, \, p_{p_{1}})$ measurement. Dashed red lines indicate the slice boundaries.}
	\label{fig:block_4_migration_matrix}
\end{figure}

\begin{figure}[!h]
	\centering
	\includegraphics[width=.80\linewidth]{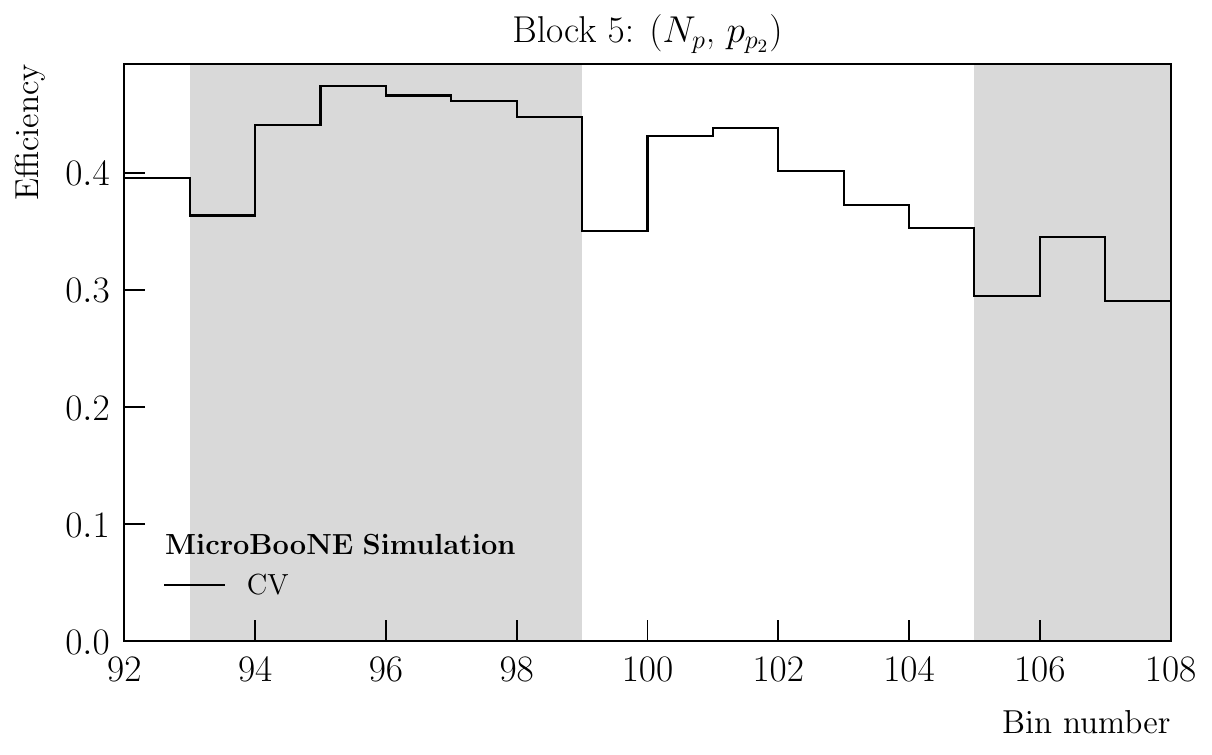}
	\caption{Selection efficiency for the true bins in block 5, used for the double-differential $(N_{p}, \, p_{p_{2}})$ measurement.}
	\label{fig:block_5_efficiency}
\end{figure}

\begin{figure}[!h]
	\centering
	\includegraphics[width=.80\linewidth]{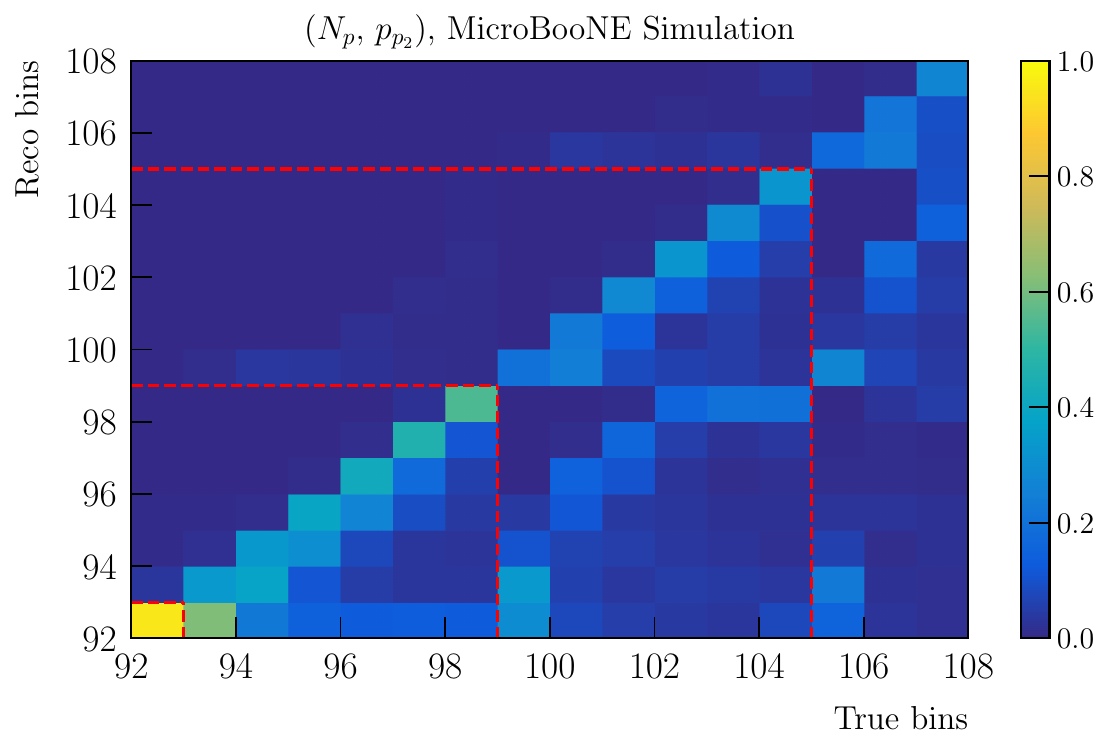}
	\caption{Migration matrix for block 5, used for the double-differential $(N_{p}, \, p_{p_{2}})$ measurement. Dashed red lines indicate the slice boundaries.}
	\label{fig:block_5_migration_matrix}
\end{figure}

\begin{figure}[!h]
	\centering
	\includegraphics[width=.80\linewidth]{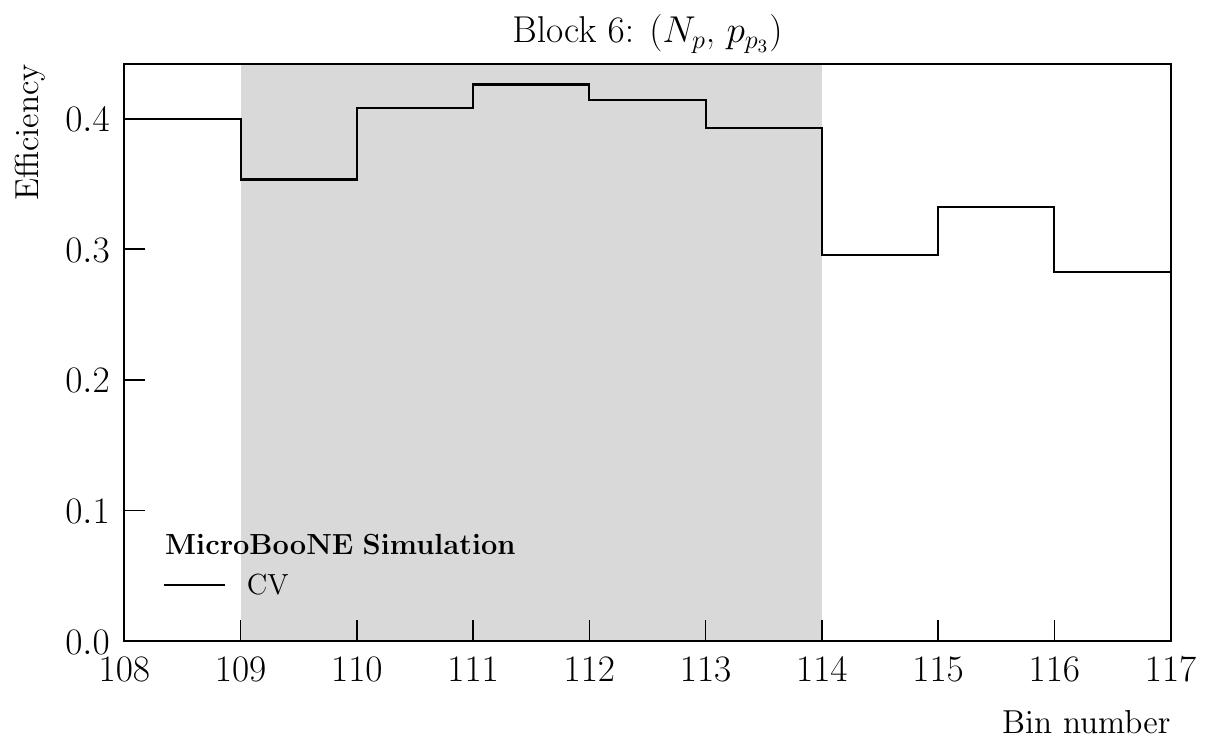}
	\caption{Selection efficiency for the true bins in block 6, used for the double-differential $(N_{p}, \, p_{p_{3}})$ measurement.}
	\label{fig:block_6_efficiency}
\end{figure}

\begin{figure}[!h]
	\centering
	\includegraphics[width=.80\linewidth]{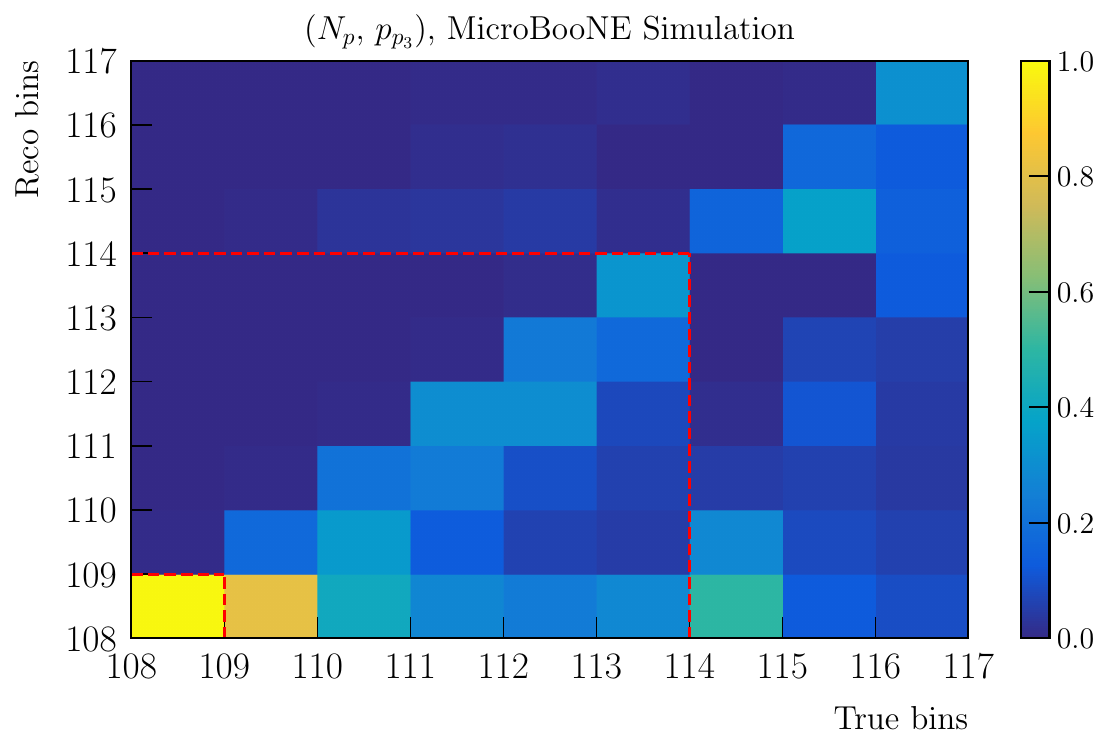}
	\caption{Migration matrix for block 6, used for the double-differential $(N_{p}, \, p_{p_{3}})$ measurement. Dashed red lines indicate the slice boundaries.}
	\label{fig:block_6_migration_matrix}
\end{figure}

\begin{figure}[!h]
	\centering
	\includegraphics[width=.80\linewidth]{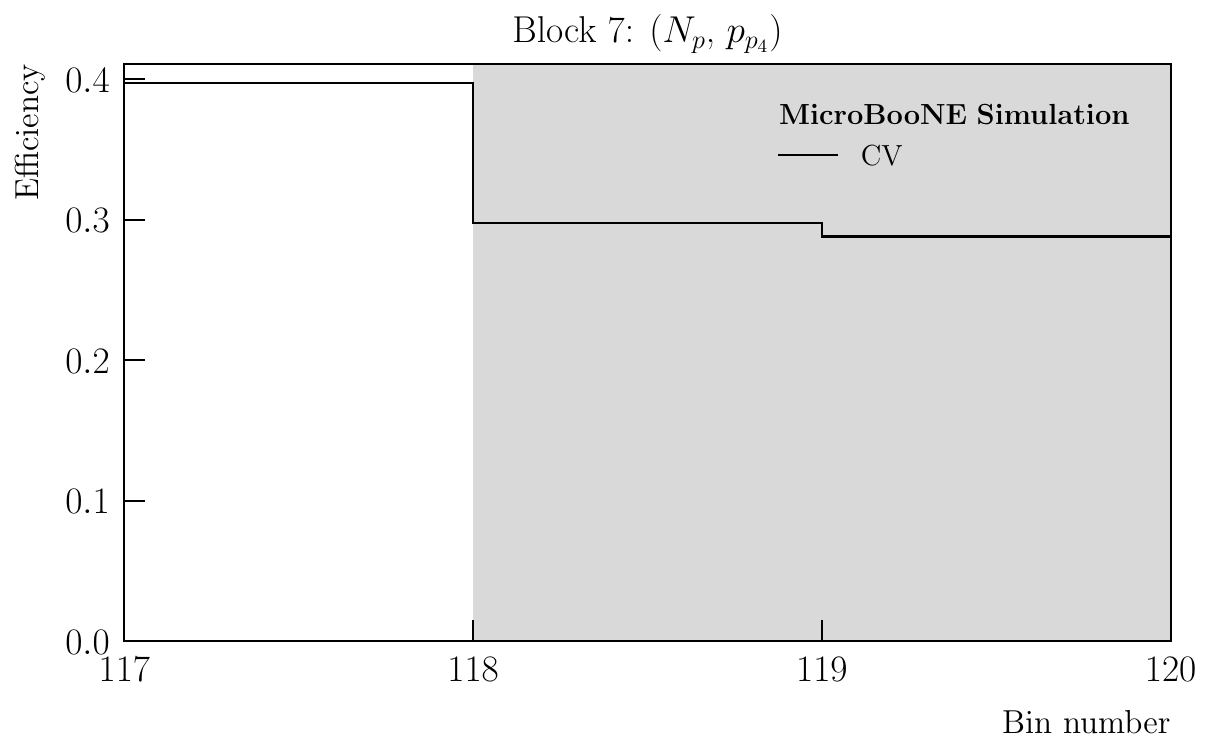}
	\caption{Selection efficiency for the true bins in block 7, used for the double-differential $(N_{p}, \, p_{p_{4}})$ measurement.}
	\label{fig:block_7_efficiency}
\end{figure}

\begin{figure}[!h]
	\centering
	\includegraphics[width=.80\linewidth]{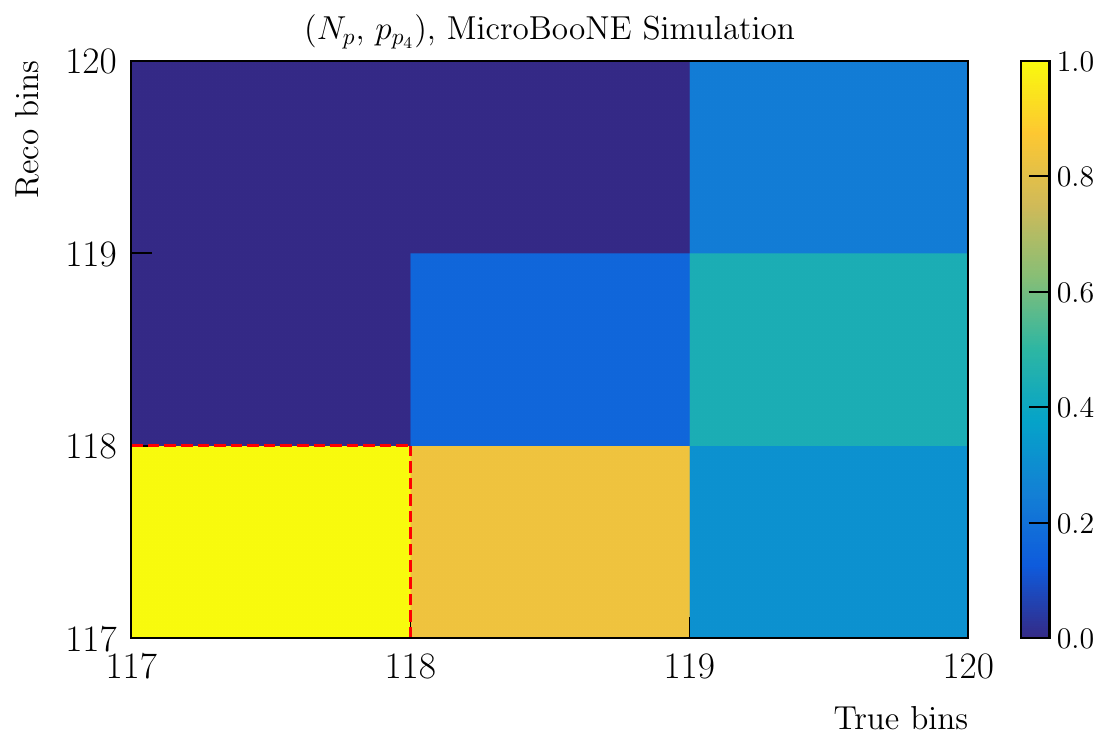}
	\caption{Migration matrix for block 7, used for the double-differential $(N_{p}, \, p_{p_{4}})$ measurement. Dashed red lines indicate the slice boundaries.}
	\label{fig:block_7_migration_matrix}
\end{figure}

\begin{figure}[!h]
	\centering
	\includegraphics[width=.80\linewidth]{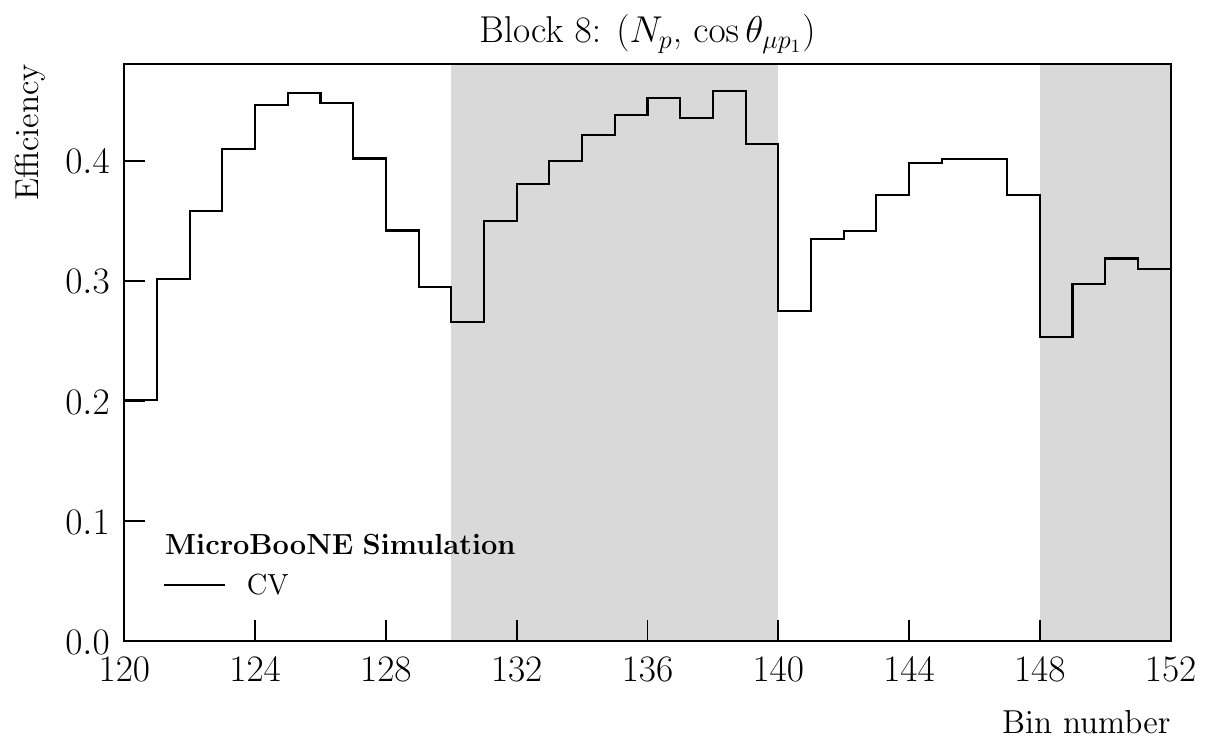}
	\caption{Selection efficiency for the true bins in block 8, used for the double-differential $(N_{p}, \, \mathrm{cos} \, \theta_{\mu p_{1}})$ measurement.}
	\label{fig:block_8_efficiency}
\end{figure}

\begin{figure}[!h]
	\centering
	\includegraphics[width=.80\linewidth]{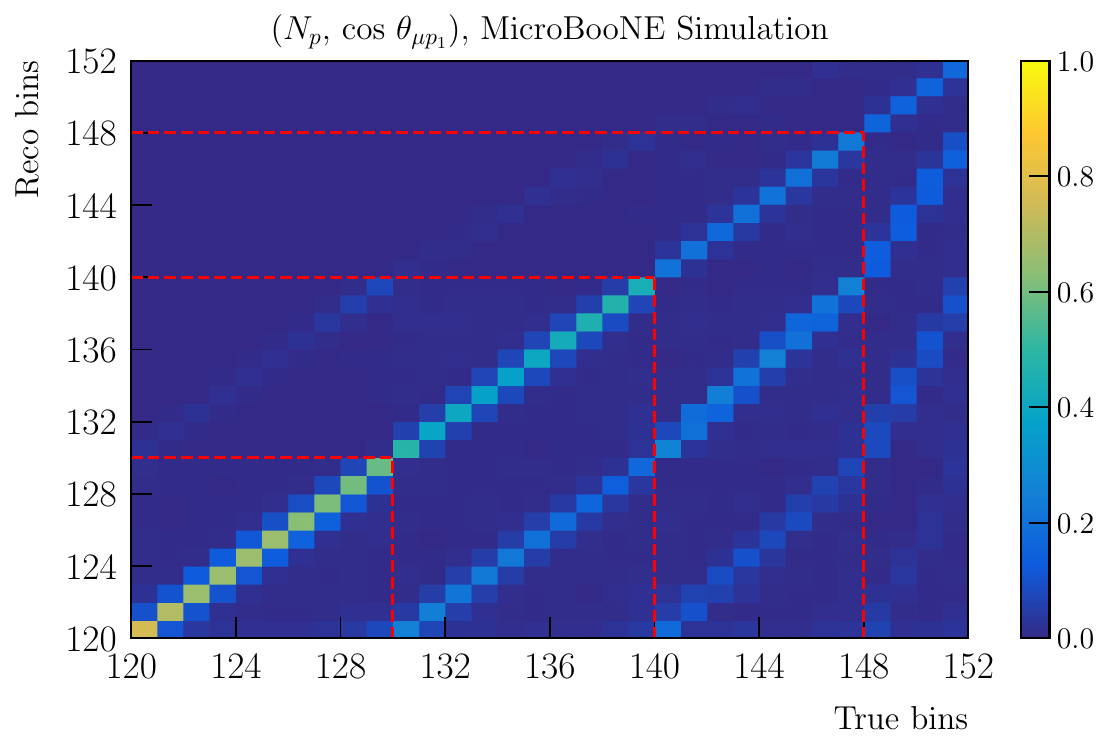}
	\caption{Migration matrix for block 8, used for the double-differential $(N_{p}, \, \mathrm{cos} \, \theta_{\mu p_{1}})$ measurement. Dashed red lines indicate the slice boundaries.}
	\label{fig:block_8_migration_matrix}
\end{figure}

\begin{figure}[!h]
	\centering
	\includegraphics[width=.80\linewidth]{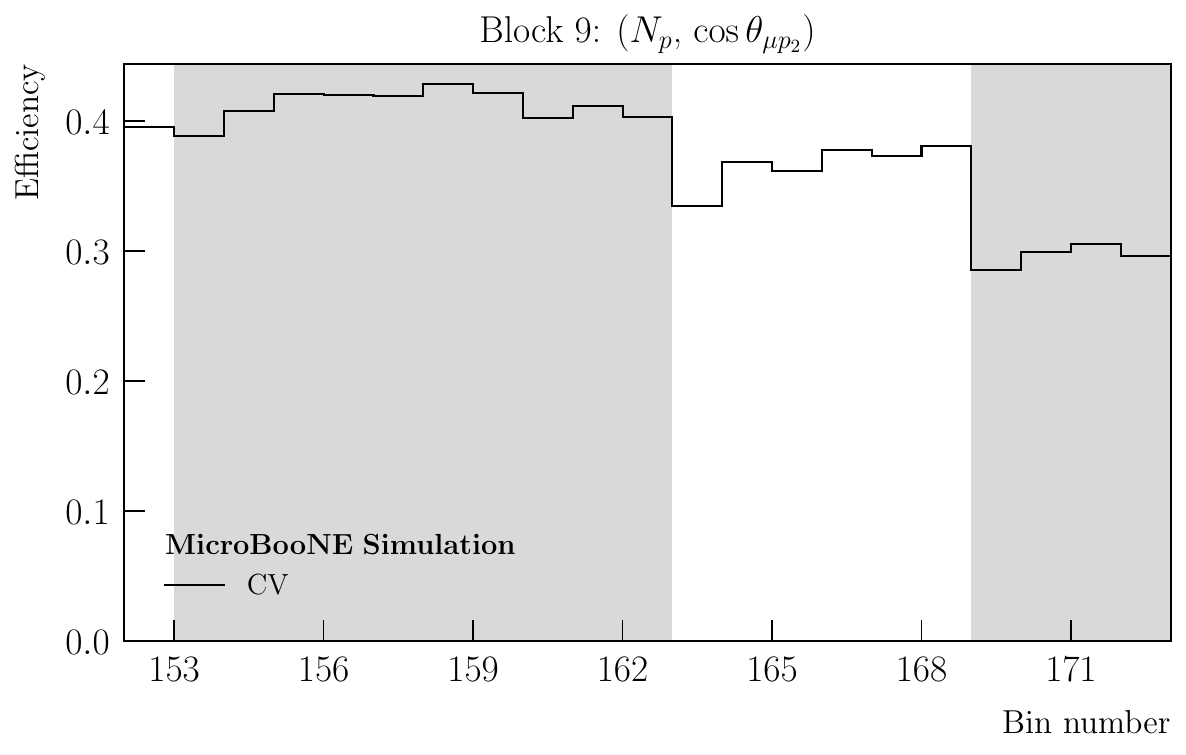}
	\caption{Selection efficiency for the true bins in block 9, used for the double-differential $(N_{p}, \, \mathrm{cos} \, \theta_{\mu p_{2}})$ measurement.}
	\label{fig:block_9_efficiency}
\end{figure}

\begin{figure}[!h]
	\centering
	\includegraphics[width=.80\linewidth]{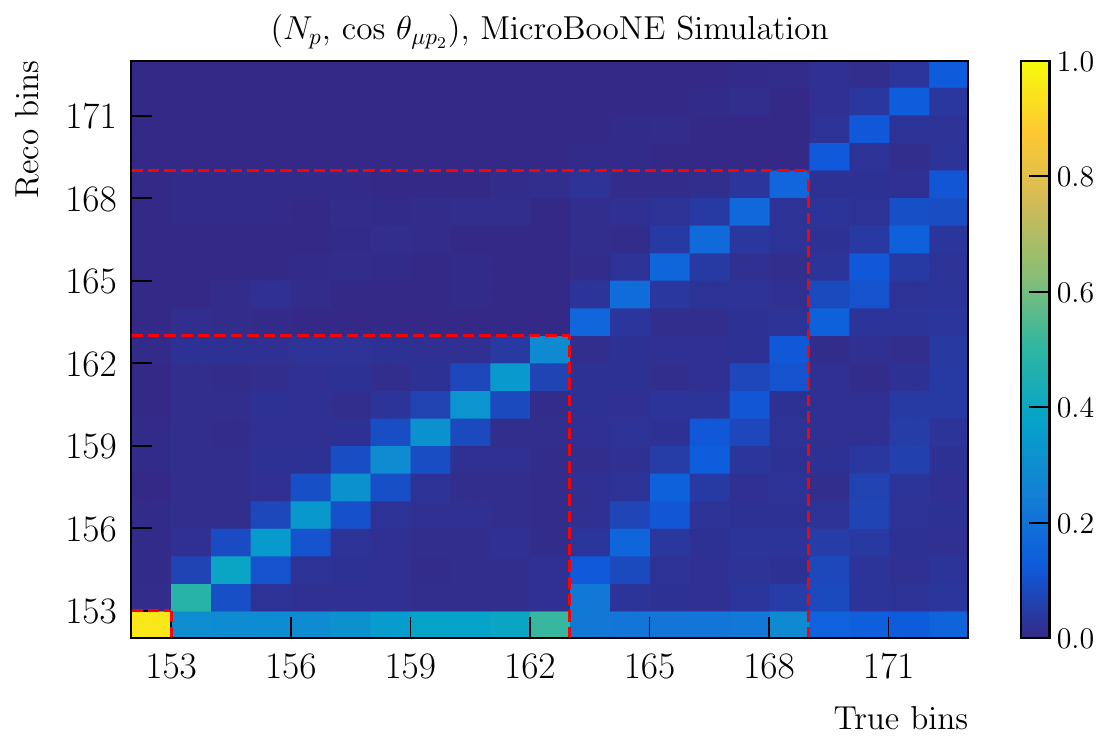}
	\caption{Migration matrix for block 9, used for the double-differential $(N_{p}, \, \mathrm{cos} \, \theta_{\mu p_{2}})$ measurement. Dashed red lines indicate the slice boundaries.}
	\label{fig:block_9_migration_matrix}
\end{figure}

\begin{figure}[!h]
	\centering
	\includegraphics[width=.80\linewidth]{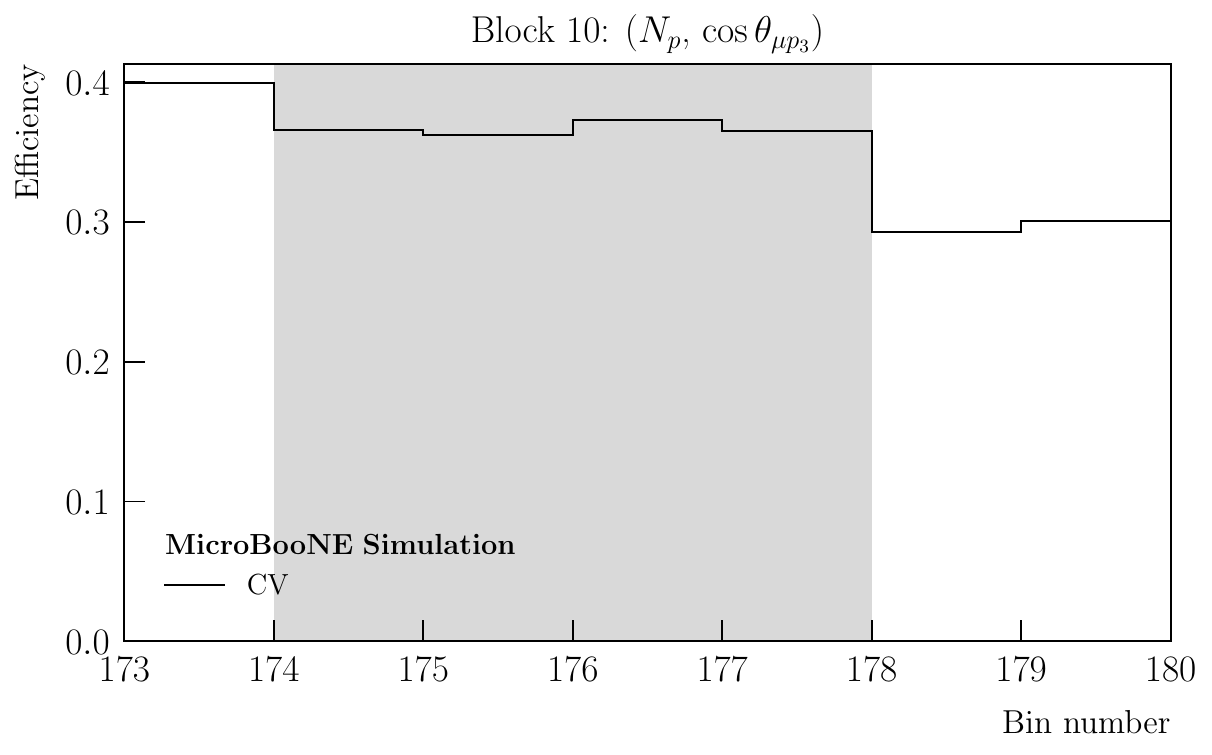}
	\caption{Selection efficiency for the true bins in block 10, used for the double-differential $(N_{p}, \, \mathrm{cos} \, \theta_{\mu p_{3}})$ measurement.}
	\label{fig:block_10_efficiency}
\end{figure}

\begin{figure}[!h]
	\centering
	\includegraphics[width=.80\linewidth]{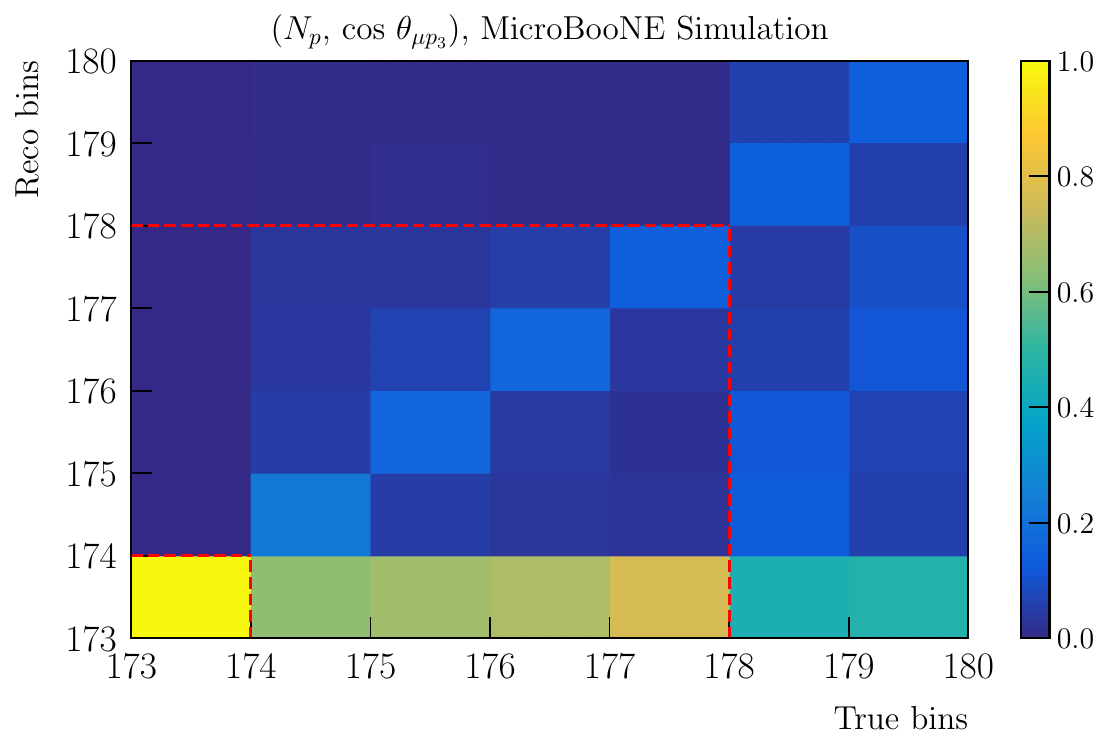}
	\caption{Migration matrix for block 10, used for the double-differential $(N_{p}, \, \mathrm{cos} \, \theta_{\mu p_{3}})$ measurement. Dashed red lines indicate the slice boundaries.}
	\label{fig:block_10_migration_matrix}
\end{figure}

\begin{figure}[!h]
	\centering
	\includegraphics[width=.80\linewidth]{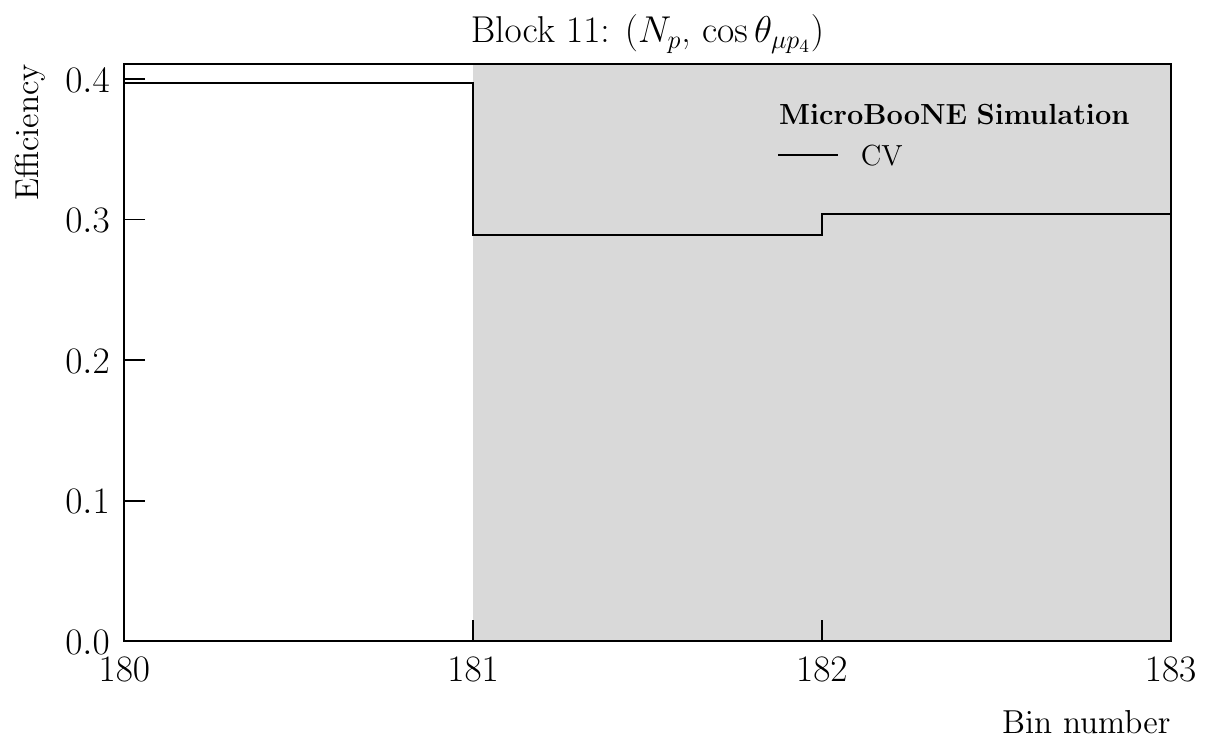}
	\caption{Selection efficiency for the true bins in block 11, used for the double-differential $(N_{p}, \, \mathrm{cos} \, \theta_{\mu p_{4}})$ measurement.}
	\label{fig:block_11_efficiency}
\end{figure}

\begin{figure}[!h]
	\centering
	\includegraphics[width=.80\linewidth]{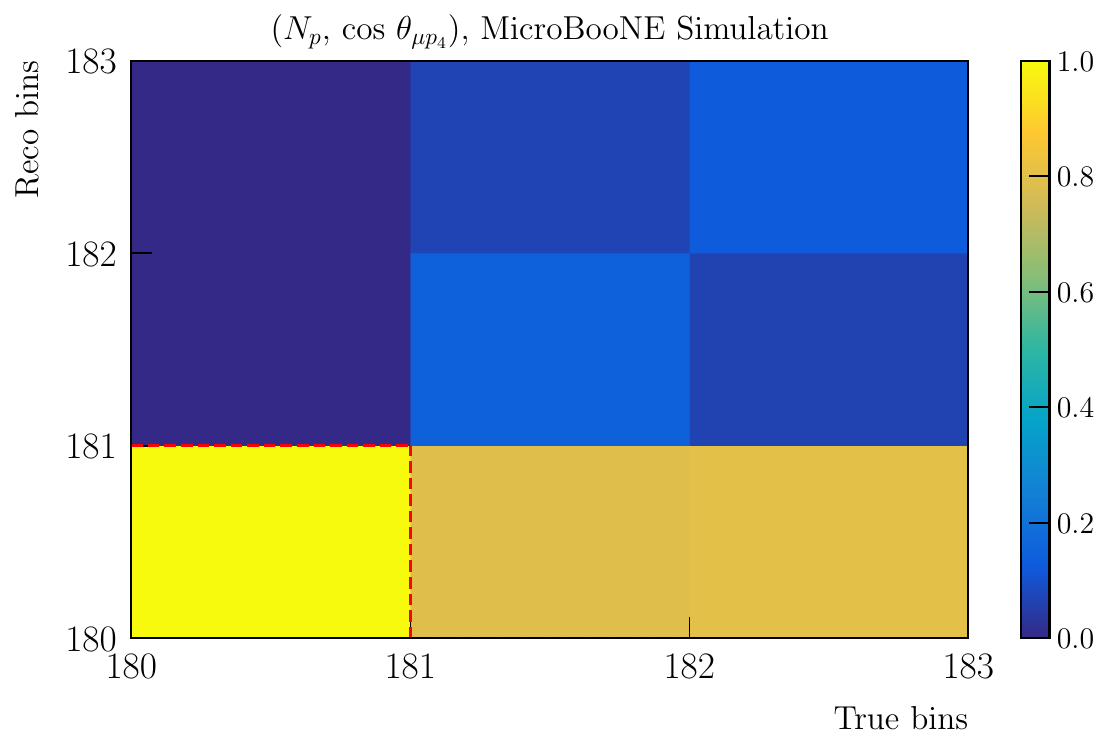}
	\caption{Migration matrix for block 11, used for the double-differential $(N_{p}, \, \mathrm{cos} \, \theta_{\mu p_{4}})$ measurement. Dashed red lines indicate the slice boundaries.}
	\label{fig:block_11_migration_matrix}
\end{figure}

The selection efficiency and migration matrix, introduced in Eq. (3) of the main text, quantify the detector effects on the observables used in this analysis. Figures \ref{fig:block_1_efficiency} through \ref{fig:block_11_migration_matrix} show the per-bin selection efficiency and migration matrix for each of the 11 blocks in the analysis.

\FloatBarrier

\section{Event rates by interaction mode}

\begin{figure*}[!h]
	\centering
	\includegraphics[width=0.83\linewidth]{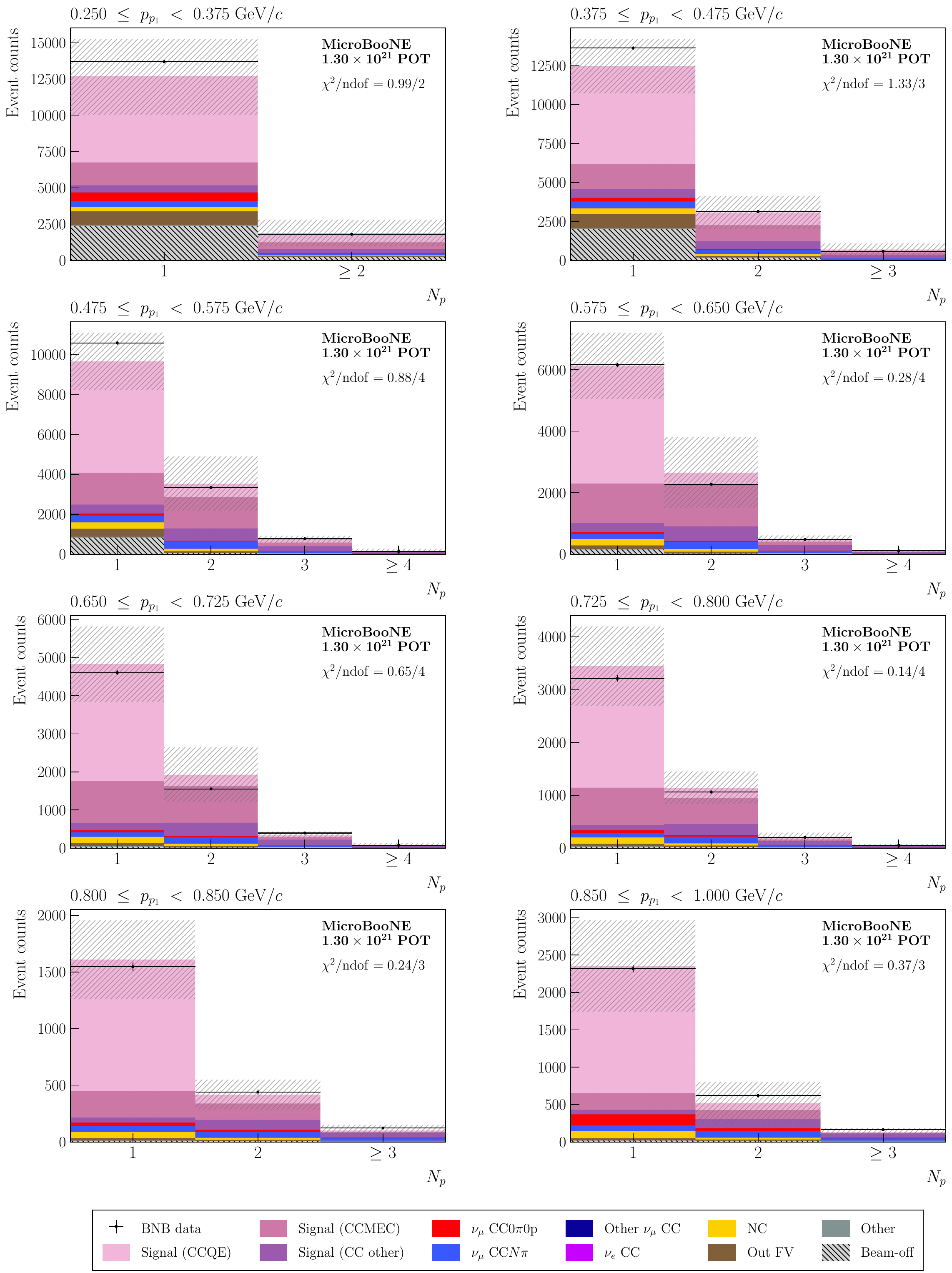}
	\caption{Reconstructed event rate distributions for block 1, with the signal broken down by interaction mode, corresponding to the double-differential measurement $(p_{p_{1}}, \, N_{p})$. The total uncertainty on the prediction is indicated by the hatched boxes.}
	\label{fig:event_rates_int_block_1}
\end{figure*}

\begin{figure*}[!htbp]
	\centering
	\includegraphics[width=0.83\linewidth]{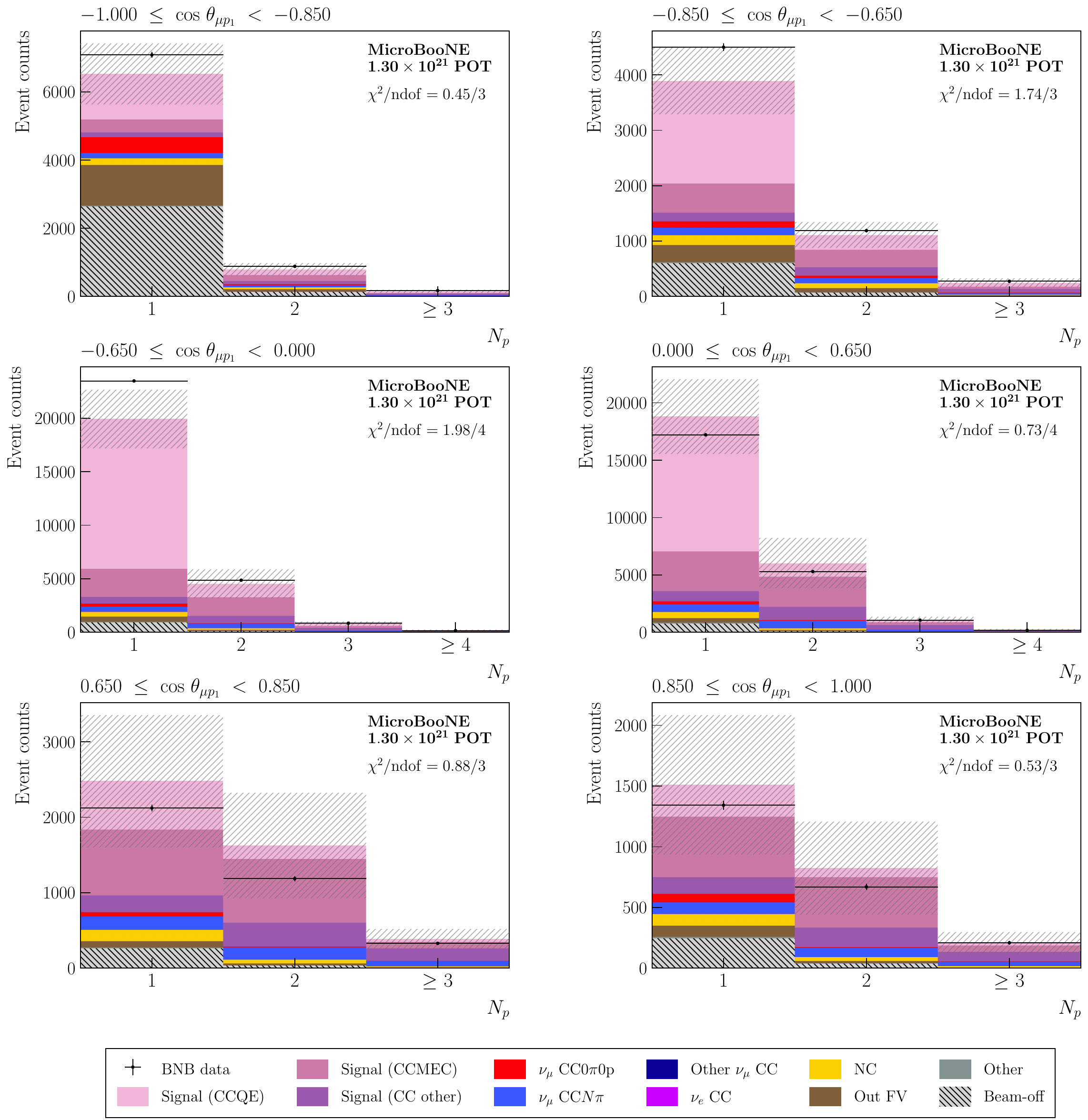}
	\caption{Reconstructed event rate distributions for block 2, with the signal broken down by interaction mode, corresponding to the double-differential measurement $(\mathrm{cos} \, \theta_{\mu p_{1}}, \, N_{p})$. The total uncertainty on the prediction is indicated by the hatched boxes.}
	\label{fig:event_rates_int_block_2}
\end{figure*}

\begin{figure*}[!htbp]
	\centering
	\includegraphics[width=0.83\linewidth]{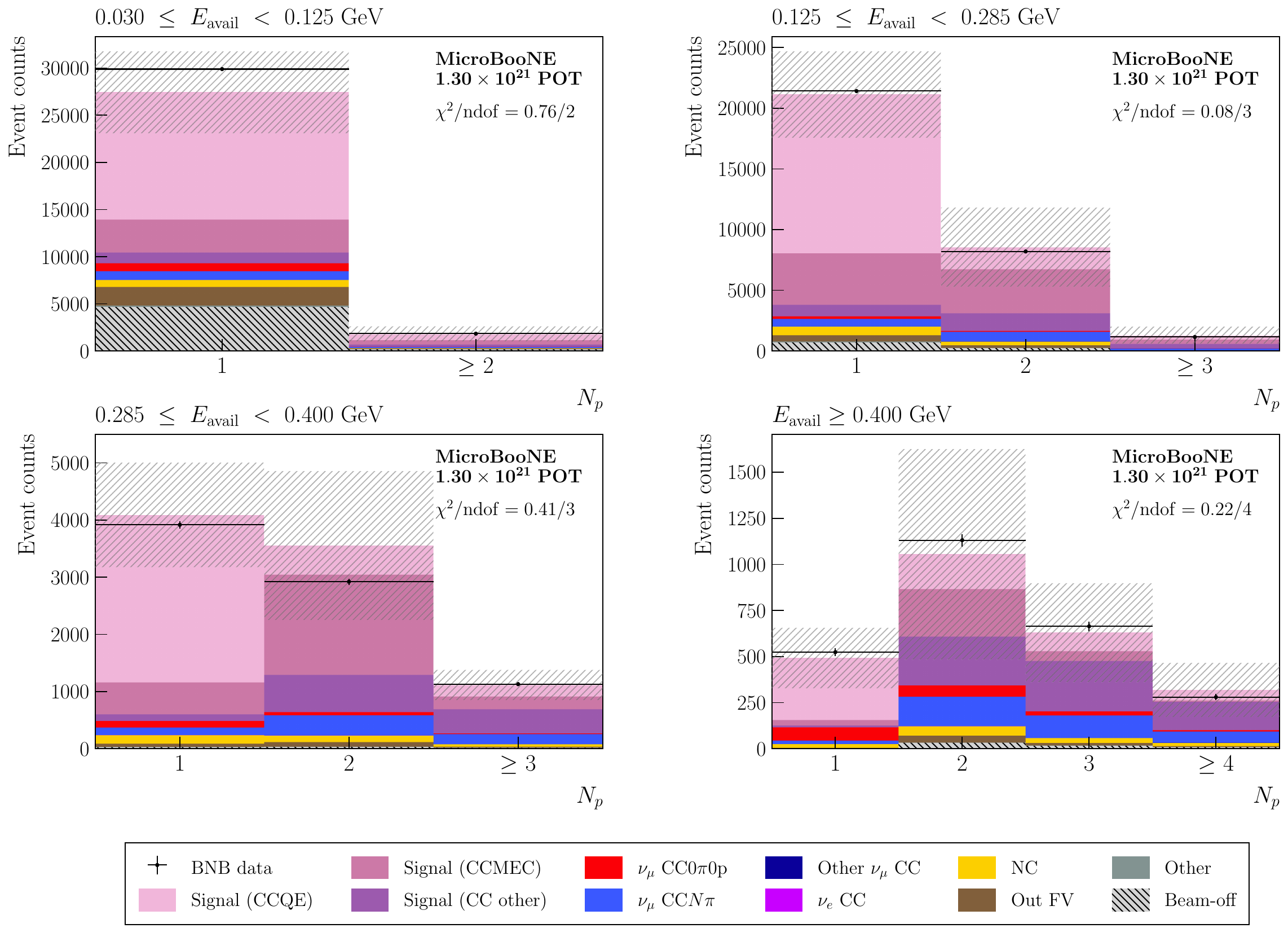}
	\caption{Reconstructed event rate distributions for block 3, with the signal broken down by interaction mode, corresponding to the double-differential measurement $(E_{\mathrm{avail}}, \, N_{p})$. The total uncertainty on the prediction is indicated by the hatched boxes.}
	\label{fig:event_rates_int_block_3}
\end{figure*}

\begin{figure*}[!htbp]
	\centering
	\includegraphics[width=1.00\linewidth]{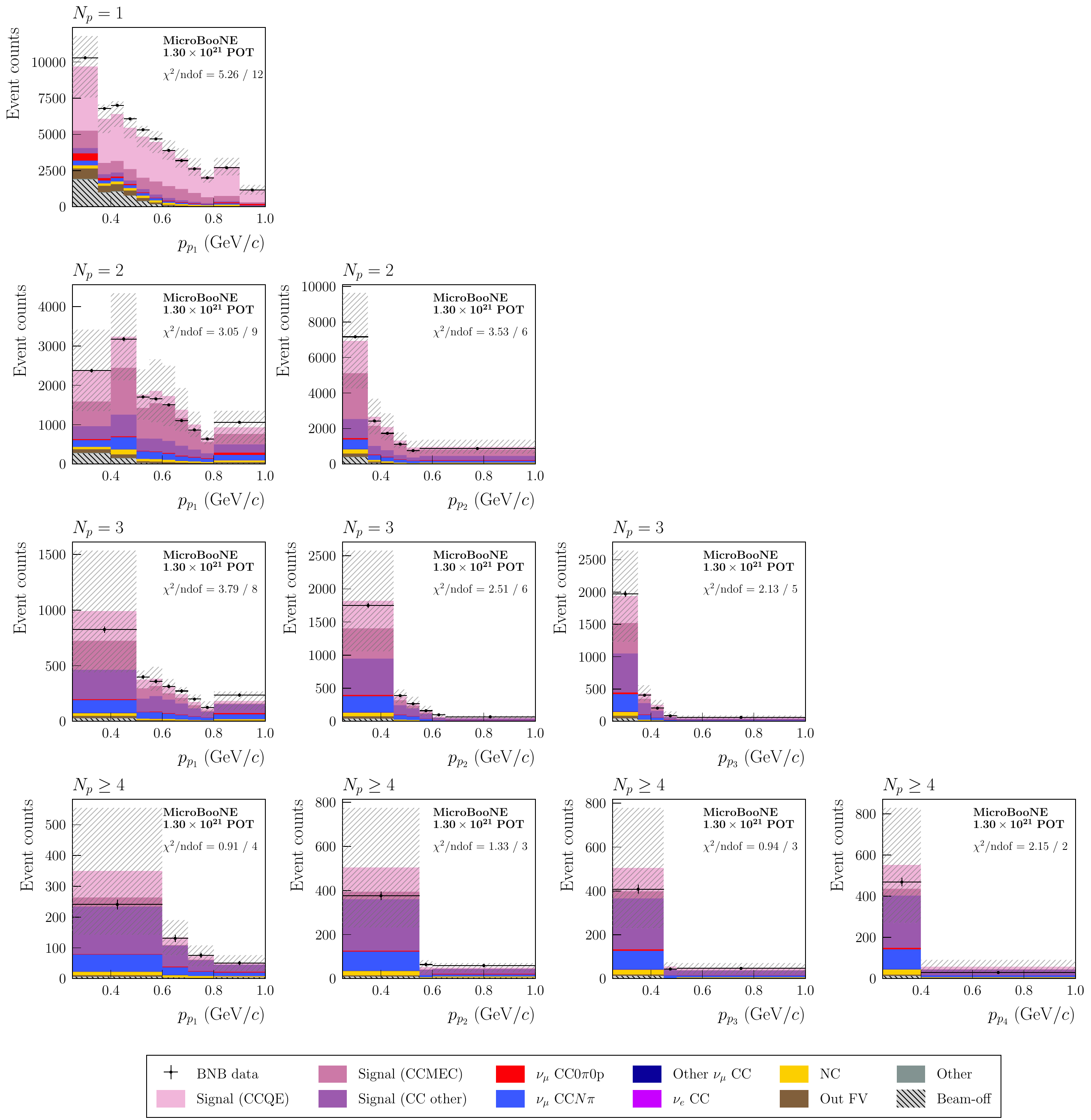}
	\caption{Reconstructed event rate distributions for blocks 4, 5, 6, and 7, with the signal broken down by interaction mode, corresponding to the double-differential measurements $(N_{p}, \, p_{p_{i}})$. The total uncertainty on the prediction is indicated by the hatched boxes.}
	\label{fig:event_rates_int_blocks_4_5_6_7}
\end{figure*}

\begin{figure*}[!htbp]
	\centering
	\includegraphics[width=1.00\linewidth]{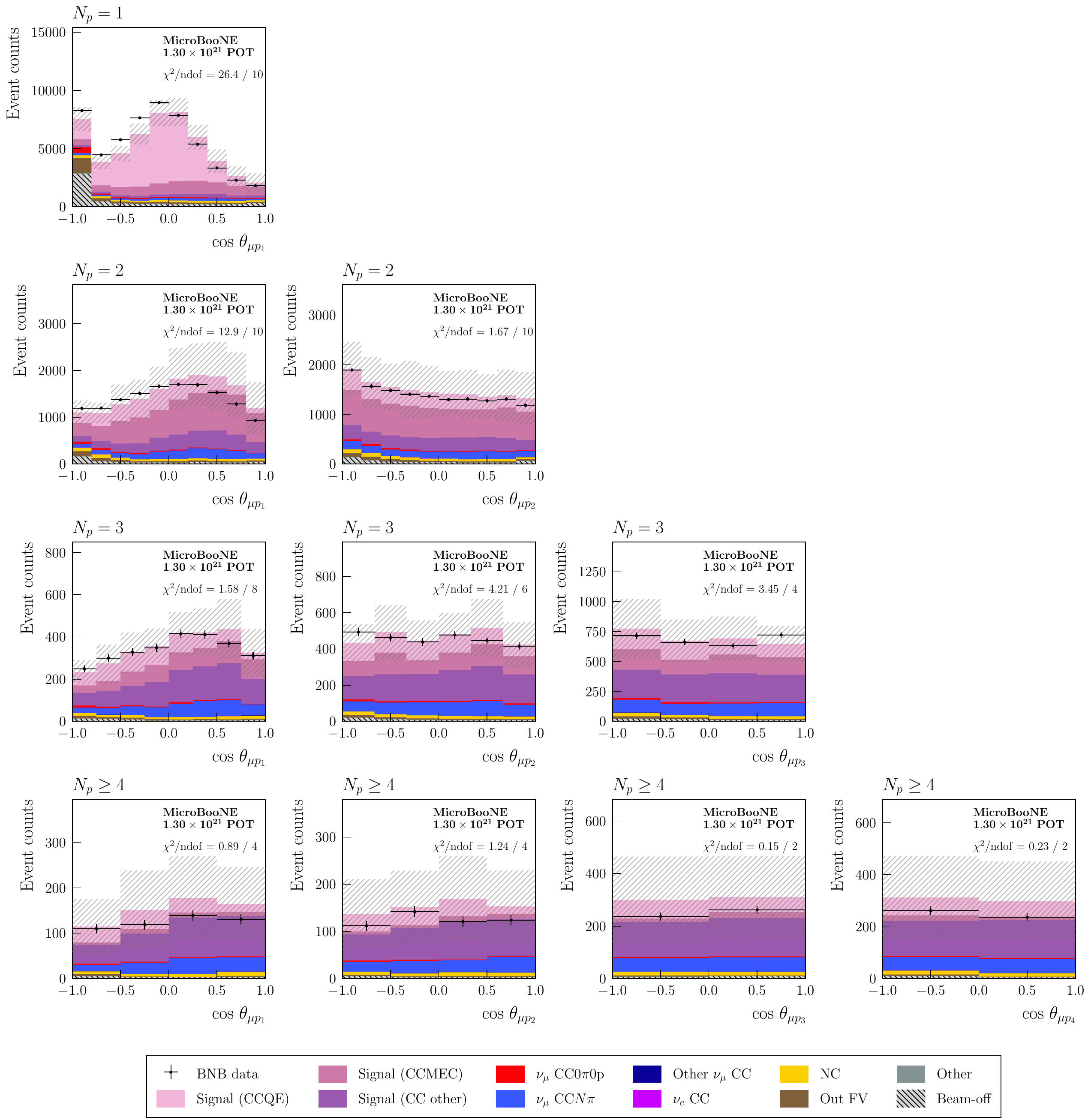}
	\caption{Reconstructed event rate distributions for blocks 8, 9, 10, and 11, with the signal broken down by interaction mode, corresponding to the double-differential measurements $(N_{p}, \, \mathrm{cos} \, \theta_{\mu p_{i}})$. The total uncertainty on the prediction is indicated by the hatched boxes.}
	\label{fig:event_rates_int_blocks_8_9_10_11}
\end{figure*}

Figures \ref{fig:event_rates_int_block_1} through \ref{fig:event_rates_int_blocks_8_9_10_11} show the reconstructed event rate distributions in the signal region decomposed by primary interaction mode (QE, MEC, and other) for each block.

\FloatBarrier
\newpage

\section{Sideband event rate distributions}

\begin{figure*}[!hbp]
	\centering
	\includegraphics[width=0.83\linewidth]{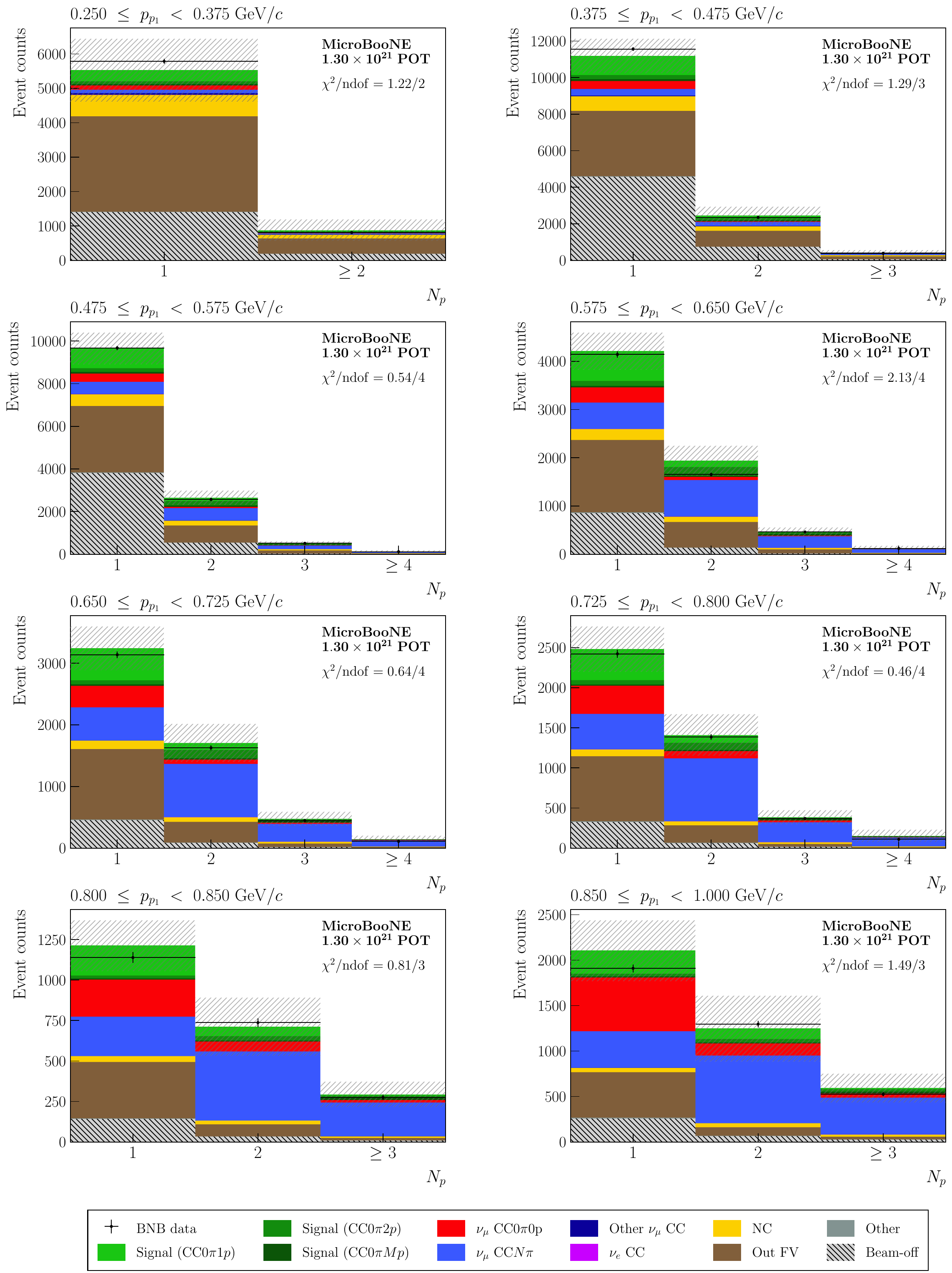}
	\caption{Reconstructed event rate distributions for block 1, corresponding to the double-differential measurement $(p_{p_{1}}, \, N_{p})$. The total uncertainty on the prediction is indicated by the hatched boxes.}
	\label{fig:sideband_event_rates_block_1}
\end{figure*}

\begin{figure*}[!t]
	\centering
	\includegraphics[width=0.83\linewidth]{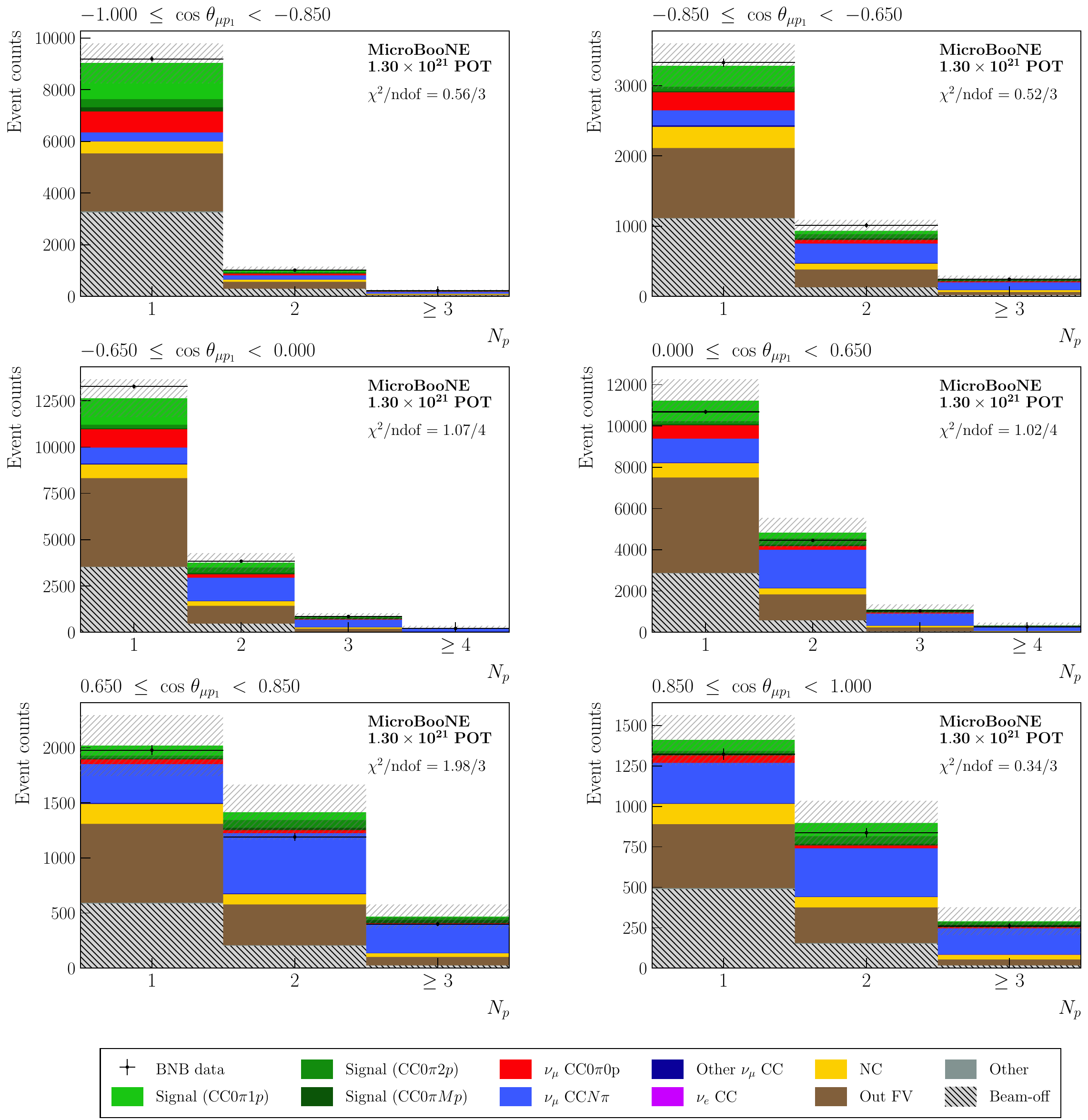}
	\caption{Reconstructed event rate distributions for block 2, corresponding to the double-differential measurement $(\mathrm{cos} \, \theta_{\mu p_{1}}, \, N_{p})$. The total uncertainty on the prediction is indicated by the hatched boxes.}
	\label{fig:sideband_event_rates_block_2}
\end{figure*}

\begin{figure*}[!htbp]
	\centering
	\includegraphics[width=0.83\linewidth]{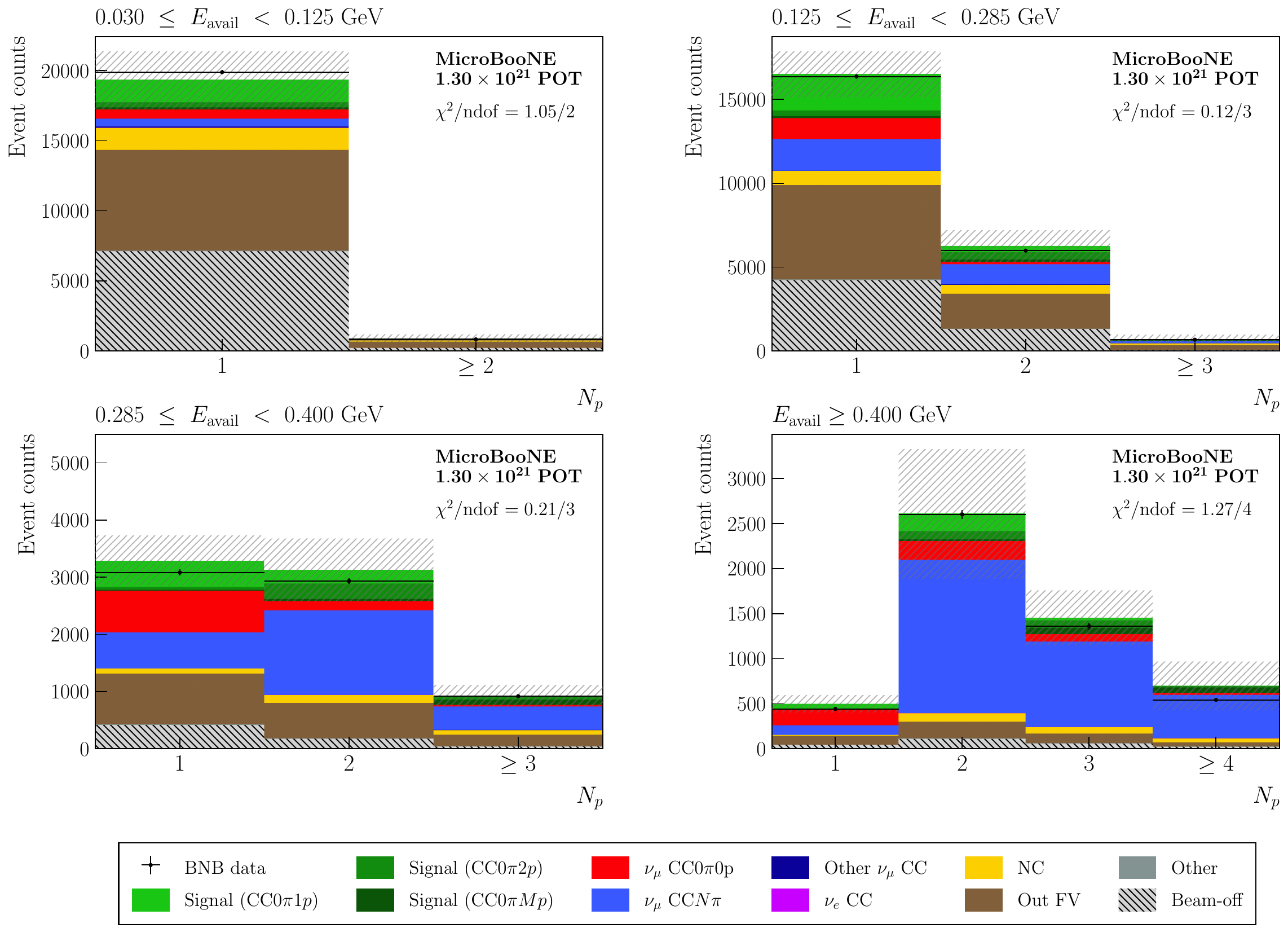}
	\caption{Reconstructed event rate distributions for block 3, corresponding to the double-differential measurement $(E_{\mathrm{avail}}, \, N_{p})$. The total uncertainty on the prediction is indicated by the hatched boxes.}
	\label{fig:sideband_event_rates_block_3}
\end{figure*}

\begin{figure*}[!htbp]
	\centering
	\includegraphics[width=1.0\linewidth]{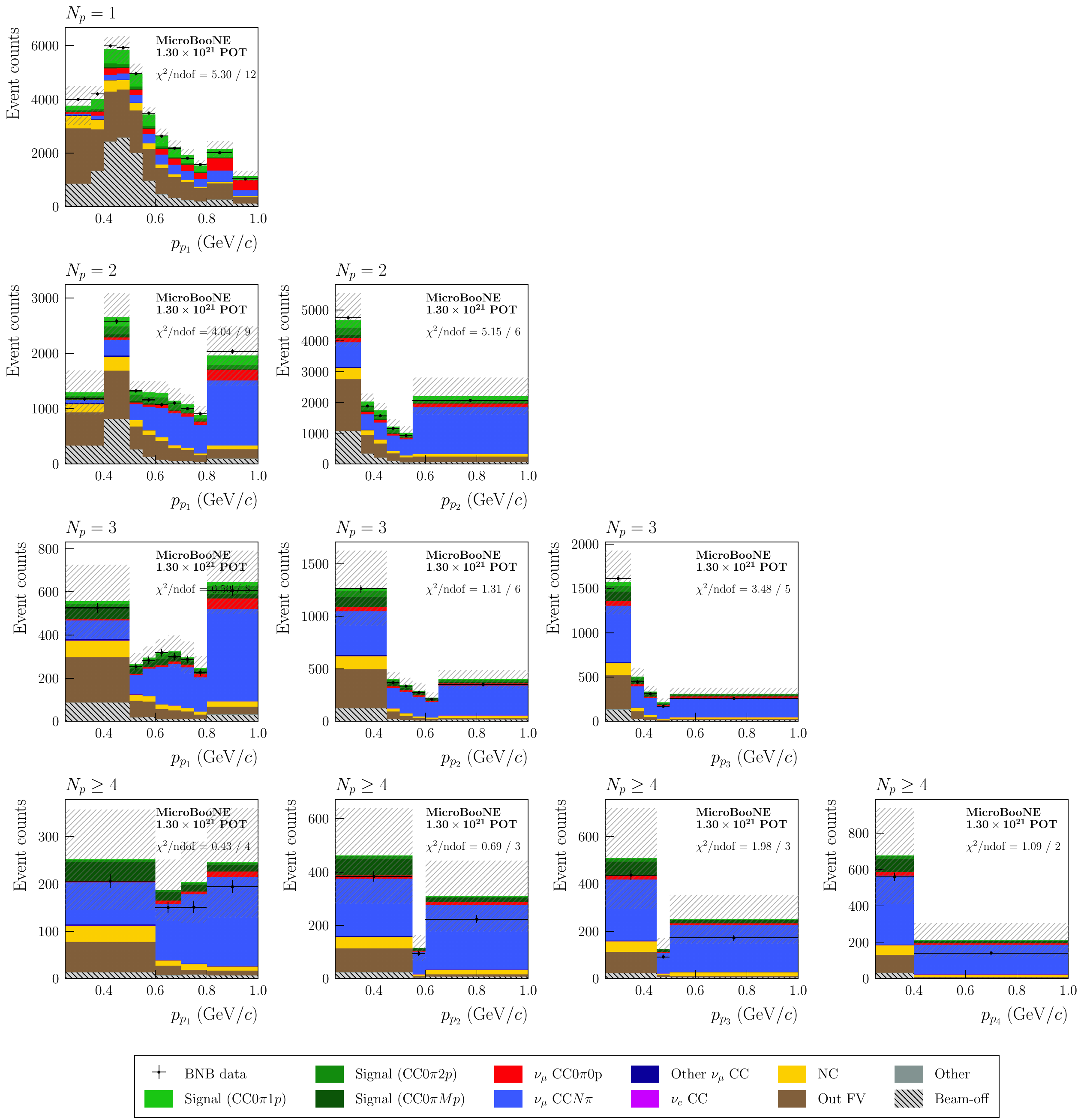}
	\caption{Reconstructed event rate distributions for blocks 4, 5, 6, and 7, corresponding to the double-differential measurements $(N_{p}, \, p_{p_{i}})$. The total uncertainty on the prediction is indicated by the hatched boxes.}
	\label{fig:sideband_event_rates_blocks_4_5_6_7}
\end{figure*}

\begin{figure*}[!htbp]
	\centering
	\includegraphics[width=1.0\linewidth]{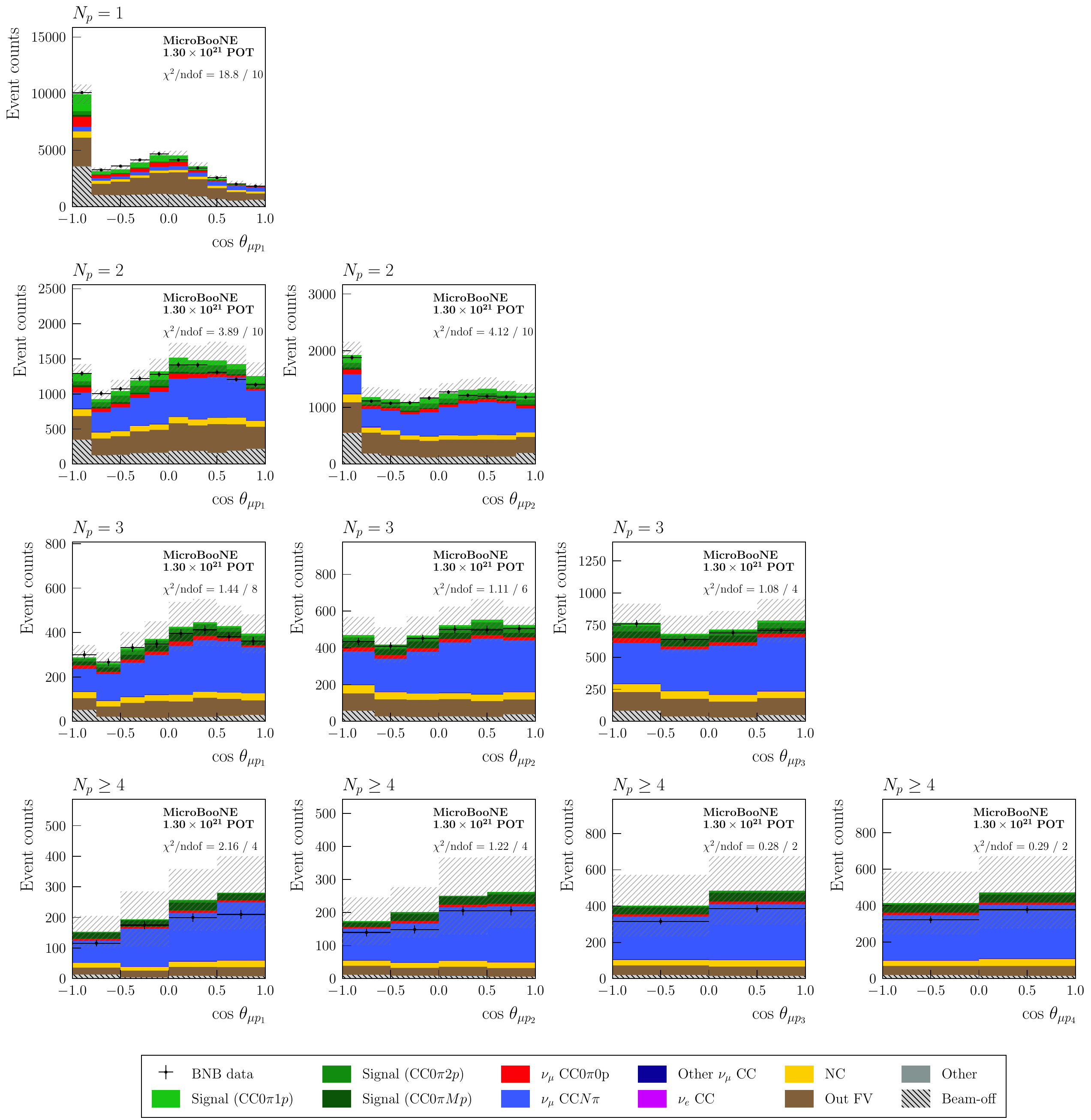}
	\caption{Reconstructed event rate distributions for blocks 8, 9, 10, and 11, corresponding to the double-differential measurements $(N_{p}, \, \mathrm{cos} \, \theta_{\mu p_{i}})$. The total uncertainty on the prediction is indicated by the hatched boxes.}
	\label{fig:sideband_event_rates_blocks_8_9_10_11}
\end{figure*}

Figures \ref{fig:sideband_event_rates_block_1} through \ref{fig:sideband_event_rates_blocks_8_9_10_11} show the reconstructed event rate distributions for the combined sideband selection described in Sec. IV A of the main text.

\FloatBarrier
\newpage

\section{Validation studies}
\subsection{NuWro fake data study}

\begin{figure*}[!hbp]
	\centering
	\includegraphics[width=0.76\linewidth]{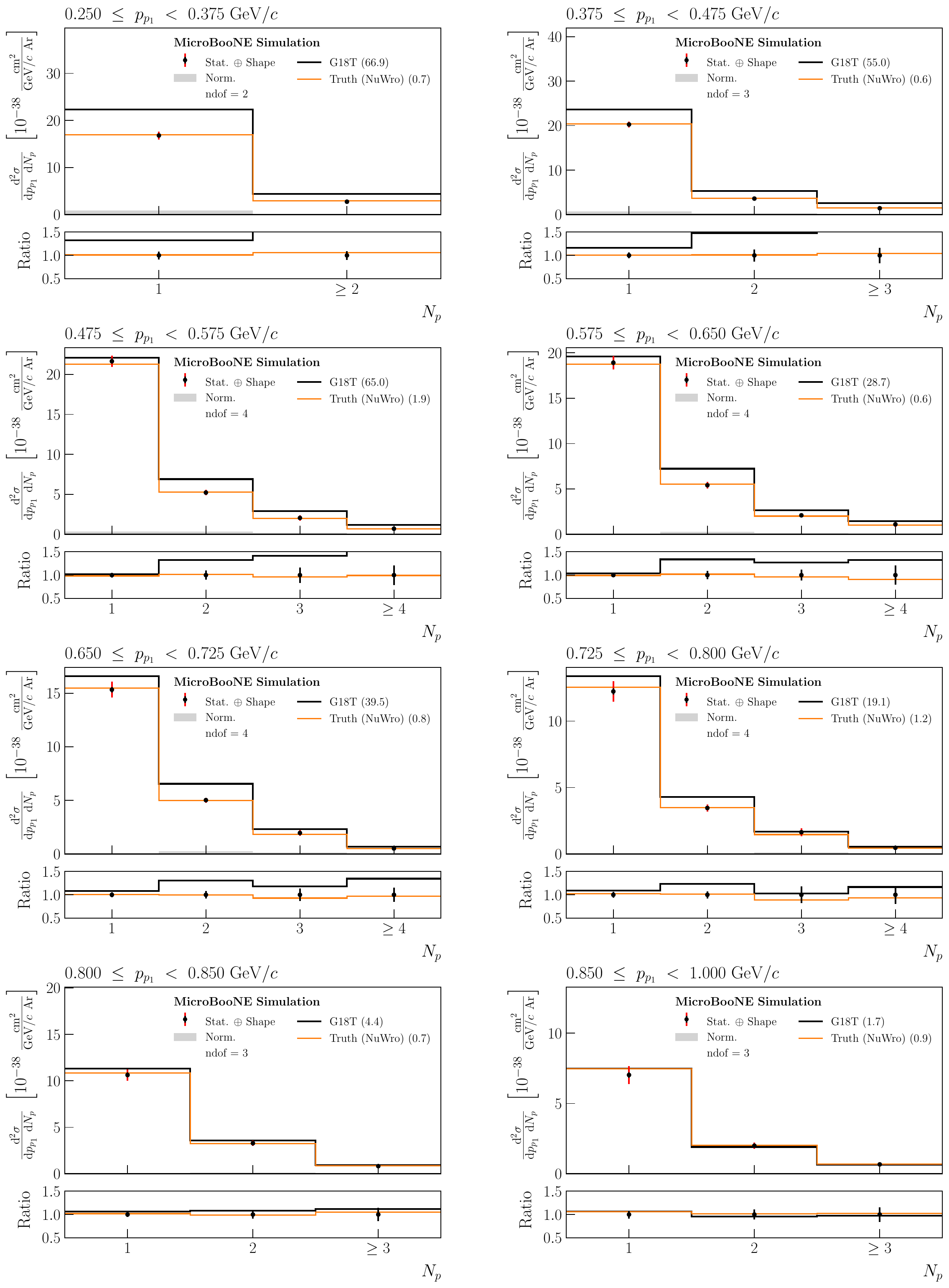}
	\caption{Unfolded cross sections extracted from NuWro fake data for block 1, corresponding to the double-differential measurement $(p_{p_{1}}, \, N_{p})$ (black data points), with the corresponding MicroBooNE GENIE tune (black lines) and NuWro (orange lines) predictions.}
	\label{fig:fake_dagostini_block_1}
\end{figure*}

\begin{figure*}[!htbp]
	\centering
	\includegraphics[width=0.76\linewidth]{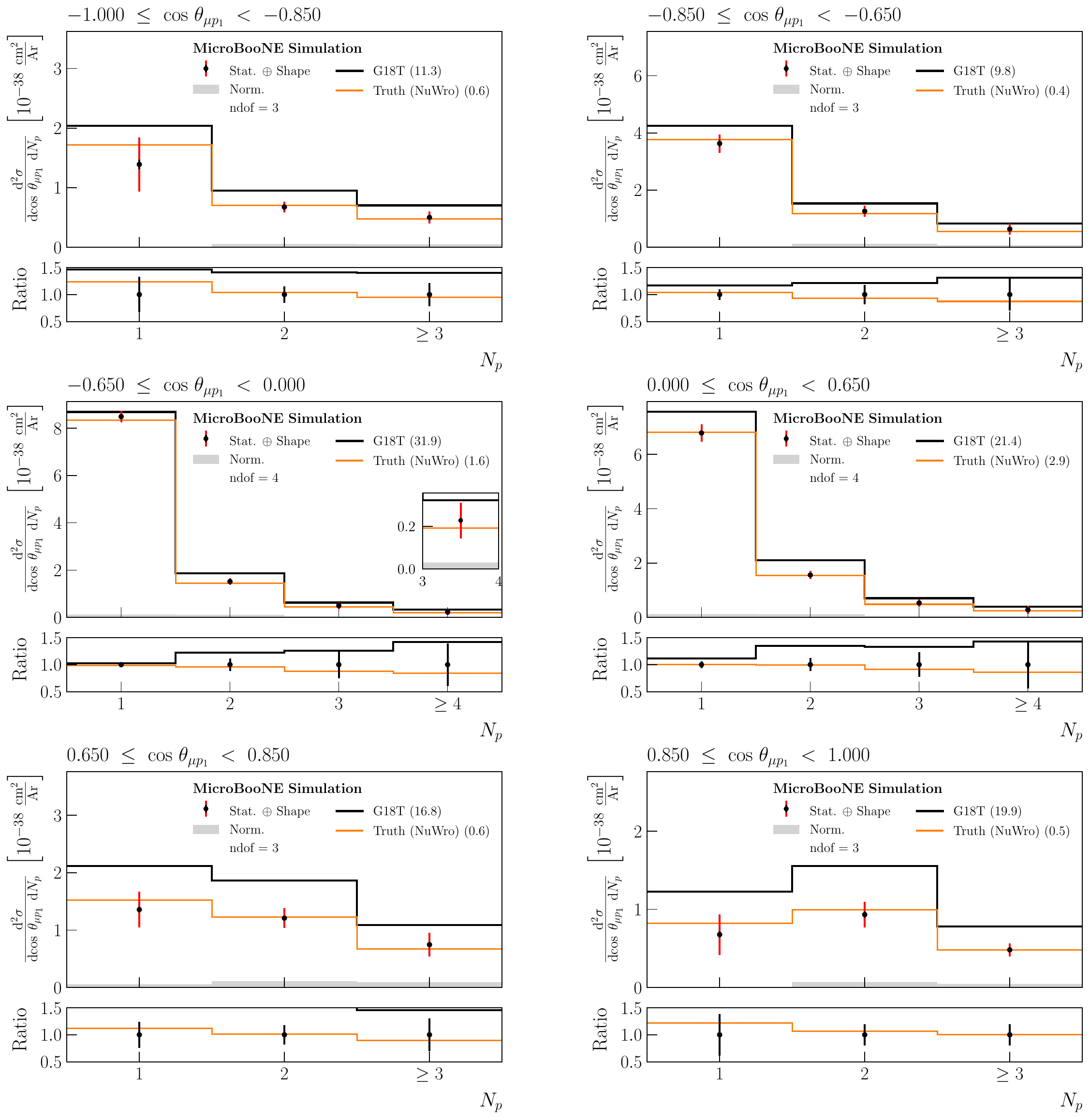}
	\caption{Unfolded cross sections extracted from NuWro fake data for block 2, corresponding to the double-differential measurement $(\cos\theta_{\mu p_{1}}, \, N_{p})$ (black data points), with the corresponding MicroBooNE GENIE tune (black lines) and NuWro (orange lines) predictions.}
	\label{fig:fake_dagostini_block_2}
\end{figure*}

\begin{figure*}[!htbp]
	\centering
	\includegraphics[width=0.76\linewidth]{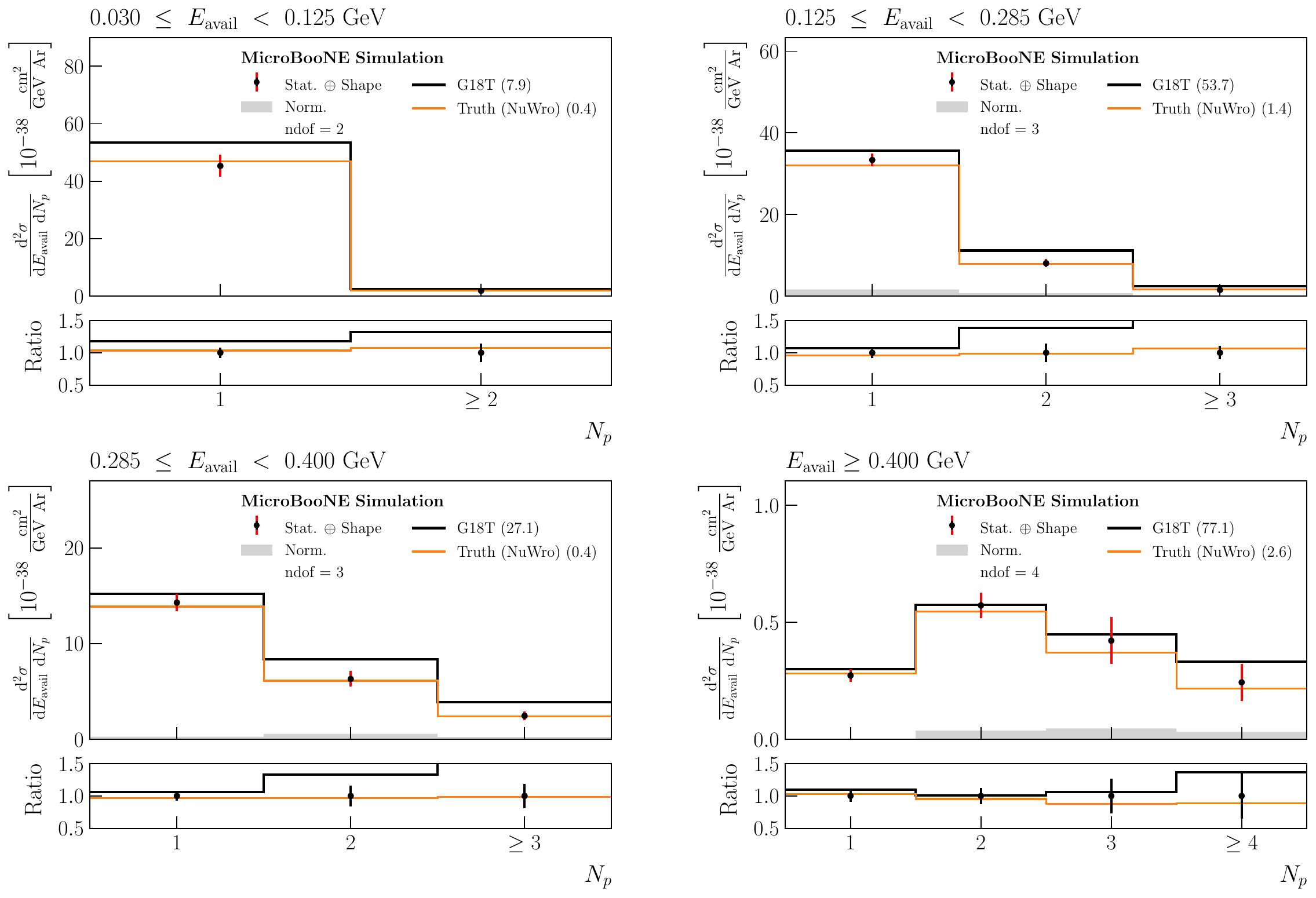}
	\caption{Unfolded cross sections extracted from NuWro fake data for block 3, corresponding to the double-differential measurement $(E_{\mathrm{avail}}, \, N_{p})$ (black data points), with the corresponding MicroBooNE GENIE tune (black lines) and NuWro (orange lines) predictions.}
	\label{fig:fake_dagostini_block_3}
\end{figure*}

\begin{figure*}[!htbp]
	\centering
	\includegraphics[width=1.00\linewidth]{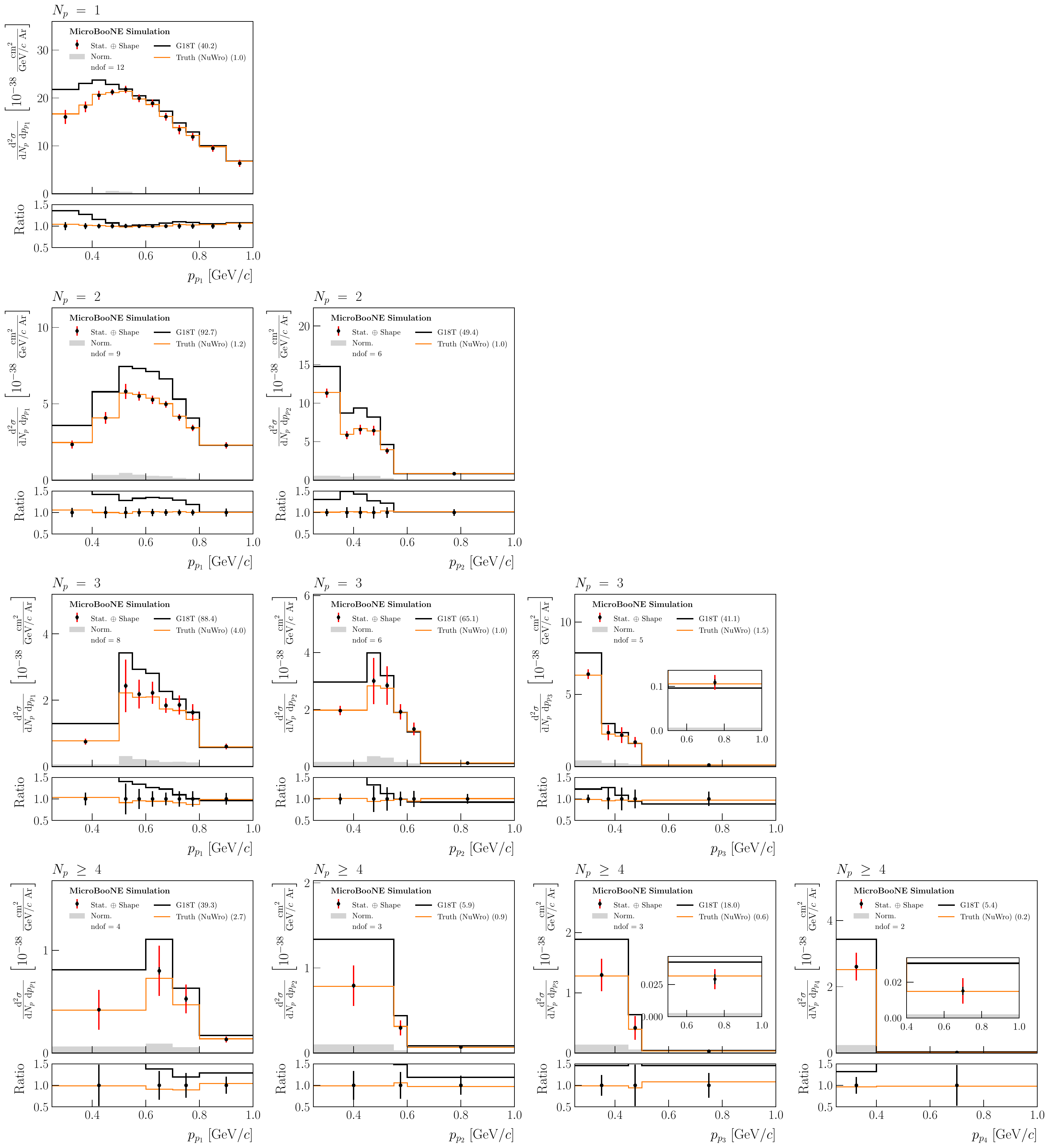}
	\caption{Unfolded cross sections extracted from NuWro fake data for blocks 4, 5, 6, and 7, corresponding to the double-differential measurements $(N_{p}, \,p_{p_{i}})$ (black data points), with the corresponding MicroBooNE GENIE tune (black lines) and NuWro (orange lines) predictions.}
	\label{fig:fake_dagostini_blocks_4_5_6_7}
\end{figure*}

\begin{figure*}[!htbp]
	\centering
	\includegraphics[width=1.00\linewidth]{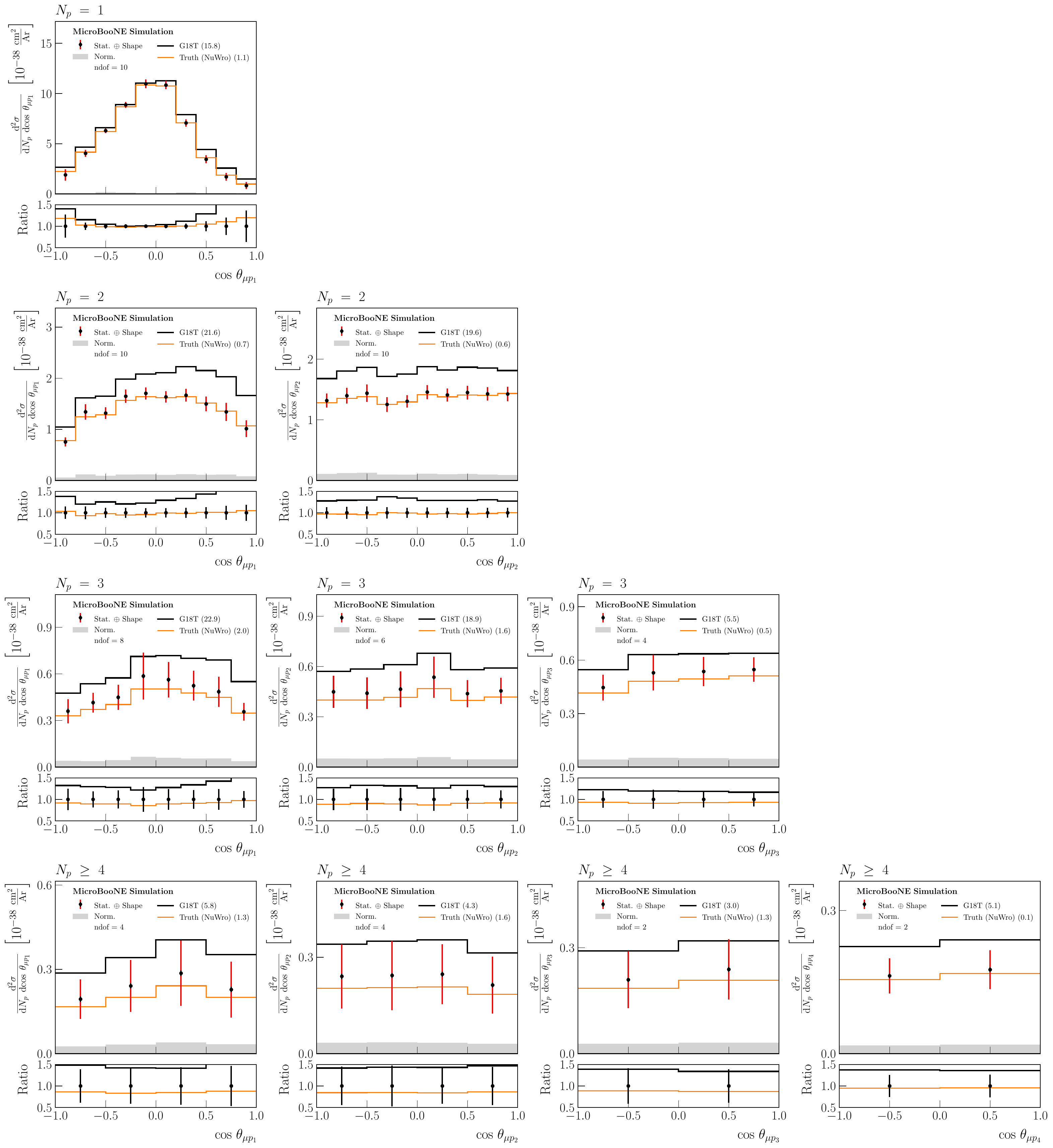}
	\caption{Unfolded cross sections extracted from NuWro fake data for blocks 8, 9, 10, and 11, corresponding to the double-differential measurements $(N_{p}, \,\cos\theta_{\mu p_{i}})$ (black data points), with the corresponding MicroBooNE GENIE tune (black lines) and NuWro (orange lines) predictions.}
	\label{fig:fake_dagostini_blocks_8_9_10_11}
\end{figure*}

Figures \ref{fig:fake_dagostini_block_1} through \ref{fig:fake_dagostini_blocks_8_9_10_11} show the unfolded cross sections extracted from the NuWro fake data study described in Sec. IV C of the main text.

\FloatBarrier
\pagebreak

\subsection{Data-driven validation}

\begin{figure*}[!hbp]
	\centering
	\includegraphics[width=0.77\linewidth]{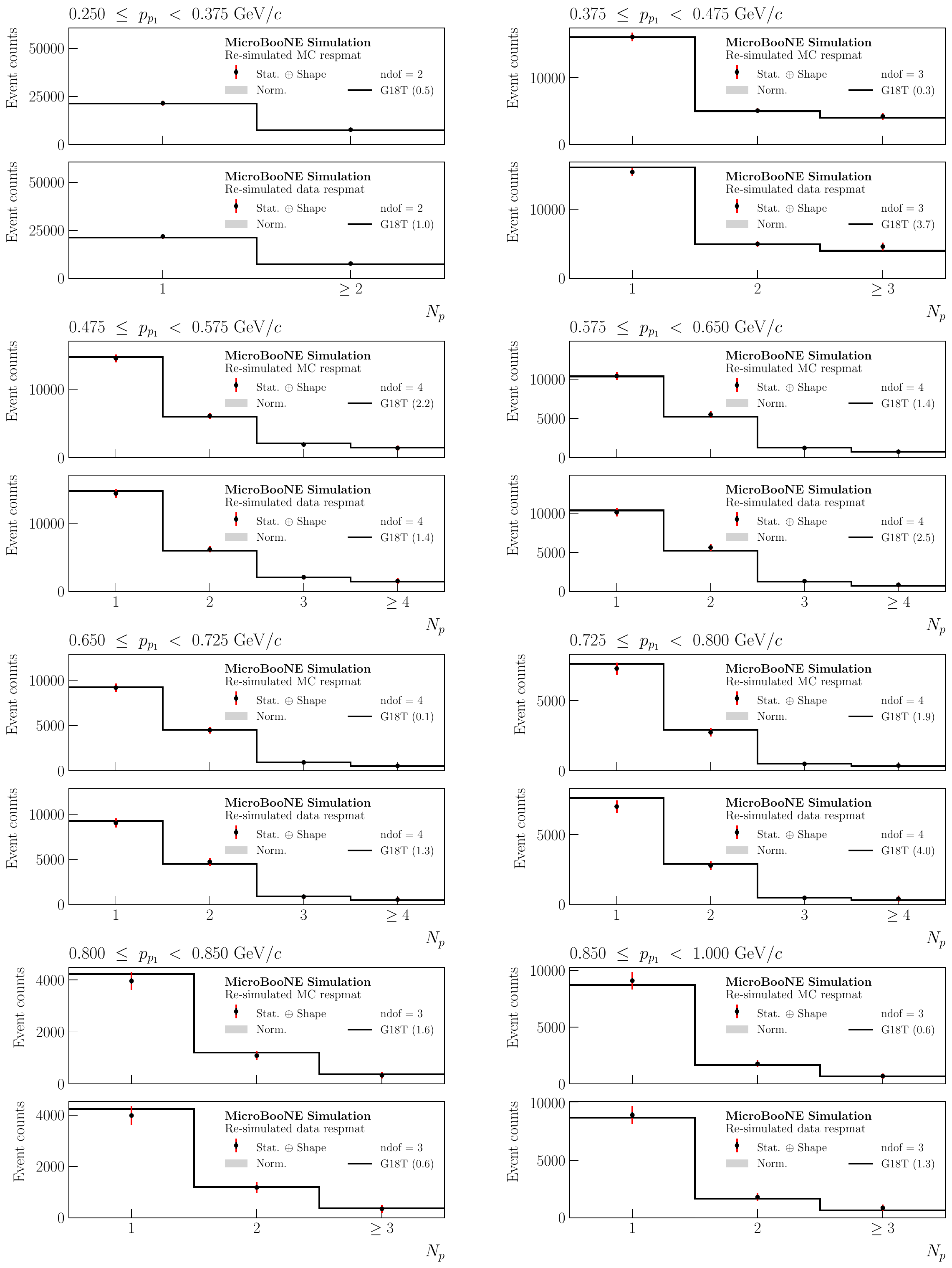}
	\caption{Unfolded event counts extracted from GENIE fake data for block 1 (black data points), with the corresponding MicroBooNE GENIE tune predictions (black lines). The top and bottom subpanels use the re-simulated data and MC response matrices, respectively.}
	\label{fig:resim_closure_dagostini_block_1}
\end{figure*}

\begin{figure*}[!htbp]
	\centering
	\includegraphics[width=0.77\linewidth]{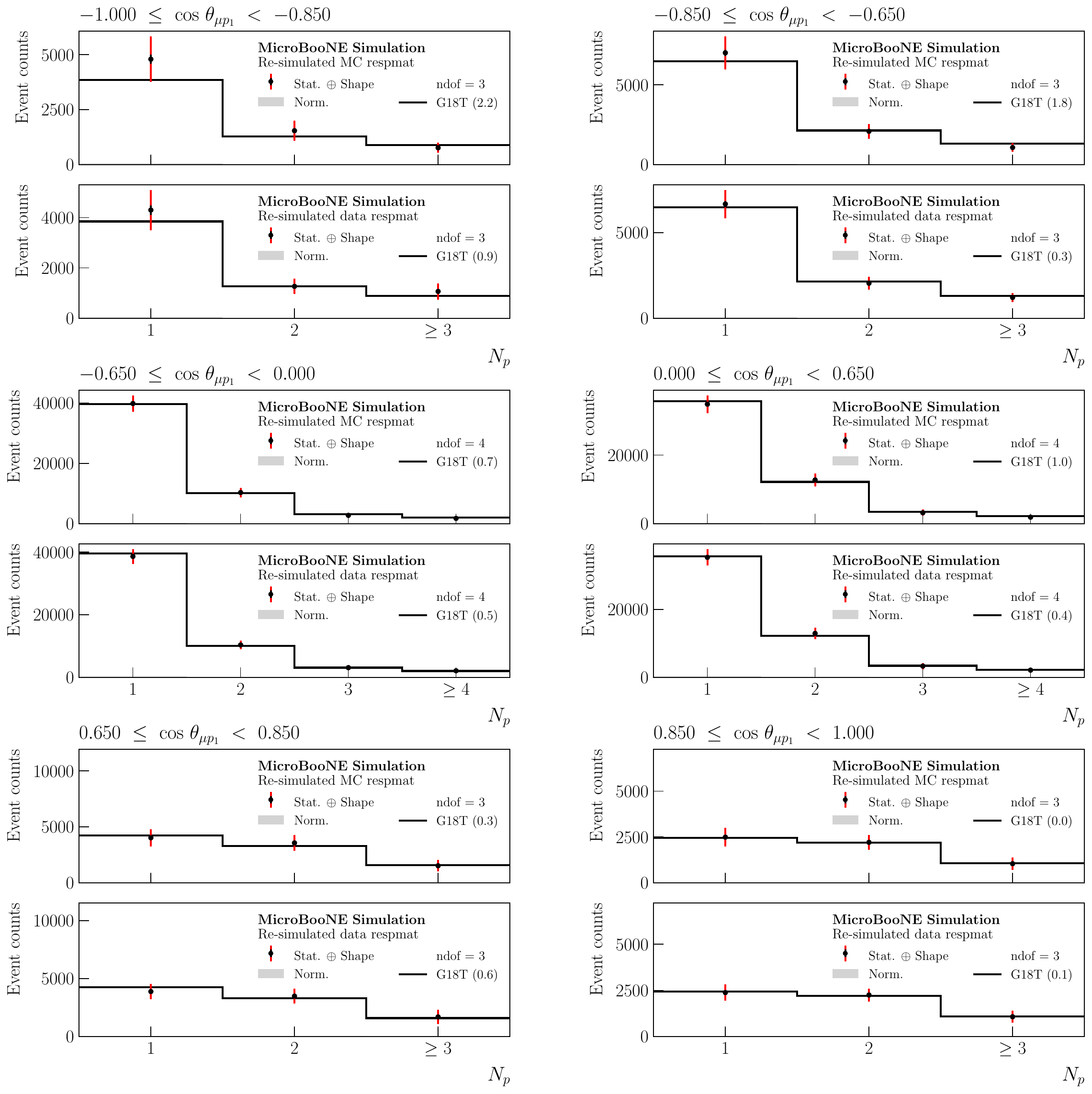}
	\caption{Unfolded event counts extracted from GENIE fake data for block 2 (black data points), with the corresponding MicroBooNE GENIE tune predictions (black lines). The top and bottom subpanels use the re-simulated data and MC response matrices, respectively.}
	\label{fig:resim_closure_dagostini_block_2}
\end{figure*}

\begin{figure*}[!htbp]
	\centering
	\includegraphics[width=0.77\linewidth]{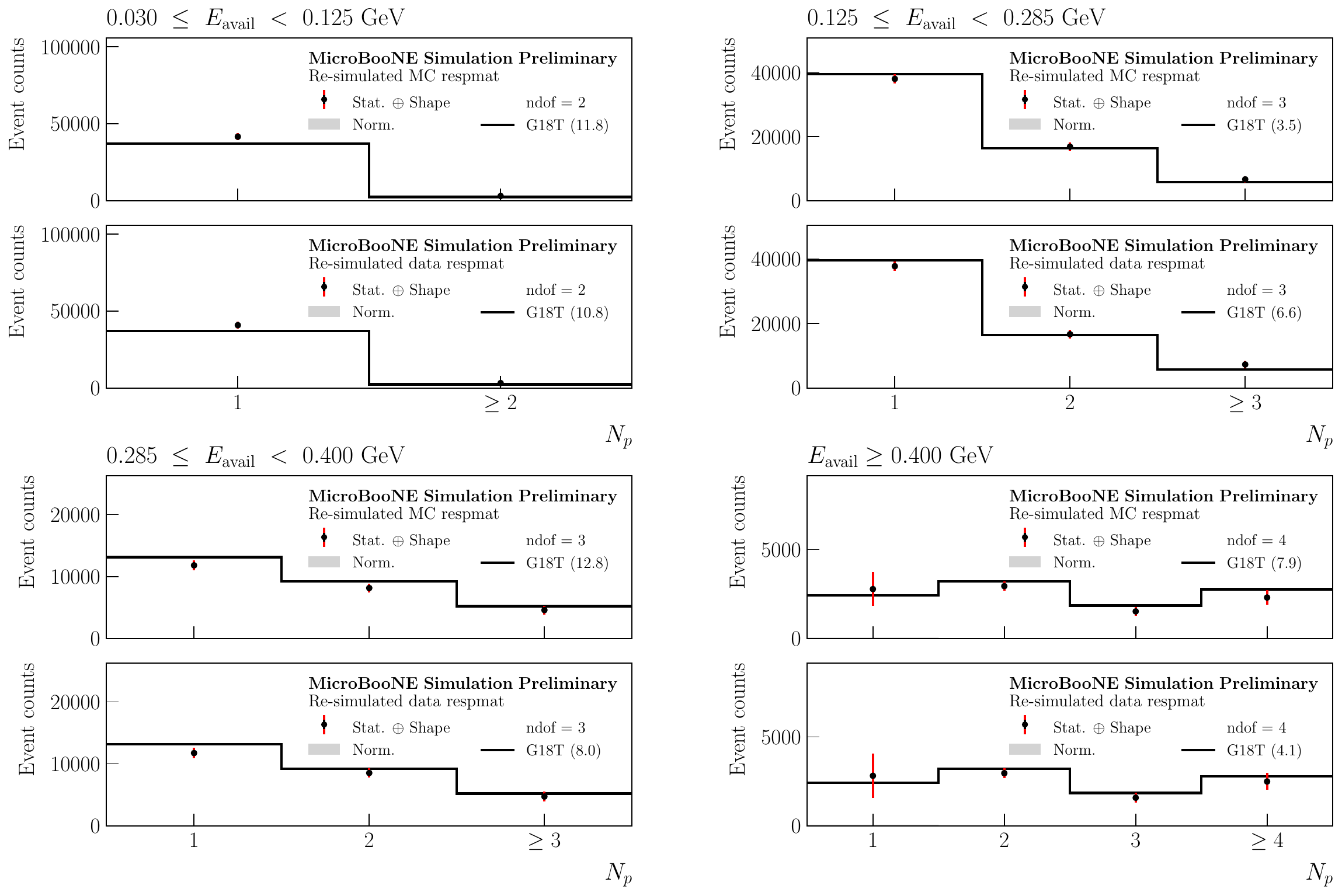}
	\caption{Unfolded event counts extracted from GENIE fake data for block 3 (black data points), with the corresponding MicroBooNE GENIE tune predictions (black lines). The top and bottom subpanels use the re-simulated data and MC response matrices, respectively.}
	\label{fig:resim_closure_dagostini_block_3}
\end{figure*}

\begin{figure*}[!htbp]
	\centering
	\includegraphics[width=1.00\linewidth]{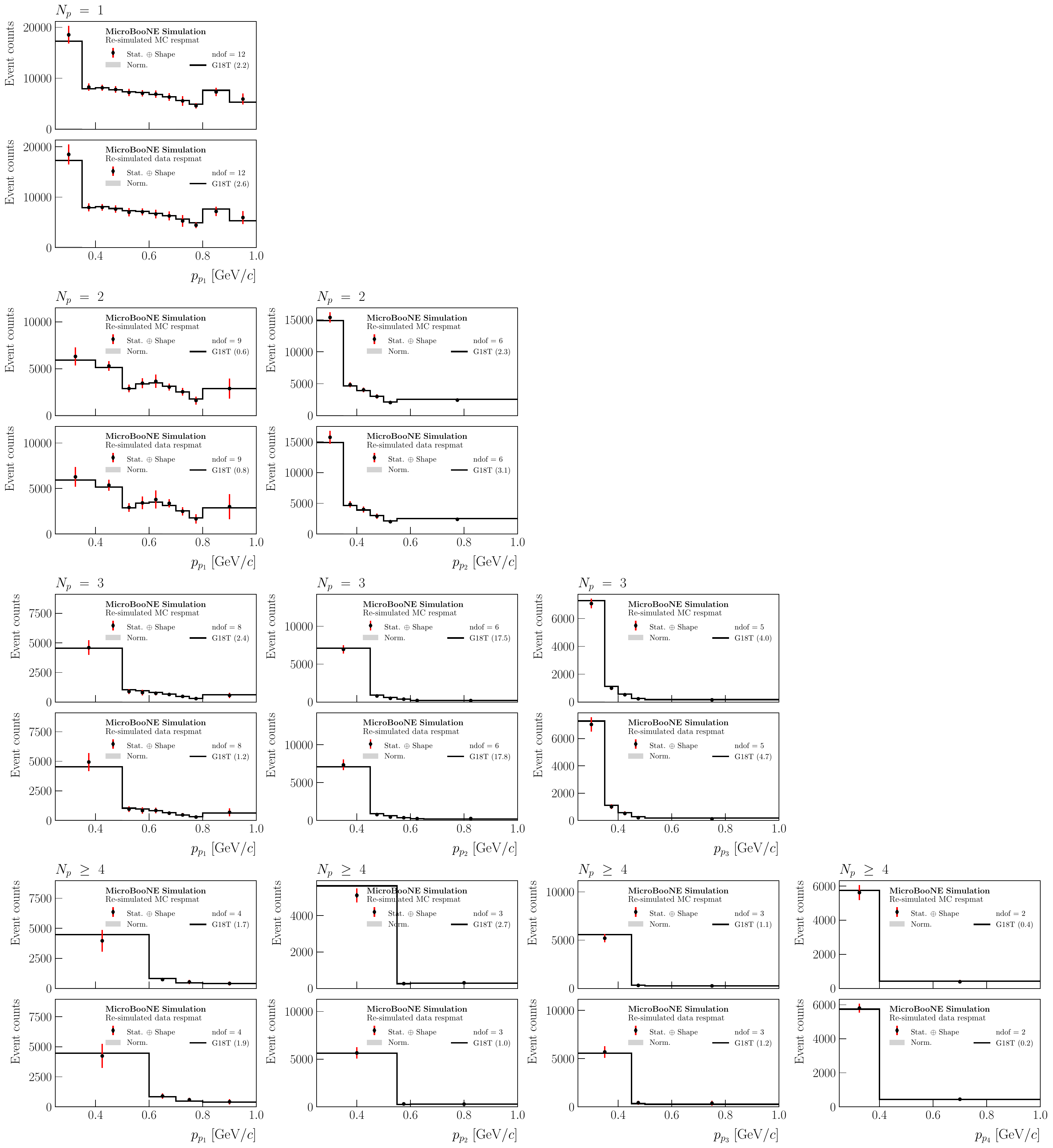}
	\caption{Unfolded event counts extracted from GENIE fake data for blocks 4, 5, 6, and 7 (black data points), with the corresponding MicroBooNE GENIE tune predictions (black lines). The top and bottom subpanels use the re-simulated data and MC response matrices, respectively.}
	\label{fig:resim_closure_dagostini_blocks_4_5_6_7}
\end{figure*}

\begin{figure*}[!htbp]
	\centering
	\includegraphics[width=1.00\linewidth]{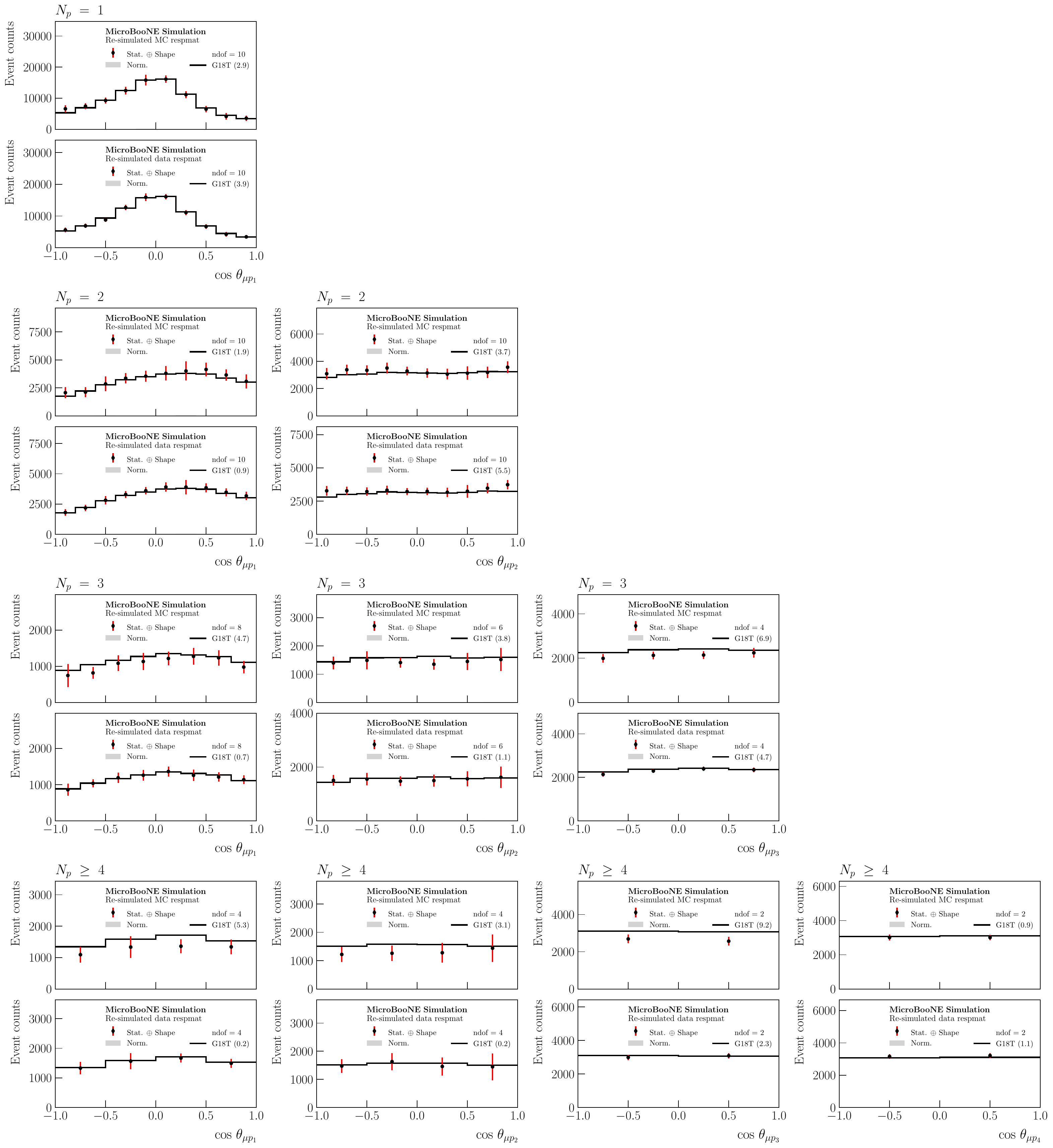}
	\caption{Unfolded event counts extracted from GENIE fake data for blocks 8, 9, 10, and 11 (black data points), with the corresponding MicroBooNE GENIE tune predictions (black lines). The top and bottom subpanels use the re-simulated data and MC response matrices, respectively.}
	\label{fig:resim_closure_dagostini_blocks_8_9_10_11}
\end{figure*}

Figures \ref{fig:resim_closure_dagostini_block_1} through \ref{fig:resim_closure_dagostini_blocks_8_9_10_11} show the unfolded event counts obtained using the re-simulated data and MC response matrices, as described in Sec. IV C and Appendix B of the main text.

\FloatBarrier

\section{Additional smearing matrix}

\begin{figure*}[!hbp]
	\centering
	\includegraphics[width=0.85\linewidth]{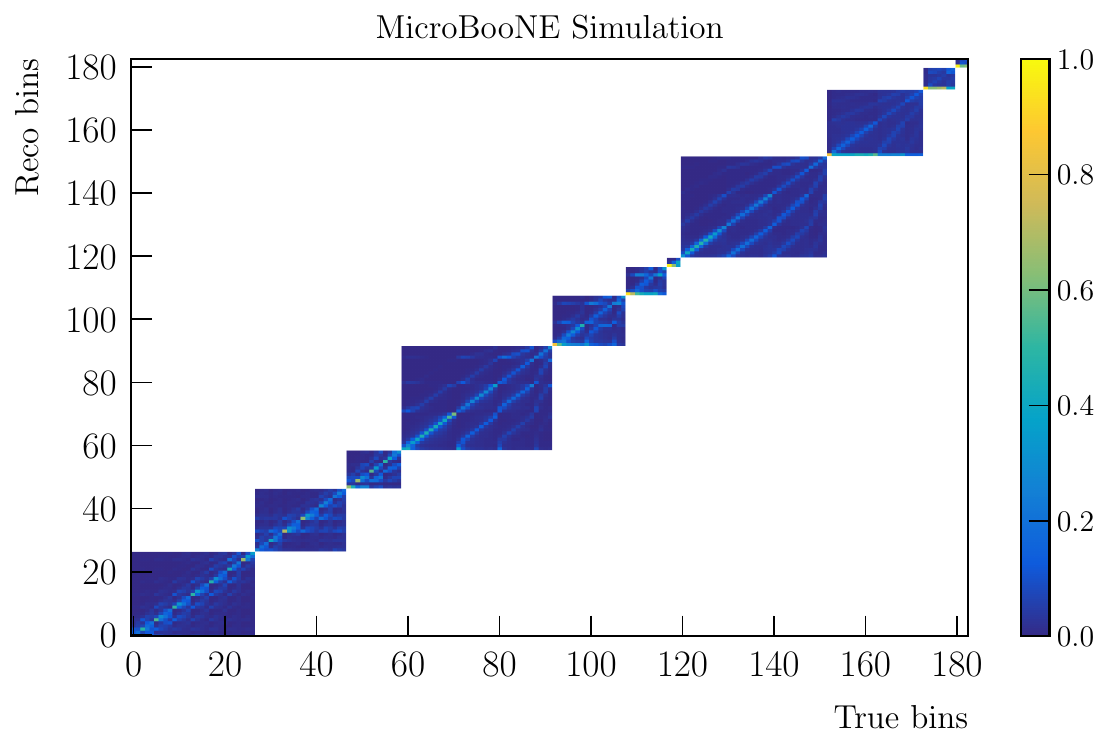}
	\caption{Additional smearing matrix $A_C$ for each bin block, as defined in Eq. (9) of the main text.}
	\label{fig:additional_smearing}
\end{figure*}

Figure \ref{fig:additional_smearing} shows the additional smearing matrix used to account for regularization bias in the goodness-of-fit comparisons, as described in Sec. IV B of the main text. Note that the elements of the additional smearing matrix are dimensionless.

\section{Additional generator comparisons}

This section presents comparisons of the unfolded cross section results to additional generator configurations not included in the main text. These configurations are briefly described below.

The untuned GENIE \texttt{G18\_10a\_02\_11a} configuration, generated with \texttt{v3.6.2}, uses the same underlying physics models as the MicroBooNE tune but without the parameter adjustments \cite{MicroBooNE:2021ccs}. The \texttt{02\_11a} label denotes a preliminary version of the GENIE nucleon cross-section tune to bubble chamber data \cite{GENIE:2021zuu}. The GENIE \texttt{G21\_11b\_00\_000} configuration, generated with \texttt{v3.2.0}, replaces the Nieves QE and Valencia 2p2h models with the SuSAv2 model \cite{Gonzalez-Jimenez:2014eqa} for both QE and MEC interactions. Additionally, the hN intranuclear cascade model is used for the treatment of FSI, instead of the effective hA model.

NuWro \texttt{v19.02.2} differs from \texttt{v25.11.1} in several respects. The LFG model is used for all interaction channels including QE, the MEC contribution is described by the original Valencia 2p2h model \cite{Nieves:2011pp} which provides only inclusive cross sections, and single pion production uses a dedicated $\Delta(1232)$ resonance model \cite{Graczyk:2009qm} with PYTHIA \cite{Sjostrand:2006za} for higher resonances. The intranuclear cascade does not include the convolution scheme used in \texttt{v25.11.1}, but does include effective nucleon density corrections for nucleon-nucleon correlations \cite{Niewczas:2019fro}. NuWro \texttt{v25.03.1} shares the same interaction models as \texttt{v25.11.1} but uses the LFG model for all channels, without the spectral function or the QE convolution scheme. An alternative \texttt{v25.03.1} sample is produced using the SuSAv2 model \cite{Gonzalez-Jimenez:2014eqa} for MEC in place of the Valencia 2020 model.

\begin{figure*}[!htbp]
	\centering
	\includegraphics[width=0.88\linewidth]{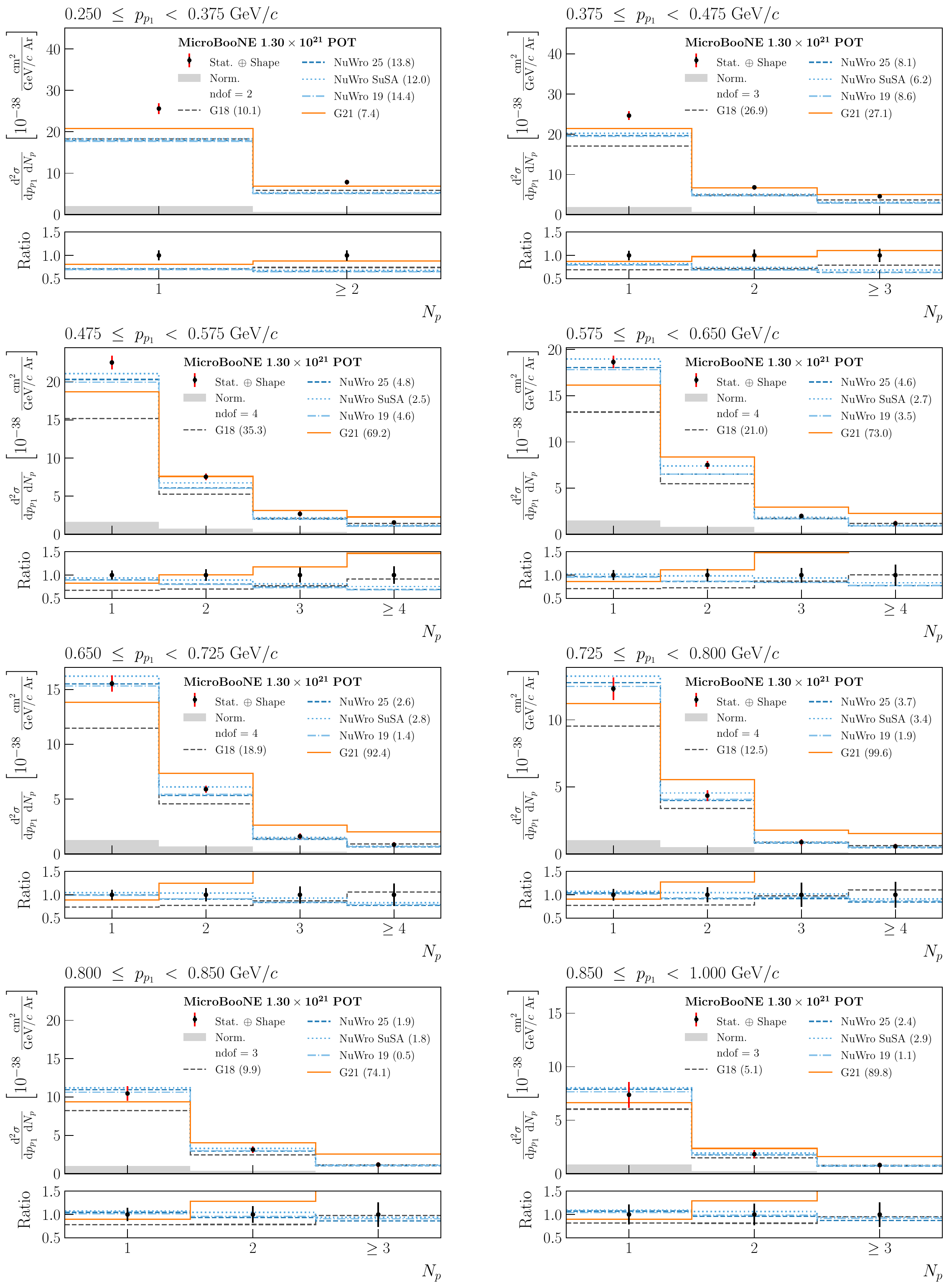}
	\caption{Flux-integrated double-differential cross sections for block 1 $(p_{p_{1}}, \, N_{p})$, extracted from the full MicroBooNE BNB dataset (black data points). Error bars indicate statistical and shape-only uncertainties, while the gray band indicates the normalization uncertainty. Predictions from the GENIE \texttt{G18\_10a\_02\_11a} and \texttt{G21\_11b\_00\_000} tunes, NuWro \texttt{v19.02.2}, NuWro \texttt{25.03.1} (default), and NuWro \texttt{25.03.1} (SuSAv2) are overlaid as indicated in the legend. The $\chi^{2}$ for each prediction is indicated in parentheses. The bottom panels show the ratio of the predictions to the unfolded data.}
	\label{fig:unfolded_xsec_other_block_1}
\end{figure*}

\begin{figure*}[!htbp]
	\centering
	\includegraphics[width=0.88\linewidth]{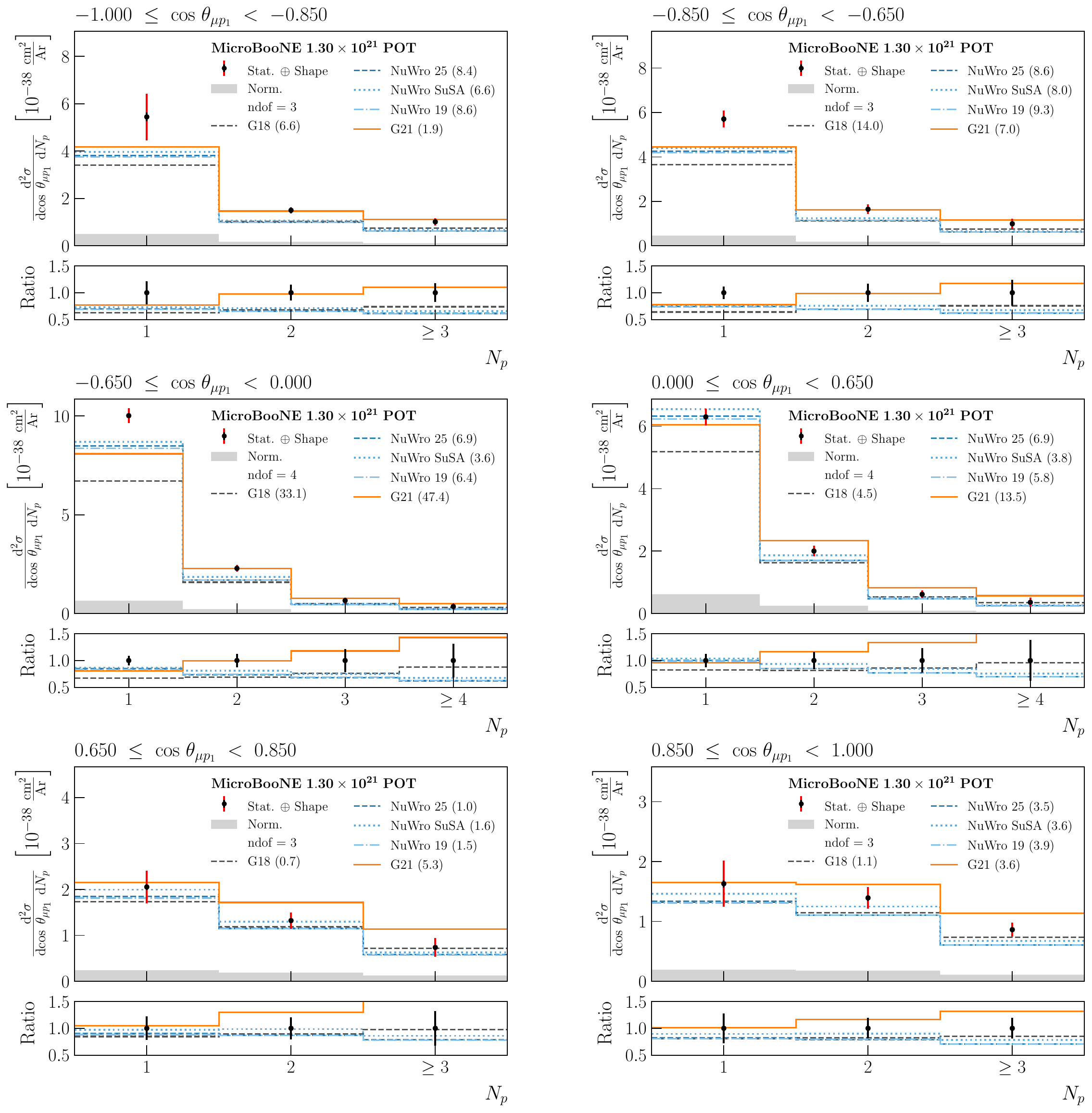}
	\caption{Flux-integrated double-differential cross sections for block 2 $(\mathrm{cos} \, \theta_{\mu p_{1}}, \, N_{p})$, extracted from the full MicroBooNE BNB dataset (black data points). Error bars indicate statistical and shape-only uncertainties, while the gray band indicates the normalization uncertainty. Predictions from the GENIE \texttt{G18\_10a\_02\_11a} and \texttt{G21\_11b\_00\_000} tunes, NuWro \texttt{v19.02.2}, NuWro \texttt{25.03.1} (default), and NuWro \texttt{25.03.1} (SuSAv2) are overlaid as indicated in the legend. The $\chi^{2}$ for each prediction is indicated in parentheses. The bottom panels show the ratio of the predictions to the unfolded data.}
	\label{fig:unfolded_xsec_other_block_2}
\end{figure*}

\begin{figure*}[!htbp]
	\centering
	\includegraphics[width=0.88\linewidth]{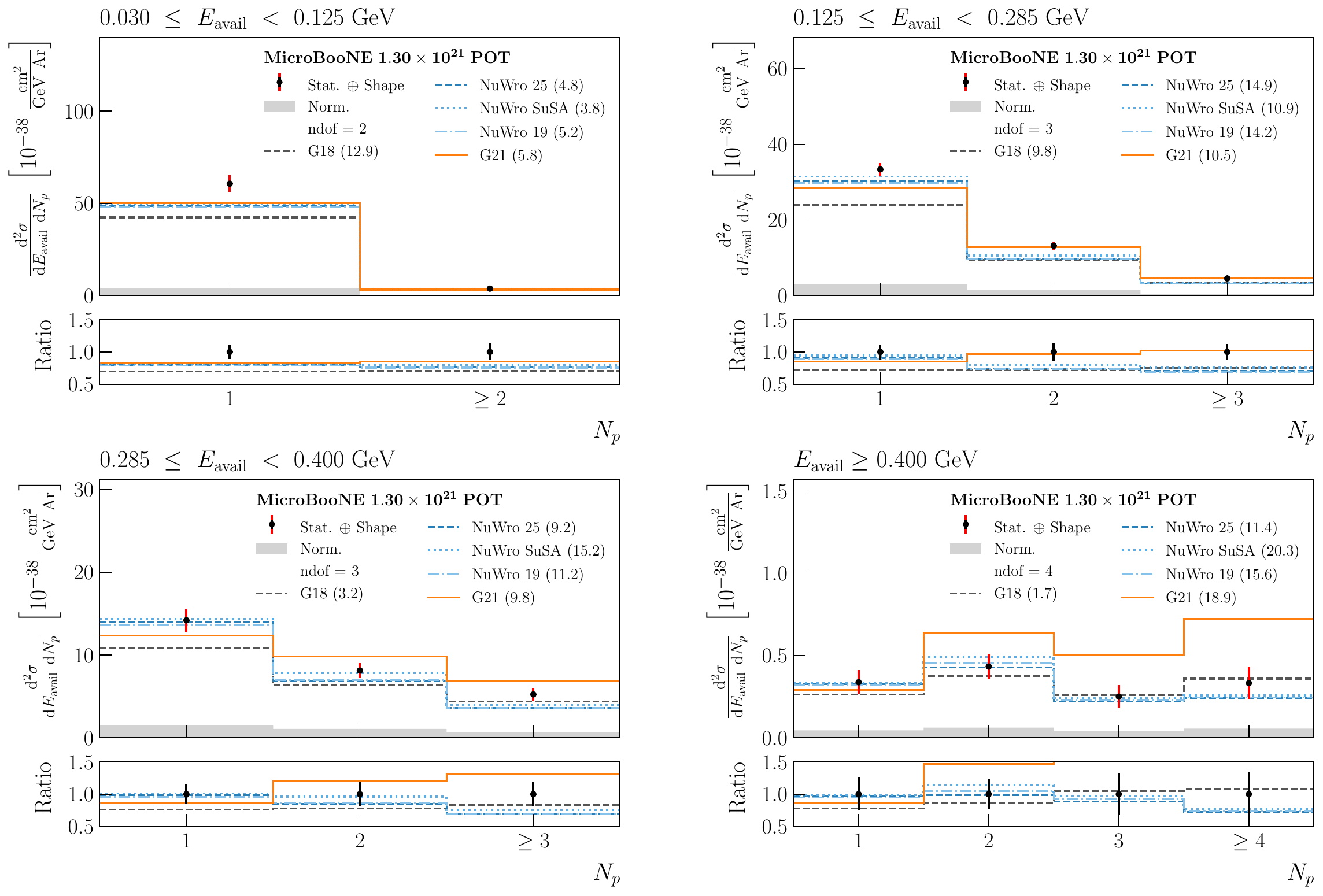}
	\caption{Flux-integrated double-differential cross sections for block 3 $(E_{\mathrm{avail}}, \, N_{p})$, extracted from the full MicroBooNE BNB dataset (black data points). Error bars indicate statistical and shape-only uncertainties, while the gray band indicates the normalization uncertainty. Predictions from the GENIE \texttt{G18\_10a\_02\_11a} and \texttt{G21\_11b\_00\_000} tunes, NuWro \texttt{v19.02.2}, NuWro \texttt{25.03.1} (default), and NuWro \texttt{25.03.1} (SuSAv2) are overlaid as indicated in the legend. The $\chi^{2}$ for each prediction is indicated in parentheses. The bottom panels show the ratio of the predictions to the unfolded data.}
	\label{fig:unfolded_xsec_other_block_3}
\end{figure*}

\begin{figure*}[!htbp]
	\centering
	\includegraphics[width=1.00\linewidth]{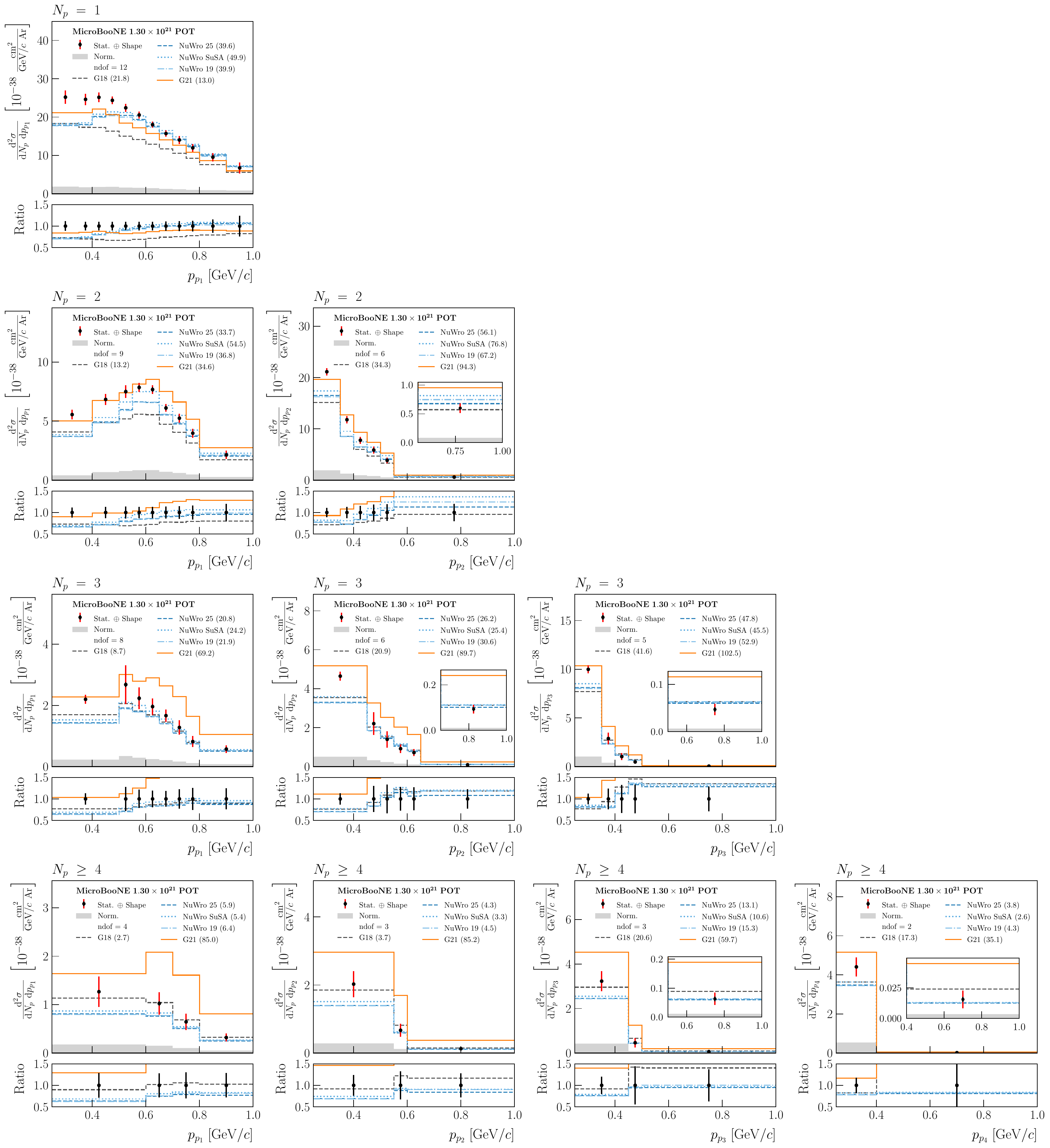}
	\caption{Flux-integrated double-differential cross sections for blocks 4, 5, 6, and 7 $(N_{p}, \, p_{p_{i}})$, extracted from the full MicroBooNE BNB dataset (black data points). Error bars indicate statistical and shape-only uncertainties, while the gray band indicates the normalization uncertainty. Predictions from the GENIE \texttt{G18\_10a\_02\_11a} and \texttt{G21\_11b\_00\_000} tunes, NuWro \texttt{v19.02.2}, NuWro \texttt{25.03.1} (default), and NuWro \texttt{25.03.1} (SuSAv2) are overlaid as indicated in the legend. The $\chi^{2}$ for each prediction is indicated in parentheses. The bottom panels show the ratio of the predictions to the unfolded data.}
	\label{fig:unfolded_xsec_other_blocks_4_5_6_7}
\end{figure*}

\begin{figure*}[!htbp]
	\centering
	\includegraphics[width=1.00\linewidth]{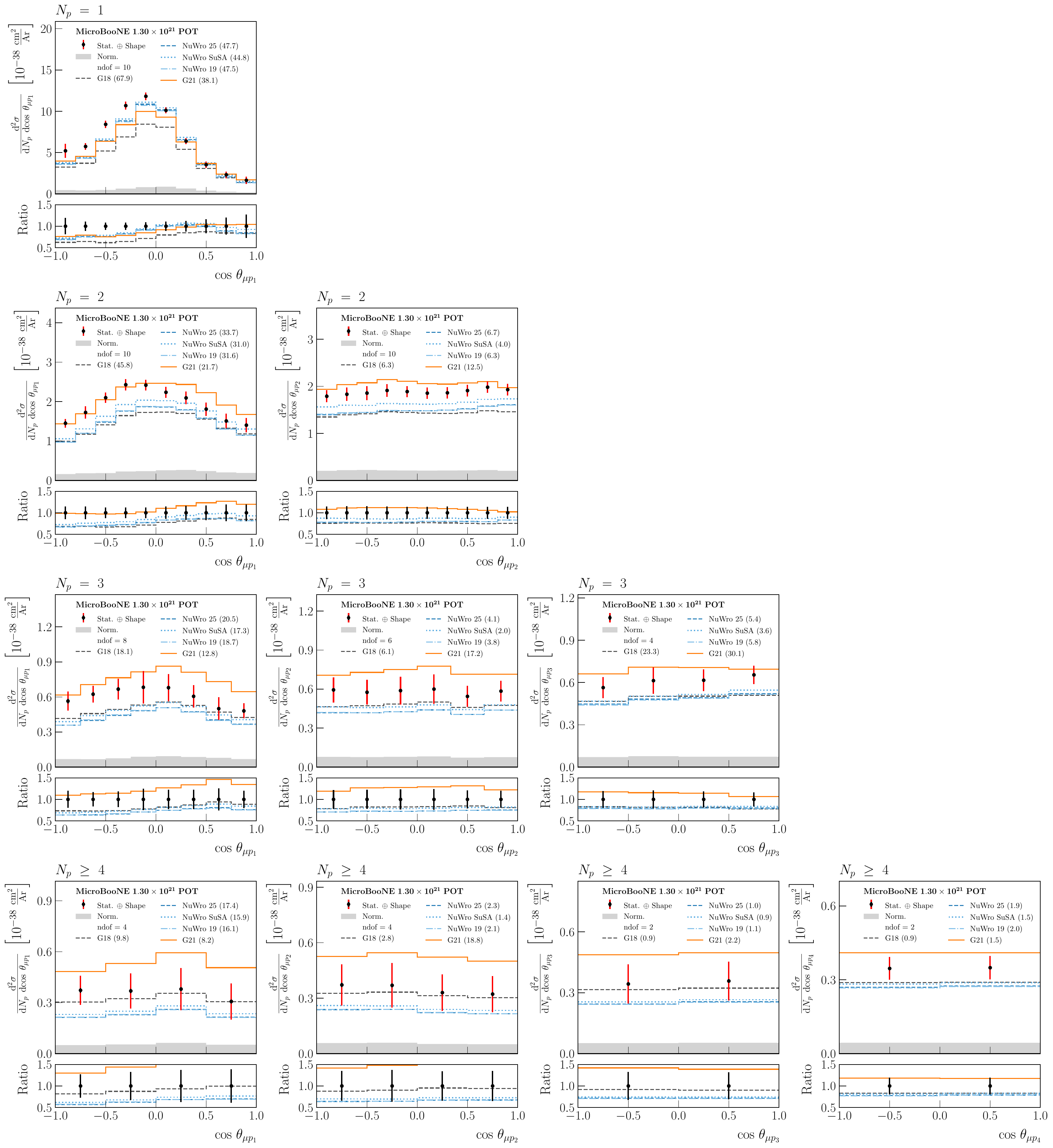}
	\caption{Flux-integrated double-differential cross sections for blocks 8, 9, 10, and 11 $(N_{p}, \, \mathrm{cos} \, \theta_{\mu p_{i}})$, extracted from the full MicroBooNE BNB dataset (black data points). Error bars indicate statistical and shape-only uncertainties, while the gray band indicates the normalization uncertainty. Predictions from the GENIE \texttt{G18\_10a\_02\_11a} and \texttt{G21\_11b\_00\_000} tunes, NuWro \texttt{v19.02.2}, NuWro \texttt{25.03.1} (default), and NuWro \texttt{25.03.1} (SuSAv2) are overlaid as indicated in the legend. The $\chi^{2}$ for each prediction is indicated in parentheses. The bottom panels show the ratio of the predictions to the unfolded data.}
	\label{fig:unfolded_xsec_other_blocks_8_9_10_11}
\end{figure*}

Figures \ref{fig:unfolded_xsec_other_block_1} through \ref{fig:unfolded_xsec_other_blocks_8_9_10_11} show the unfolded cross section results compared to these configurations. Table \ref{tab:unfold_data_summary_others} shows the per-block and global $\chi^{2}$ values for these generator comparisons.

\begin{table*}[!htbp]
    \caption{Summary of the $\chi^{2}$ values obtained for each bin block and globally for the GENIE \texttt{G18\_10a\_02\_11a} and \texttt{G21\_11b\_00\_000} tunes, NuWro \texttt{v19.02.2}, and NuWro \texttt{25.03.1} (default and SuSAv2 configurations) predictions compared to the unfolded cross section data.}
    \begin{center}
        \begin{small}
            \begin{tabular}{lcccccc}
                \multirow{2}{9em}{Block}                             & \multirow{2}{3em}{\centering ndof} & \multirow{2}{9em}{\centering \texttt{G18\_10a\_02\_11a}} &  \multirow{2}{9em}{\centering \texttt{G21\_11b\_00\_000}} &  \multirow{2}{9em}{\centering NuWro \texttt{v19.02.2}} &      \multicolumn{2}{c}{\parbox{9em}{\centering NuWro \texttt{v25.03.1}}} \\
                                                                     &                                    &                                                          &                                                           &                                                        &   \parbox{4.5em}{\centering Default} &  \parbox{4.5em}{\centering SuSAv2} \\[1mm] \hline
                \strutlike 1: $(p_{p_{1}}, \, N_{p})$                &                                 27 &                                                   100.06 &                                                    445.93 &                                                  97.08 &                                98.32 &                             129.68 \\[1.5mm]
                2: $(\mathrm{cos} \, \theta_{\mu p_{1}}, \, N_{p})$  &                                 20 &                                                    98.24 &                                                     92.08 &                                                  66.53 &                                70.05 &                              61.96 \\[1.5mm]
                3: $(E_{\mathrm{avail}}, \, N_{p})$                           &                                 12 &                                                    58.09 &                                                    289.64 &                                                  48.80 &                                40.69 &                              56.99 \\[1.5mm]
                4: $(N_{p}, \, p_{p_{1}})$                           &                                 33 &                                                    95.40 &                                                    484.38 &                                                  93.63 &                                92.92 &                             129.22 \\[1.5mm]
                5: $(N_{p}, \, p_{p_{2}})$                           &                                 15 &                                                    52.43 &                                                    263.94 &                                                 112.42 &                                95.88 &                             129.72 \\[1.5mm]
                6: $(N_{p}, \, p_{p_{3}})$                           &                                  8 &                                                    43.78 &                                                    113.67 &                                                  65.79 &                                60.05 &                              59.64 \\[1.5mm]
                7: $(N_{p}, \, p_{p_{4}})$                           &                                  2 &                                                    21.69 &                                                     36.27 &                                                   4.57 &                                 4.03 &                               2.56 \\[1.5mm]
                8: $(N_{p}, \, \mathrm{cos} \, \theta_{\mu p_{1}})$  &                                 32 &                                                   109.31 &                                                    105.67 &                                                  88.90 &                                97.76 &                              83.29 \\[1.5mm]
                9: $(N_{p}, \, \mathrm{cos} \, \theta_{\mu p_{2}})$  &                                 20 &                                                    38.74 &                                                     93.08 &                                                  11.42 &                                11.89 &                               9.80 \\[1.5mm]
                10: $(N_{p}, \, \mathrm{cos} \, \theta_{\mu p_{3}})$ &                                  6 &                                                    29.14 &                                                     40.64 &                                                   6.27 &                                 5.80 &                               4.39 \\[1.5mm]
                11: $(N_{p}, \, \mathrm{cos} \, \theta_{\mu p_{4}})$ &                                  2 &                                                    10.46 &                                                      6.69 &                                                   2.71 &                                 2.47 &                               1.61 \\[1.5mm] \hline
                \strutlike Global                                    &                                145 &                                                    516.1 &                                                    1582.0 &                                                  828.5 &                                817.7 &                              919.4 
            \end{tabular} 
        \end{small}
    \end{center}
    \label{tab:unfold_data_summary_others}
\end{table*}

\FloatBarrier

\section{Uncertainty breakdown}

\begin{figure*}[!htbp]
	\centering
	\includegraphics[width=0.85\linewidth]{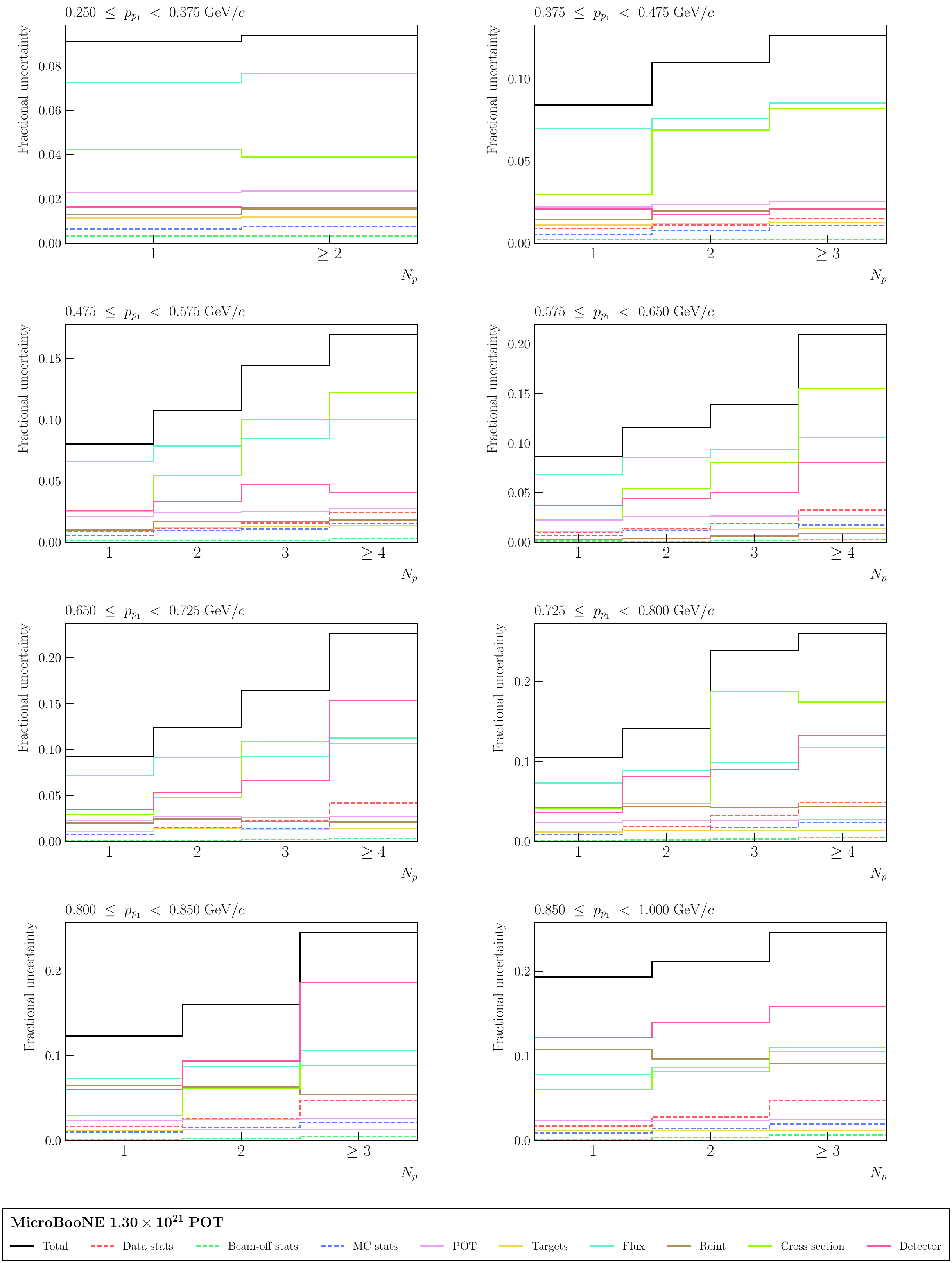}
	\caption{Fractional uncertainties on the unfolded cross section for block 1, corresponding to the double-differential measurement $(p_{p_{1}}, \, N_{p})$.}
	\label{fig:unfolded_xsec_unc_block_1}
\end{figure*}

\begin{figure*}[!htbp]
	\centering
	\includegraphics[width=0.85\linewidth]{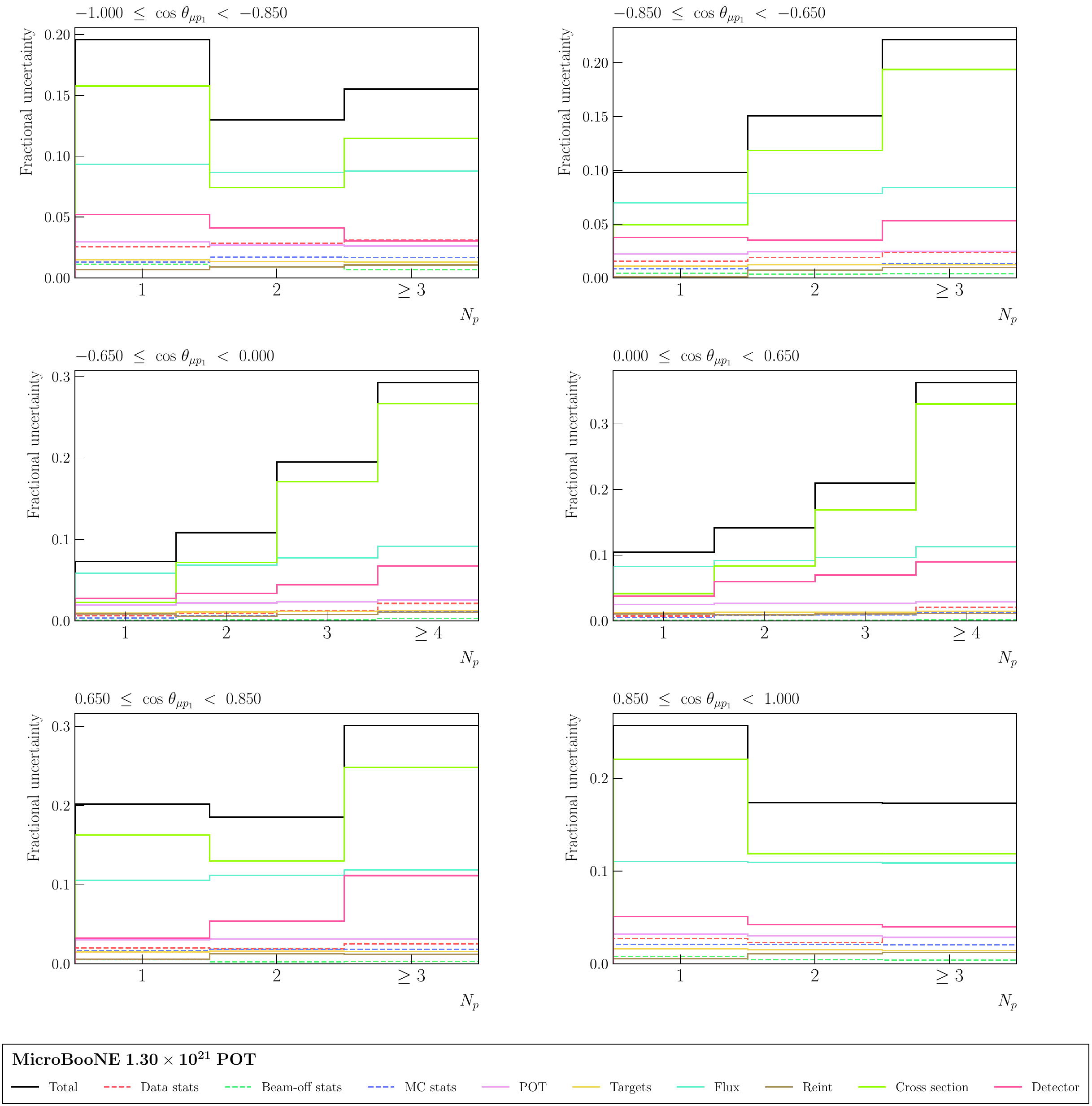}
	\caption{Fractional uncertainties on the unfolded cross section for block 2, corresponding to the double-differential measurement $(\mathrm{cos} \, \theta_{\mu p_{1}}, \, N_{p})$.}
	\label{fig:unfolded_xsec_unc_block_2}
\end{figure*}

\begin{figure*}[!htbp]
	\centering
	\includegraphics[width=0.85\linewidth]{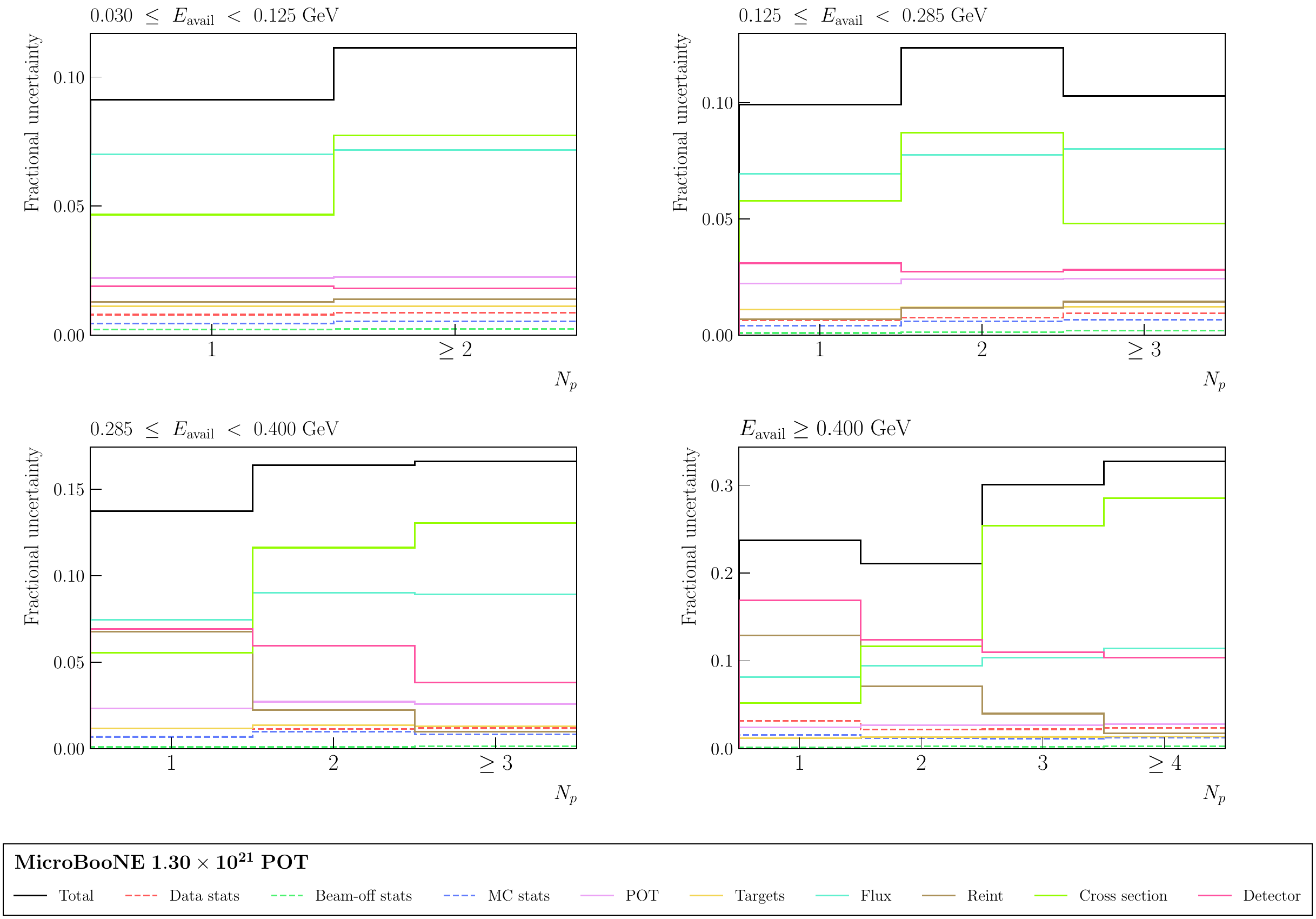}
	\caption{Fractional uncertainties on the unfolded cross section for block 3, corresponding to the double-differential measurement $(E_{\mathrm{avail}}, \, N_{p})$.}
	\label{fig:unfolded_xsec_unc_block_3}
\end{figure*}

\begin{figure*}[!htbp]
	\centering
	\includegraphics[width=1.00\linewidth]{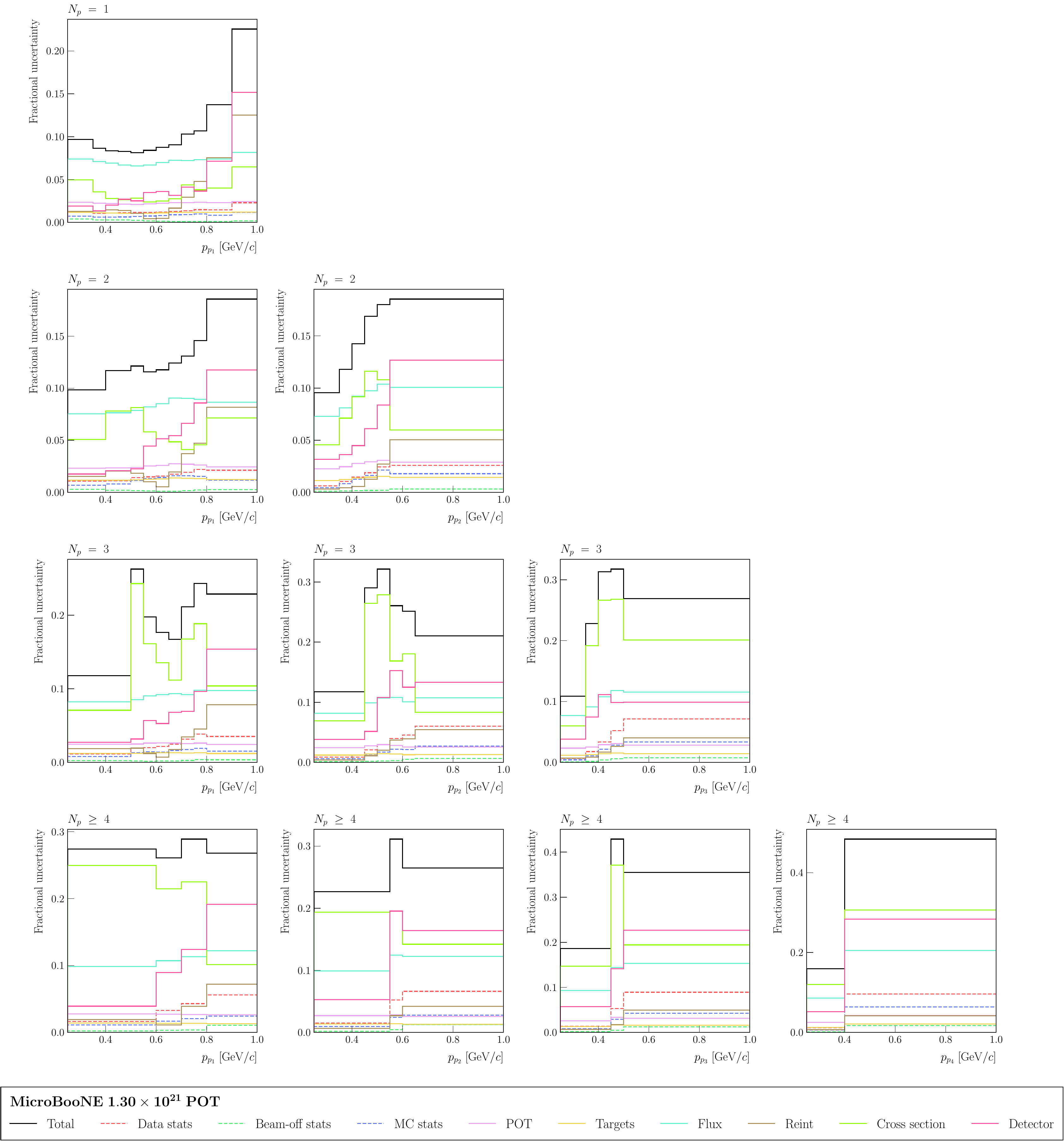}
	\caption{Fractional uncertainties on the unfolded cross section for blocks 4, 5, 6, and 7, corresponding to the double-differential measurements $(N_{p}, \, p_{p_{i}})$.}
	\label{fig:unfolded_xsec_unc_blocks_4_5_6_7}
\end{figure*}

\begin{figure*}[!htbp]
	\centering
	\includegraphics[width=1.00\linewidth]{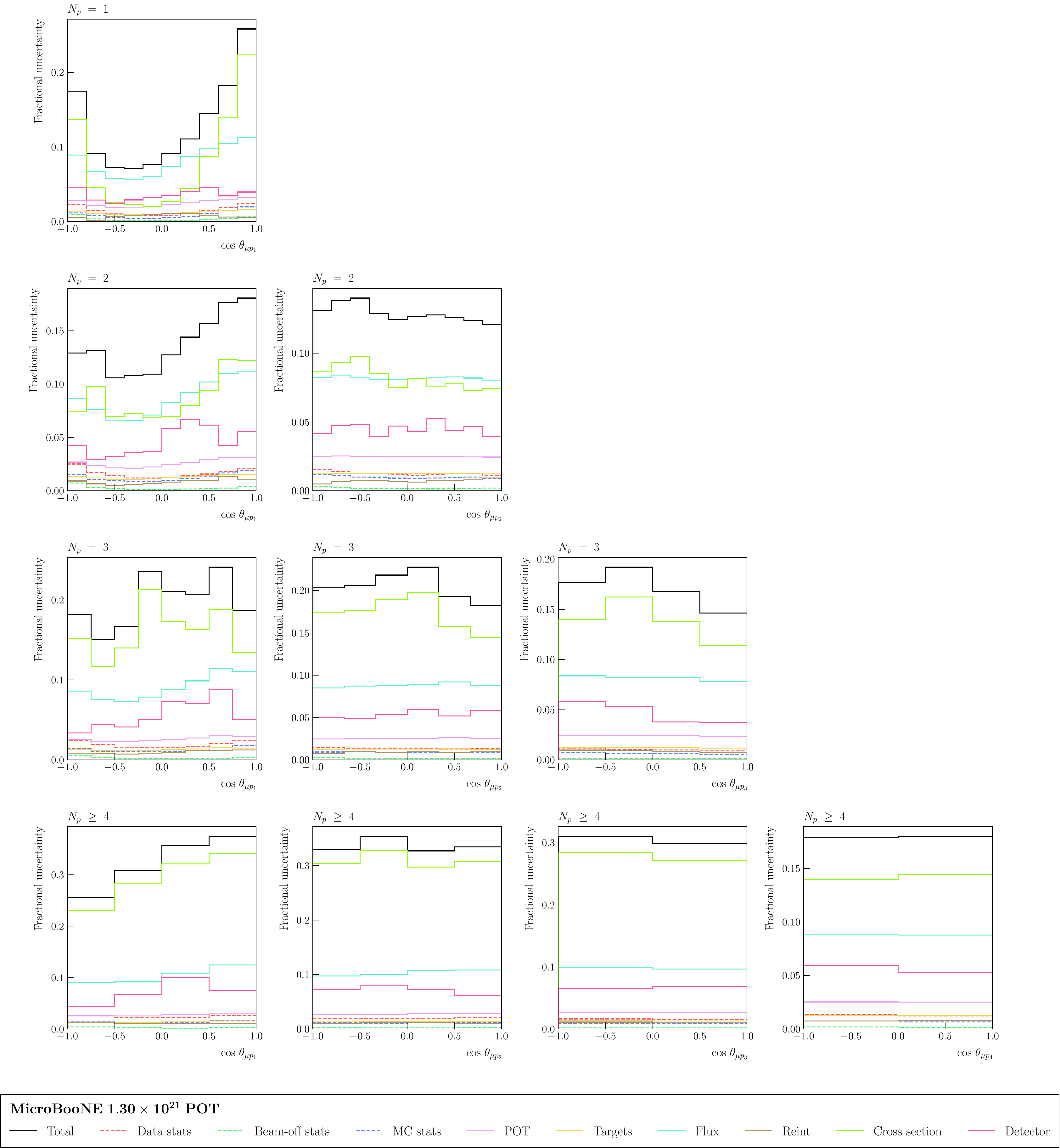}
	\caption{Fractional uncertainties on the unfolded cross section for blocks 8, 9, 10, and 11, corresponding to the double-differential measurements $(N_{p}, \, \mathrm{cos} \, \theta_{\mu p_{i}})$.}
	\label{fig:unfolded_xsec_unc_blocks_8_9_10_11}
\end{figure*}

Figures \ref{fig:unfolded_xsec_unc_block_1} through \ref{fig:unfolded_xsec_unc_blocks_8_9_10_11} show the fractional uncertainty with respect to the extracted data from each systematic source for all bin blocks in the analysis. The individual contributions shown are described in Sec. III D of the main text.

\FloatBarrier

\bibliography{refs}